\documentclass[12pt,a4paper]{article}
\usepackage[english]{babel}
\usepackage{geometry}

\usepackage[T1]{fontenc}

\usepackage{multirow}
\usepackage{lmodern}
\usepackage{amsfonts, amsmath, amssymb}
\usepackage{mathtools}
\usepackage{scalerel,stackengine}
\usepackage{dsfont}
\usepackage{mathrsfs}
\usepackage{rotating}
\usepackage{bm}
\usepackage{graphicx}
\usepackage{tabularx}
\usepackage{soul}
\usepackage{nicefrac}
\usepackage{float}
\usepackage{booktabs}
\usepackage{makecell}
\usepackage{csquotes}
\usepackage{comment}
\usepackage[flushleft]{threeparttable}
\usepackage{bbm}
\usepackage{array}
\usepackage{lscape}
\usepackage{xurl}
\usepackage{wrapfig}
\usepackage{caption}
\usepackage{mathpazo}

\usepackage[onehalfspacing]{setspace}
\DeclareUnicodeCharacter{1EF3}{\`y}
\DeclareMathOperator*{\argmax}{arg\,max}

\newcolumntype{L}[1]{>{\raggedright\let\newline\\\arraybackslash\hspace{0pt}}m{#1}}
\newcolumntype{C}[1]{>{\centering\let\newline\\\arraybackslash\hspace{0pt}}m{#1}}
\newcolumntype{R}[1]{>{\raggedleft\let\newline\\\arraybackslash\hspace{0pt}}m{#1}}
\usepackage{rotating}
\usepackage[flushleft]{threeparttable}

\usepackage{subcaption}
\usepackage{amsthm}
\usepackage{calc}
\makegapedcells
\usepackage[table]{xcolor}
\usepackage{tocvsec2}
\usepackage{xpatch}
\makeatletter
\xpatchcmd{\sv@part}{\huge \bfseries \partname \nobreakspace \thepart \par \vskip 20\p@ \fi \Huge \bfseries #2}{\fi \Huge \bfseries \thepart. #2}{}{}
\makeatother

\usepackage{afterpage}
\usepackage{titlesec}
\titlespacing{\paragraph}{%
	0pt}{
	0.5\baselineskip}{
	1em}

\titlespacing{\subsection}{%
	0pt}{
	0.4\baselineskip}{
	0.4em}

\titlespacing{\section}{%
	0pt}{
	0.5\baselineskip}{
	0.7em}

\usepackage{amsthm}
\usepackage[toc,page,header]{appendix}
\usepackage{minitoc}

\renewcommand \thepart{}
\renewcommand \partname{}

\usepackage[backend=biber, style=apa,giveninits=true,uniquename=init,natbib=true, doi=false,isbn=false,url=false, eprint=false, date=year, maxcitenames=2,uniquelist=false]{biblatex}

\AtEveryBibitem{\clearfield{note}}
\AtEveryBibitem{\clearlist{language}}
\AtEveryBibitem{\clearlist{series}}
\renewbibmacro{in:}{}

\DeclareSourcemap{
	\maps[datatype=bibtex]{
		\map{
			\step[typesource=techreport, typetarget=report]
		}
	}
}

\usepackage{tikz}
\usetikzlibrary{positioning,chains,fit,shapes,calc}
\usetikzlibrary{fit,shapes}

\usetikzlibrary{shadows.blur}  %
\tikzset{
  invisible/.style={opacity=0},
  visible on/.style={alt=#1{}{invisible}},
  alt/.code args={<#1>#2#3}{%
    \alt<#1>{\pgfkeysalso{#2}}{\pgfkeysalso{#3}} 
    } ,
}

\theoremstyle{definition}

\theoremstyle{plain}
\newtheorem{prop}{Proposition}

\newtheorem{result}{Result}

\newcommand{\bi}{\begin{itemize}}
\newcommand{\ei}{\end{itemize}}
\newcommand{\be}{\begin{enumerate}}
\newcommand{\ee}{\end{enumerate}}
\newcommand{\bd}{\begin{description}}
\newcommand{\ed}{\end{description}}
\newcommand{\beqa}{\begin{eqnarray}}
\newcommand{\eeqa}{\end{eqnarray}}
\newcommand{\beq}{\begin{equation}}
\newcommand{\eeq}{\end{equation}}

\stackMath
\newcommand\reallywidehat[1]{%
	\savestack{\tmpbox}{\stretchto{%
			\scaleto{%
				\scalerel*[\widthof{\ensuremath{#1}}]{\kern-.6pt\bigwedge\kern-.6pt}%
				{\rule[-\textheight/2]{1ex}{\textheight}}
			}{\textheight}%
		}{0.5ex}}%
	\stackon[1pt]{#1}{\tmpbox}%
}
\begin{document}

\doparttoc 
\faketableofcontents 


\selectlanguage{english}

\title{
\Large{\textsc{Profit Shifting and International Tax Reforms}}\footnote{We wish to thank James Albertus, Alan Auerbach, Pierre Boyer, Fotis Dellis, Michael Devereux, Rafael Dix-Carneiro, Peter Egger, Clemens Fuest, Philipp Kircher, Marko K\"othenb\"urger, Niels Johannesen, Isabella Manelici, Julien Martin, Philippe Martin, Nora Paulus, Javier Quintana, Nadine Riedel, Florian Scheuer, Juan Carlos Suarez-Serrato, José P. Vasquez, Gabriel Zucman and seminar participants at Banque de France, City University of London, CESifo, CREST, ECARES, ETH Zurich, European University Institute, Geneva, Honk Kong University, Kiel, Moscow Higher School of Economics, Nottingham, OECD, Paris 1, Paris School of Economics, UC Irvine, UC San Diego, UPF-CREi, Carlos III, Bank of Portugal and at the ERWIT conference, the Villars CEPR Workshop on International Trade, the Mannheim Taxation Conference, the NTA Congress, the CESIfo Area Conference on Public Economics, the 31st FIW Workshop, the Mainz ``Shaping Globalization'' Workshop, the ECARES Workshop on International Corporate Taxation, Bank of Italy ``Trade, value chains
and financial linkages in the global economy'' Conference, the NBER Business Taxation in a Federal System Conference, the NBER Summer Institute Macro Public Finance, the ``The Regulation of Multi-Firm Groups'' Workshop, and the CEPR's ``Globalisation: What's Next?'' conference for useful comments and discussions. We thank Baptiste Souillard for his contributions to the earlier stages of this project. Felix Samy Soliman provided excellent research assistance.}}

 	\author{
		Alessandro Ferrari\footnote{Department of Economics, University of Zurich \& CEPR, \textit{alessandro.ferrari@econ.uzh.ch}.}
 		\and 
 		S\'ebastien Laffitte\footnote{THEMA, CY Cergy Paris University, \textit{sebastien.laffitte@cyu.fr}.} 
 		\and 
 		Mathieu Parenti\footnote{Corresponding author. Paris School of Economics, INRAE, CESifo \& CEPR, \textit{mathieu.parenti@psemail.eu}.}
 		\and
 		Farid Toubal\footnote{Université Paris-Dauphine, CEPII, CESifo \& CEPR, \textit{farid.toubal@dauphine.psl.eu }.}
 	}
\date{
December 2024}

\maketitle	
\vspace{-1.2cm}
\abstract{

International taxation rules are outdated, allowing multinationals to shift profits to tax havens. This paper examines how tax reforms affect profit shifting and cross-country welfare. We propose a model that separates real economic profits from paper profits, introducing 'triangle identities' to estimate bilateral profit-shifting flows. Using macro- and firm-level data, paper profits' elasticity is three times that of the tax base. Global minimum tax reforms improve welfare by increasing public goods funding and reducing tax competition. We also identify optimal minimum rates under various taxing-right scenarios and demonstrate that unilateral destination-based-cash-flow-tax reforms' welfare effects depend highly on trade imbalances.


}

\newpage

\section*{}

\begin{refsection}
\onehalfspacing
\section{Introduction}
The international tax system, rooted in principles established by the League of Nations in 1928, treats multinational corporations (MNCs) as separate legal entities across different jurisdictions. 
Mounting evidence shows that MNCs exploit these outdated international tax rules to shift profits to low- or no-tax jurisdictions, and avoid taxes.
In response, international taxation is undergoing reform through the OECD/G20 Inclusive Framework on Base Erosion and Profit Shifting (\citealp{oecd_2021}). 
The success of such a reform effort fundamentally depends on how multinational corporations adjust the location of their real activities and profit shifting. Despite its policy relevance, we still lack a comprehensive framework accounting for firms' endogenous responses to international tax reforms with regards to their production location and tax-engineering choices.

In this paper, we build a general equilibrium model of international taxation to examine the influence of tax reforms on multinational corporations' decisions of production and profit allocation across countries, including tax havens.
We use our model to quantify the impact of corporate tax reforms on several outcomes, including GDP, tax revenues, profit shifting, and welfare.

We model MNCs' joint decisions on production, investment, and profit shifting as influenced by the structure of the international tax system. 
This system is characterized by two elements: the allocation of taxing rights and the level of tax rates. 
We model tax havens as jurisdictions recording shifted profits, leading to a mismatch between the location of value creation and the location of taxing rights.

Our model incorporates profit-shifting frictions (\textit{i.e.}, the cost of moving profits from a source country to a low-tax jurisdiction), which impact the location choices of MNCs, as well as common endogenous country characteristics (market potential, production costs) and bilateral trade and investment costs. 
Profit-shifting frictions are bilateral to capture variations in profit-shifting technologies, communication costs, and the compatibility of tax and legal systems between source and haven countries.
The model delivers gravity equations for bilateral real and reported (paper) profits, which we use to calibrate key tax elasticities. 
These elasticities, combined with profit-shifting frictions, determine how tax reforms affect firms' profitability and reshape the global distribution of production and profit shifting. 




Next, to quantify our model, we need three previously unavailable inputs: i) trilateral profit-shifting flows, ii) tax base and profit-shifting elasticities, and iii) profit-shifting frictions. 
We start by estimating trilateral profit-shifting flows. These shifted profits are indexed by the residence country of tax-avoiding plants, the source country where they produce, and the tax haven where profits are eventually booked. 
To estimate profit-shifting flows, we introduce a novel methodology that uses data on bilateral FDI income, corrected for conduit FDI to avoid double counting. 
Estimating a gravity model for FDI income, we incorporate tax havens as predictors and compute counterfactual flows in their absence. From this we obtain bilateral profit-shifting flows between tax havens and residence countries. Next, through a set of model-implied relations, which we refer to as ``triangle identities'', between residence countries, source countries, and tax havens, we can compute profit-shifting flows from source countries to tax havens. 
The intuition for these identities is simple: the total profits shifted from a headquarter country to a tax haven must consistently match the sum of the profits that each its production plants in any source country books in the tax haven. 
This accounting constraint, combined with structural equations defining the level of multinational production in each source country, allows us to identify trilateral profit-shifting flows.
We find that profit shifting amounts to \$350 billion in 2017, representing 33\% of multinational profits. 
Moreover, the allocation of profit-shifting flows shows the importance of geography: tax havens are more likely to host profits from nearby source countries.

Next, we estimate the two key tax elasticities from the model's gravity equations: one for the (reported) tax base and one for profit shifting.
Using firm-level data from Orbis Historical, we estimate these elasticities with appropriate controls and fixed effects, leveraging within-firm variation.
We find that the elasticity of profit shifting is about three times larger than that of the tax base. 
 To validate our approach, we conduct a series of robustness checks with respect to profit shifting flows and their elasticity to taxes using alternative datasets such as OECD's Country-by-country reports, profit shifting estimations from \citet{torslov_missing_2022} (TWZ), \citet{cortax2016}, or \citet{tjn_2020}, and bilateral service trade from the OECD-WTO's BATIS database, all of which converge consistently.

Finally, we estimate profit-shifting frictions and find them to be substantial: moving profits from a residence country to a tax haven through a source country raises production costs by an average of 12\%.\footnote{Anecdotal evidence underscores these profit-shifting costs. For example, the U.S. Permanent Subcommittee on Investigations reported that Caterpillar paid over \$10 million annually to PwC for Swiss tax planning (\citealp{levin_caterpillars_2014}, p.42). 
These consulting fees represent set-up costs, not the full cost of tax planning.} These estimates underscore the importance of our modeling of the endogenous selection into profit shifting:  despite these frictions being substantial, we observe profit-shifting firms. This is because shifting-firms are not random: they are the ones with the best-shifting technology and they are the largest, consistently with the empirical evidence (\citealp{davies_knocking_2018}, \citealp{bilicka_comparing_2019}, \citealp{wier_2020jpube}, \citealp{wier_dominant_2023}).
The profit-shifting frictions include a bilateral component based on the source-haven pair and a unilateral component reflecting the residence country's capacity to reduce shifting costs.\footnote{Our findings align with literature showing that the U.S. and some European countries better minimize profit-shifting costs, highlighting the ``aggressiveness" of their firms (\citealp{garcia2021multinational}; \citealp{torslov_missing_2022}).} 
We find that our novel bilateral component accounts for 39\% of the variation in profit-shifting costs.

 
In the last part of the paper, we conduct counterfactual policy experiments to evaluate the impact of key international tax reforms.
First, we assess the effects of minimum taxation, examining both unilateral implementation and the 15\% global minimum tax (GMT), in line with the OECD’s Pillar Two, adopted by several countries from 2024.
We then explore the implications of an alternative design that may be more effective at curbing profit-shifting: Destination-Based Cash Flow Taxation (DBCFT). This approach shifts taxing rights to destination countries and offers the advantage of potential unilateral implementation.

The effects of introducing a global minimum tax hinge on the allocation of taxing rights.
We simulate three scenarios: assigning taxing rights to the residence country (OECD's \textit{Income Inclusion Rule}), to the source country (OECD's \textit{Under-Taxed Profit Rule}), and to the country where profits are booked (OECD's \textit{Qualified Domestic Minimum Top-up Tax}). 
While the global tax rate distribution remains unchanged, the country that collects the tax differs in each case.
In all cases, the minimum tax raises revenues from MNCs through two channels: reduced profit shifting, which expands the domestic tax base, and direct collection by the country to which taxing rights are assigned. 
For the U.S., residence- and source-based minimum taxes increase corporate tax revenues by about 4\%. 
Although responses from tax havens -- such as matching the minimum rate -- can halve these gains, the reform still reduces the incentives for profit shifting, ultimately strengthening the tax base of non-haven countries.\footnote{Recent announcements from the Irish government (\citealp{irish_department_of_finance_consultation_2022}) and the Government of Bermuda (\citealp{government_of_bermuda_2023}) and the adoption of the QDMTT by many tax havens suggest this response is already underway.}

The welfare impact of the reform balances increased tax revenues with potential effects on private consumption. Higher revenues support public goods, but higher effective tax rates can lead firms to relocate or exit, affecting private consumption. Anti-inversion laws can also mitigate these relocation pressures. 

We conclude our analysis of global minimum tax reforms with three additional results. First, we study the distributional effects of the reform. We find that when havens adjust their tax rates, the global gains are small and very heterogeneous. In particular, tax havens stand to obtain substantial gains, while non-haven countries have more muted or negative effects. Second, we ask what is the globally optimal minimum tax rate. We find that it lies between 17\% and 40\%, depending on the allocation of taxing rights, with positive welfare gains in all cases for rates up to 24\%. Finally, we note that the implementation of a global minimum tax reduces tax competition by stabilizing tax bases, prompting countries to raise their tax rates and incentivizing firms to choose locations based on economic fundamentals, such as market potential and wages, rather than tax-rate differentials.

This exercise reveals a trade-off between efficiency (higher real income) and profit-shifting reduction. Is it possible to find an alternative tax system that can eliminate profit shifting without compromising efficiency? 
To answer this question, we turn to DBCFT as it received attention both in policy circles and in the economic discussion for its theoretical and practical features (\citealp{auerbach2017destination}, \citealp{barbiero2019macroeconomics}, \citealp{costinot2019lerner}, \citealp{becker_unilateral_2020}, \citealp{devereux_taxing_2021}). In particular, its implementation, even unilaterally, should drastically limit tax avoidance as it is done in the current international tax system (\citealp{auerbach_international_2017}). 
Practically, the reform replaces the corporate income tax with a border adjustment tax (BAT). Since firms' tax burden becomes entirely determined by the location of their sales, most standard profit-shifting strategies become ineffective. 
Our quantitative analysis of the unilateral implementation of DBCFT shows highly heterogeneous effects across countries. On the one hand, the elimination of shifted incomes induces an efficiency gain as firms locate across space based on productivity considerations. On the other hand, the effect on tax revenues and public good provision is ex-ante ambiguous. The elimination of CIT and its replacement with a BAT implies that countries with large trade deficits may gain while countries with large trade surpluses lose. To exemplify this point, if the U.S. were to implement DBCFT at a 40\% CIT-equivalent rate (pure BAT), it would experience a 1.8\% increase in welfare, thanks mostly to the revenue gains driven by its large trade deficit. Conversely, if Japan were to implement it, it would lose 2.4\% of welfare due to the significant public good consumption loss driven by its large trade surplus. If the U.S. and Japan implemented their optimal DBCFT rate (33.3\% and 25\% CIT-equivalent), they would obtain a 2.1\% and -2.2\% change in welfare, respectively. Depending on the initial trade balance, the optimal level of DBCFT can induce significant welfare gains or losses.


\paragraph{Related Literature.}
First, our paper is related to the literature estimating profit shifting. To the best of our knowledge, \citet{torslov_missing_2022} is the only paper that provides estimates of bilateral profit-shifting flows for several country pairs.\footnote{A large literature focuses on the profit shifting of U.S. multinational firms (\citealp{hines_fiscal_1994}, \citealp{UNCTAD_world_2015}, \citealp{clausing_effect_2016, clausing_2020},  \citealp{wright_exorbitant_2018}, \citealp{Laffitte_LTPS_2020}, \citealp{guvenenetal_2022}, \citealp{blouin_double_2021}), or provide estimates at a global scale (\citealp{Jansky2019}, \citealp{garcia2021multinational}, \citealp{vicard_profit_2022}). Note that \citet{guvenenetal_2022} estimates bilateral profit shifting to several tax havens, but only when the U.S. is the source country.}
Their methodology infers profit shifting by comparing the profitability of domestic and multinational firms in tax havens. 
This country-level profit premium of MNCs, representing profit shifting, is then allocated to country pairs using mainly bilateral excess trade in services between source countries and tax havens. 
Instead, we need to compute trilateral profit shifting flows for our calibration. To do so, we rely on bilateral excess FDI income and a model-consistent allocation of profits shaped by the tax base and profit-shifting elasticities, to recover trilateral flows. 
Our methodology allows us to remain agnostic on the channel through which profit shifting occurs.
\footnote{Profit shifting can occur through various channels, including IP-related mechanisms (\citealp{santacreu_international_forthcoming}, \citealp{deng2023local}, \citealp{dyrda2023macroeconomic}), internal debt (\citealp{buettner_internal_2013}), mispricing of goods (\citealp{davies_knocking_2018}), or services (\citealp{hebous_2021}). All these mechanisms are compatible with our model but cannot individually account for the global scale of profit shifting.}

Second, our research contributes to empirical studies on the real effects of corporate tax reforms, extending the analysis beyond changes in tax revenues (see \citealp{suarez-serrato_who_2016} and \citealp{fuest_higher_2018} in the domestic context, and \citealp{hines_fiscal_1994}, \citealp{hines_lessons_1999}, \citealp{grubert_effect_1998}, \citealp{egger_impact_2015}, \citealp{serrato_2018}, \citealp{demooij_tp_2020, demooij_thin_2021},  \citealp{bilicka_real_2021} in the international context).
In line with the results from this literature, our model allows firms to adjust their location and profit-shifting strategy after a tax reform. We also show that both of these margins of adjustment quantitatively matter to estimate reforms' impact on tax revenues and welfare.

We also contribute to the burgeoning literature that evaluates international tax reforms (\citealp{hanappi2020impact}). 
The reforms of international taxation and their potential impacts are discussed, for instance, in \citet{fuest2019international}, \citet{imf_corporate_2019}, \citet{clausing_ending_2021}, and \citet{devereux_taxing_2021}.
Most of the literature evaluates the so-called Pillar II \textit{i.e.} the introduction of minimum taxation. 
\citet{assessment_oecd_2020} and \citet{collecting_eutax_2021} propose estimations of the expected tax revenue gains from implementing Pillar II.\footnote{In addition to these revenue estimations, \citet{bachas_effective_2023} explores the impact of Pillar II on developing countries, and \citet{bilicka_tax-avoidance_2023} discusses the effect of Pillar II on IP location incentives.}
None of these contributions allow for real and profit-shifting responses from multinational firms nor general equilibrium effects.
Moreover, they focus exclusively on tax revenues and do not consider country-level welfare or the worldwide optimality of the reforms. 
Our model also allows us to quantify the impact of these reforms on welfare and on the incentive for countries to adjust their tax rate post-reform. To the best of our knowledge, this is also the first effort to benchmark the current reform against the DBCFT proposal. We find that a unilateral DBCFT reform can generate substantial welfare effects.  
Our results point to the quantitative importance of violations of Lerner symmetry (\citealp{costinot2019lerner}), driven by the presence of multinationals.


On the theoretical side, the mechanisms at play are reminiscent of the papers by \citet{janeba_global_2022}, \citet{johannesen_global_2022}, and \citet{hebous_pareto_2022} who build tax competition models to investigate the impact of minimum taxation. 
In line with these models, we take into account the potential policy reaction of tax havens to the global minimum tax. In our simulations, we find that a global minimum tax is welfare-improving for the majority of non-haven countries.  Finally, we study the incentive of governments to change taxes after the implementation of minimum taxation. We find that the majority of them would gain by increasing their taxes.


Finally, this paper borrows tools from the quantitative trade and economic geography literature. 
We build our model from a multi-country Krugman-type model à la \citet{head_market_2004} augmented with multinational firms and profit shifting.\footnote{The patterns of trade and multinational production have received substantial attention (\citealp{arkolakis2018innovation}, \citealp{head2019brands}) with applications to corporate tax reforms (\citealp{wang2020multinational}, \citealp{santacreu_international_forthcoming},  \citealp{dyrda2023macroeconomic}). The importance of geography for corporate taxation is highlighted in the work of \citet{fajgelbaum_state_2019} in the domestic context.}  
Methodologically, however, the calibration of our model requires an estimation of worldwide trilateral profit-shifting flows and the elasticity of paper profits. The main contribution of our paper in this regard is to provide a model-consistent estimation of these profit-shifting flows as well as an estimate of paper-profit elasticity that we find to be three times as large as the elasticity of the reported tax base.
We also add two important features. First, we add a public good whose contribution to welfare is disciplined by the data. This feature allows a trade-off for welfare between public and private consumption as in \citet{johannesen_global_2022}. Under minimum taxation, we find that an increase in public good provision dominates the reduction in consumption. Second, we introduce a set of tax havens whose geography is embedded in bilateral profit-shifting frictions. Unlike existing approaches, this allows us to endogenize the intensity of profit-shifting: the share of shifted profits depends not only on the opportunity cost of engaging in profit-shifting but also on the cost of shifting to any other tax haven, conditional upon being a tax avoider





\section{Conceptual Framework and Empirical Strategy:\\ \mbox{ an Overview}}\label{overview}

We begin by illustrating the world of multinational profit flows and profit shifting in Figure \ref{fig:PSnetwork}. The figure captures the key relationships central to our analysis: multinational firms generate sales and profits in source countries (solid arrows), shift profits to tax havens to reduce tax liabilities (dashed arrows), and either retain the shifted profits in the tax havens or repatriate them to residence countries (dotted arrows). This framework forms the basis of our analysis of profit shifting patterns and their relationship to real economic activity.
Understanding the relationship between shifted profits and real economic activity is crucial for analyzing tax reform effects, and this relationship depends on the elasticity of profits to tax rates. While \citet{hines_fiscal_1994}'s widely-used approach estimates these elasticities, it assumes a simple bilateral relationship between source countries and tax havens, overlooking the potential for profit shifting across multiple havens. This limitation has important implications for analyzing tax reforms that target complex, multi-jurisdictional tax avoidance. In Section \ref{model}, we explicitly model the substitutability between tax havens, enabling us to estimate both the elasticity of paper profits and tax avoiders' responses to reforms.
A key challenge in estimating profit shifting flows and their associated elasticities is the lack of direct observational data on cross-border profit shifting. In Section \ref{secPSestimate}, we address this by developing model-consistent "triangle identities" between residence, source, and tax haven countries to reconstruct trilateral profit-shifting patterns. Our approach combines estimates of profits shifted to tax havens from each residence country (dotted arrows) with data on tax avoiders' investments mapped to observed multinational production flows between residence and source countries (solid arrows). We then derive the bilateral profit shifting flows from source countries to tax havens (dashed arrows) using these triangle identities.
Section \ref{sec:empirics_results} presents our estimates of real profit elasticity, paper profit elasticity, and the model-implied profit-shifting frictions. The paper concludes in Section \ref{sec_counter} with an analysis of how various international taxation reforms affect profit shifting, production, and global welfare.

\begin{figure}[htbp]
\makebox[\textwidth][c]{

\begin{tikzpicture}[
    node distance=1.5cm,
    box/.style={
        draw=gray!60,
        rounded corners=3pt,
        minimum width=2.5cm,
        minimum height=0.8cm,
        blur shadow={shadow blur steps=5},
        thick
    },every node/.style={font=\small},
    legend/.style={font=\normalsize},
    group1/.style={
        box,
        fill=orange!15,
        text=black
    },
    group2/.style={
        box,
        fill=blue!15,
        text=black
    },
    group3/.style={
        box,
        fill=green!15,
        text=black
    },
    solid_arrow/.style={
        ->,
        >=latex,
        thick,
        draw=black!70
    },
    dashed_arrow/.style={
        ->,
        >=latex,
        thick,
        draw=black!70,
        dashed,
        dash pattern=on 4pt off 2pt
    },
    dotted_arrow/.style={
        ->,
        >=latex,
        thick,
        draw=black!70,
        dotted,
        dash pattern=on 2pt off 2pt
    }
]

\node[group1] (RC1) at (-5,8) {Residence Country 1};
\node[group1] (RC2) at (-7,6) {Residence Country 2};
\node[group1] (RC3) at (-9,4) {Residence Country 3};

\node[group2] (SC1) at (1,8) {Source Country 1};
\node[group2] (SC2) at (3,6) {Source Country 2};
\node[group2] (SC3) at (5,4) {Source Country 3};

\node[group3] (TH1) at (-2,3) {Tax Haven 1};
\node[group3] (TH2) at (-2,-0.3) {Tax Haven 2};

\foreach \i in {1,2,3} {
    \foreach \j in {1,2,3} {
        \draw[solid_arrow] (RC\i) -- (SC\j);
    }
}

\foreach \i in {1,2,3} {
    \foreach \j in {1,2} {
        \draw[dashed_arrow] (SC\i) -- (TH\j);
    }
}

\foreach \i in {1,2} {
    \foreach \j in {1,2,3} {
        \draw[dotted_arrow] (TH\i) -- (RC\j);
    }
}

\draw[gray!60, rounded corners=3pt] (-10.2,2) rectangle (-6.3,0);  
\path (-9,1.7) node (leg1) [right] {\small MP sales};
\draw[solid_arrow] (-10,1.7) -- (-9.2,1.7);
\path (-9,1.0) node (leg2) [right] {\small Profit Shifting};
\draw[dashed_arrow] (-10,1.0) -- (-9.2,1.0);
\path (-9,0.3) node (leg3) [right] {\small Repatriation};
\draw[dotted_arrow] (-10,0.3) -- (-9.2,0.3);

\end{tikzpicture}
}

\medskip
\caption{\small{Interactions between residence countries, source countries and tax havens.}}\label{fig:PSnetwork}

\caption*{\footnotesize Note: This figure represents countries involved in the international tax system and the corresponding financial flows. Firms headquartered in one country invest in source countries, where profits are generated. A share of these profits is subsequently shifted to tax havens.}
\end{figure}





\newpage \section{Model} \label{model}

This section describes the model we use to guide our calibration of profit-shifting flows, their response to tax reforms, and for our counterfactual analysis. 

\subsection{Set-up}

\paragraph{Structure of the Model.} 
The world economy comprises $n=1,\hdots,N$ countries, among which $h=1,\hdots,H$ are labeled ``tax havens''. 
Each country is endowed with labor, the unique factor of production. 
The $L_{n}$ workers are immobile across countries.
Each worker inelastically supplies one unit of labor paid $w_{n}$.
An endogenous number of corporations $\mathcal{N}_i$ with tax residence in country $i$ operate under monopolistic competition. Each corporation is a collection of $M$ affiliates where $M$ is a random variable and each affiliate $m$ designs and produces a single variety in a source-country $l$. This variety can be sold in any country $n$. The profits occurring to each affiliate are booked either where it produces (territorial taxation) or in a tax-haven $h$ through profit-shifting.  

\paragraph{Demand and Pricing.}
The set of varieties supplied in country $n$ is $\Omega_n$.  The demand for any variety in $\Omega_n$ at price $p_{n}$ is given by $d_{n}(p_{n})=Y_{n}\frac{p_{n}^{-\sigma}}{P_{n}^{1-\sigma}}$. The price-elasticity of demand is $\sigma>1$; $Y_{n}$ denotes total expenditures; $P_n$ is the price index given by
$P_{n}=\left(\int_{\Omega_n}p_{n}(\omega)^{1-\sigma}d\omega\right)^{\frac{1}{1-\sigma}}.
$ Real expenditure is given by $Y_n/P_n$.  Monopolistic competition and CES preferences jointly imply that the profit-maximizing markup equals $\frac{\sigma}{\sigma-1}$ and is independent of the destination market.

\paragraph{Welfare.} We define the welfare of country $n$ as: $$ U_n=(B_n/P_n)^{\beta_n} Y_n/P_n,$$ where $B_n$ are nominal tax revenues that are used to finance a public good and $\beta_n\geq 0$ a country-specific preference parameter.

\paragraph{Frictions and Taxation.}
Consider a plant producing in $l$ belonging to a firm headquartered in $i$.
When the source country $l$ and the residence country $i$ differ, the cost to produce abroad involves a friction $\gamma_{il}\geq1$, which reflects a technology transfer from the headquarter. 
Serving foreign destination markets $n\neq l$ comes with trade frictions $\tau_{ln}\geq 1$ for iceberg transport costs. 
Neither producing nor serving the destination market $n$ requires the payment of a fixed cost.  
Therefore, plants serve all markets and $\Omega_n\equiv \Omega$. 
The geography of a source country $l$, its economic size, and that of its trade partners adjusted by trade frictions are summarized by the endogenous market potential of country $l$, $\Xi_{l}^{1-\sigma}=\sum_n \Xi_{ln}^{1-\sigma}=\sum_{n}\tau_{ln}^{1-\sigma}Y_{n}P_{n}^{\sigma-1}$ \citep{head_market_2004}. 
Corporations producing in a non-haven country $l$ can choose to book their profits in a tax haven $h$.  A tax haven can host and tax profits of foreign firms at the rate $t_{lh}<t_{ll}$ without requiring their physical presence, i.e., a production site. 
When shifting their profits, we assume that firms incur a bilateral cost $\alpha_{lh}$. 
There are various reasons to expect these costs to be heterogeneous across production countries or tax havens. 
For example, these costs can subsume heterogeneity across production countries $l$, e.g., different sector composition and sectoral differences in profit-shifting abilities, which we do not model. 
Similarly, they can capture differences across tax havens $h$. 
Tax havens differ in characteristics that facilitate profit shifting, like communications infrastructures or the legal technologies they offer to foreign firms (e.g., reduced incorporation time and costs, opacity and secrecy, accounting rules, and treaty network; see \citet{torslov_missing_2022} and \citet{laffitte_market_2024} who show that tax havens differ from other jurisdictions). 
Our reduced-form friction $\alpha_{lh}$ goes further by allowing these determinants to be bilateral, so the cost of shifting profits to a tax haven differs whether they stem from production that is sourced in the U.S. or, for instance, in France.
This approach is consistent with evidence on the sectoral and geographical specialization of tax havens (\citealp{garcia-bernardo_uncovering_2017}, \citealp{bilicka_geographical_2020}, and \citealp{Laffitte_LTPS_2020}).

\paragraph{Profits.}
The idiosyncratic profitability of an affiliate $m$ is subsumed in $\varphi_m$: this variable encompasses both the affiliate idiosyncratic physical productivity as well as its tax avoidance ability.
The global post-tax profits of an affiliate $m$ that belongs to a corporation domiciled in $i$, produces in $l$ and books it profits in $h$ are given by
\begin{align}
 \pi_{ilh}(\varphi_m)  =  \left(1-t_{ilh}\right)\frac{\iota_l}{\sigma }\left(\frac{\sigma}{\sigma-1}\frac{\gamma_{il}\alpha_{lh}}{\varphi_m}w_{l}\Xi_{l}\right)^{1-\sigma}. \label{eq:prof}
\end{align}
The term $\left(\frac{\sigma}{\sigma-1}\frac{\gamma_{il}\alpha_{lh}}{\varphi_m}w_{l}\Xi_{l}\right)^{1-\sigma}$ denotes the global revenues of an affiliate in the triplet $ilh$.
These revenues turn into pre-tax profits with the standard relationship that the sales-to-profit ratio is governed by the elasticity of demand $\sigma$.  This parameter simultaneously governs profitability and the curvature of demand.  Anticipating our calibration, we disentangle them by introducing a production-country-specific wedge $\iota_l \leq \sigma$ between sales and profits so that firms producing in $l$ have a sales-to-profit ratio equal to $\iota_l/\sigma$. 

We allow the tax rate $t_{ilh}$ to be trilateral. For instance, taxing rights at the origin matter when discussing ongoing reforms, e.g., the global minimum tax reform, which gives taxing rights over the tax deficits in tax havens $h$ to residence countries $i$.

The total profits of a firm  with $M$ affiliates headquartered in $i$ are given by $\pi_{i}=\sum_{m=1}^M\sum_{l,h} \pi_{ilh}(\varphi_m)$.
Importantly, we assume that each affiliate books all its profits in a single tax domicile. 
This assumption implies that at the micro level, tax-avoiding plants' profits in $l$ bunch at zero, consistent with \citet{bilicka_comparing_2019}. Aggregate bilateral profit-shifting flows then result from the aggregation of heterogeneous profit-shifting patterns across plants within firms and across firms. 

\subsection{From Micro to Macro}\label{micmac}

\paragraph{Firm Heterogeneity.} 
We parametrize the distribution of $\varphi$  to relate our model to bilateral macroeconomic flows, e.g., trade shares, multinational production shares, and profit shifting. 
Despite its simplicity, our model retains gravity patterns for both trade and multinational production flows. We leverage this minimal structure to incorporate profit-shifting flows to tax havens. 

We introduce firm heterogeneity as follows: in each residence country, firms decide whether to enter or not, i.e., to set up a headquarter in $i$ upon the payment of a sunk cost $w_{i}f_{E}$.\footnote{Sunk entry costs $f_{E}$ can be country-specific. Note that $T_i$ already absorbs such variations.} 
Each corporation draws a number of plants $M\sim \mathcal{H(M)},\, M>0 $ with expected value $\bar M$. For each affiliate, they find out how profitable they would be when locating their production facility in any country $l$ \emph{and} recording their profits in any country $h$ (where $h$ is equal to $l$ means that the firm does not shift profits abroad). Technically, each firm draws $M$ times 
 $\varphi$ independently and with replacement a matrix of $lh$. Last, the overall profitability of each plant depends on a deterministic component  $T_i$,  inherited from the residence country $i$.
A plant belonging to a corporation resident in $i$, sourcing production in $l$ and booking profits in $h$, makes post-tax profits $\pi_{ilh}(T_i \varphi_{lh})$. 

\paragraph{Parametrization.}

We consider a multivariate $\upsilon_{1}$-Fr\'echet distribution of productivities 
with scale parameters $A_{l}$ and a homogeneous correlation function $G_i(.)$ so that the $\varphi_{lh}$ draws by country $i$ are distributed according to the following c.d.f.:\\ $\mathbb{P}\left(\varphi_{11}\leq z_{11};\dots;\varphi_{lh}\leq z_{lh};\dots;\varphi_{NH}\leq z_{NH}\right)=e^{-G_i\left(A_{1}z_{11}^{-\upsilon_{1}}\dots,A_{l}z_{lh}^{-\upsilon_{1}},\dots,A_{N}z_{NH}^{-\upsilon_{1}}\right)}$  where
\begin{align*}
G_i({\bf x})=\sum_{l=1}^{N}x_{ll}+\theta_i^{-\upsilon_1} \left(\sum_{l=1}^{N}\sum_{h=1}^{H}   x_{lh}^{\frac{\upsilon_{2}}{\upsilon_{1}}}\right)^{\frac{\upsilon_{1}}{\upsilon_{2}}},
\end{align*}
with $\upsilon_2\geq \upsilon_1$ and $\bf x$ denotes a matrix with generic entry $x_{lh}$.  The function $G_i$ determines the substitutability across $lh$ pairs and, therefore, the mobility of the tax base and paper profits. That $\upsilon_2\geq \upsilon_1$ implies that paper profits - or, equivalently profits from tax-avoiding affiliates - are weakly more elastic to corporate taxes.  This is motivated by the idea that it is harder to move physical production than P.O. boxes for profit booking.  This allows for different residence countries $i$ to have different profit-shifting intensities. 
We also introduce a parameter $ \theta_i $ representing country-specific tax aggressiveness, reflecting the fact that headquarters in different tax domiciles have different abilities to engage in profit shifting.

\paragraph{Sourcing and Profit-shifting Decisions.}
After observing the $\varphi_{lh}$ draws for each of the $M$ affiliates, firms from $i$ select a unique pair $lh$ that maximizes their profits for each plant, as given by (\ref{eq:prof}) . 

The probability for an affiliate of a firm from country $i$ to locate its production in $l$ and book its profits in $h$ is:
\begin{align}\label{probafrechet}
\mathbb{P}_{ilh}=\frac{\tilde{A}_{ilh}G_{i,lh}({\bf \tilde{A}}_{i},{\bf t}_i)}{G_i({\bf \tilde{A}}_{i} , {\bf t}_i)}(1-t_{ilh})^{\frac{\upsilon_{1}}{\sigma-1}},  
\end{align}
where ${\bf {t}}_{i} =({t}_{ilh})_{1\leq l \leq N,1 \leq h \leq H}$ encompasses corporate income tax rates and other determinants of affiliates' location choices are contained in $\tilde{A}_{ilh}\coloneqq A_{l}\left(\gamma_{il}\alpha_{lh} \iota_l ^{\frac{1}{1-\sigma}}w_{l}\Xi_{l}\right)^{-\upsilon_{1}}$. 
We denote by $G_{i,lh}$ the partial derivative of $G_i$ with respect to the $lh$ term and, with a slight abuse of notation, we denote by $G_i({\bf \tilde{A}}_{i}, {\bf t_i})$ the correlation function evaluated at $\left(\tilde{A}_{ilh}(1-t_{ilh})^{\frac{\upsilon_{1}}{\sigma-1}}\right)_{l\leq N,h\leq H}$. Note that larger firms, e.g., that have more plants, are more likely to be engaged in profit shifting.  

Expression (\ref{probafrechet}) results directly from \citet{mcfadden_modelling_1978}'s discrete choice framework using Generalized Extreme Value distributions (GEV).\footnote{To obtain the above formula, note that using (\ref{eq:prof}), profits $\pi_{ilh}$ from a residence country $i$ follow a multivariate $\frac{\upsilon_{1}}{\sigma-1}$-Fr\'echet distribution with scale parameters $\tilde{A}_{ilh}(1-t_{lh})^{\frac{\upsilon_{1}}{\sigma-1}}$ and the same correlation function $G_i(.)$. }  
A schematic representation of the choices faced by MNCs in our model is provided in Appendix \ref{appendix_theory}.

\subsection{Equilibrium}
Using the properties of the GEV again, the expected post-tax profits $\bar{\pi}_i$ of an affiliate of a firm headquartered in $i$, taken across all possible pairs $lh$, are given by 
\begin{equation}\label{profitGEV}
\bar{\pi}_i=\frac{1}{\sigma T_{i}^{1-\sigma}}\left(\frac{\sigma}{\sigma-1}\right)^{1-\sigma}G_i(\tilde{{\bf A}}_{i},{\bf t}_{i})^{\frac{\sigma-1}{\upsilon_{1}}}\Gamma\left(1-\frac{\sigma-1}{\upsilon_{1}}\right).
\end{equation}
Given profits, we can build a government's tax revenue flow $B_k$. 
Denote $\mathcal M_i=\mathcal N_i \times \bar M$ the number of affiliates of firms with HQ in $i$. Then aggregate post-tax profits of firms from $i$ are $\mathcal{M}_i \bar\pi_i$.  Pre-tax profits of an affiliate are given by  $\frac{\bar\pi_i}{1-t_{ilh}}$.  Tax revenues are then given by
\begin{equation}\label{TaxRev} 
B_{k}=\sum_{i,l,h}t_{ilh}^{g_k} \mathcal{M}_{i} \mathbb{P}_{ilh} \frac{\bar{\pi}_i}{1-t_{ilh}}, 
\end{equation}
where $t^{g_k}_{ilh}$ is the tax rate relevant for country's $k$ tax authorities depending on the taxation regime in place (e.g., territorial, worldwide, with a worldwide minimum tax etc.). As an example, consider a minimum tax regime that reallocates taxing rights to country $k$, allowing it to tax worldwide profits that are i) generated by firms from $k$, ii) shifted to tax havens, and iii) taxed at a rate smaller than $t_k^{min}$. Country $k$  raise tax revenues from firms producing in $k$ and from firms headquartered in $k$ paying taxes in a tax haven with a tax rate lower than $t_k^{min}$. In this case, $B_{k}=\sum_{i}t_k \mathcal{M}_{i} \mathbb{P}_{ikk} \frac{\bar{\pi}_i}{1-t_{ikk}}+\sum_{l\neq h,h}\max \{ t_k^{min}-t_{lh},0\} \mathcal{M}_{k} \mathbb{P}_{klh} \frac{\bar{\pi}_k}{1-t_{klh}}$, where the first term describes the tax revenues generated by firms producing in $k$ and the second term by firms headquartered in $k$ booking profits in a tax haven for which the minimum tax rate binds.

Production in country $l$ aggregates multinational production from all origin countries. Using that production $Q_l$ is proportional to profits with a factor $\sigma/\iota_l$  we get: 
\begin{equation}\label{Prod} 
Q_l = \frac{\sigma}{\iota_l} \sum_{i,h}\mathcal{M}_{i}\frac{\mathbb{P}_{ilh}\bar{\pi}_i}{(1-t_{ilh})} .
\end{equation}
Setting up a headquarter in country $i$ involves a fixed entry cost $f_E w_i$ paid in labor. 
The only factor of production $L_i$ is used both for firm entry and production so that factor-market clearing reads
\begin{equation}\label{LM} 
w_iL_i=\mathcal{N}_{i}f_{E}w_i+\frac{\sigma-1}{\sigma} Q_i .
\end{equation} 
Last, the price index in country $n$ can be simplified as follows: 
\begin{equation}\label{PI}
P_{n}=\left(\sum_{l}\frac{\tau_{ln}^{1-\sigma}Q_l}{\Xi_{l}^{1-\sigma}} \right)^{\frac{1}{1-\sigma}}.
\end{equation}
The price-index is itself a CES aggregate of price indices of different source-countries $l$. For $l=n$, we can define $P_n^D=\frac{Q_n^{1-\sigma}}{\Xi_{n}}$, the price-index of domestically-produced varieties which will serve as a deflator for domestic production.
Finally, aggregate expenditure in country $i$ results from labor income and corporate income tax revenues:
\begin{equation}\label{BC} 
Y_{l}	=	w_{l}L_{l}+\mathcal{N}_l(\bar M\bar{\pi_l}-f_E w_l)+\Delta_l , 
\end{equation}
where $\bar M\bar{\pi_i}-f_E w_i$ are the profits net of entry costs, and the residual imbalances are captured by $\Delta_l$.\footnote{Whether imbalances are considered to remain constant in absolute terms instead of relative terms does not make a difference for our quantification exercises.} 
The system of equations (\ref{TaxRev})-(\ref{BC}) determines $Q_l$, $Y_n$, $w_i$, $P_n$ with a numeraire condition such that $P_1=1$. 
The long-run monopolistically competitive equilibrium determines $\mathcal{N}_i$ through a free-entry condition imposing that $\bar M\bar{\pi}_i=f_E w_i $.

\paragraph{Destination-based Cash Flow Tax.} 
We extend our model to include destination-based taxation, going beyond the conventional allocation of taxing rights to residence, source and profit location countries.
This approach aligns with several important reform proposals designed to reduce profit shifting and tax competition. 
The destination-based cash-flow taxation (DBCFT) is the most theoretically developed approach in this category and has been widely discussed in academic literature (\citealp{bond2002cash}, \citealp{auerbach_international_2017}, \citealp{auerbach2018cash}). 
We therefore adapt our model to explore the unilateral implementation of a DBCFT. 
This approach substitutes the corporate income tax with a border adjustment tax, which neutralizes common profit-shifting strategies by disregarding where profits are reported.

The DBCFT alters the tax system through three main changes: i) imposing a sales tax $tr_n$ on domestic consumption, ii) subsidizing domestic production costs with a subsidy $s_l$, and iii) eliminating the corporate income tax (CIT). While profits are given as before by (\ref{eq:prof}), the taxation structure now affects market potentials :
$$ \Xi_{l}^{1-\sigma}=\left(1+s_{l}\right)^{\sigma-1} \sum_{n}\frac{\tau_{ln}^{1-\sigma}}{\left(1+tr_{n}\right)^{\sigma}}\frac{Y_{n}}{P_{n}^{1-\sigma}}. $$
Under a unilateral DBCFT proposal with $s_l=tr_l\equiv tr$, the above expression implies that profits from domestic sales by local producers face an effective tax rate $(1+tr)^{-1}$ distinct from the former corporate tax rate $(1-t_{ilh})$.  Unlike a standard corporate income tax, imports are taxed, and exports receive subsidies. The rest of the model remains unchanged, with tax revenues now incorporating both consumption tax receipts and production subsidy expenditures.

\subsection{Tax-base and Profit-shifting Elasticities}

Denote $X_{ilh}$ the total sales of firms from $i$ whose production has been sourced in $l$ and taxed in $h$.  
Combining our specific $G_i$ function and equations (\ref{probafrechet}), and (\ref{profitGEV}), we obtain 
the following proposition.
\begin{prop}[Gravity Structure of Multinational Production and Profit Shifting]\label{gravityprop} 
The fraction of profits that remain taxable in each source country $l$ is given by:
\begin{align}\label{gravitynoPSmodel}
	\frac{X_{ill}}{X_{i}}=\frac{\tilde{A}_{ill}(1-t_{ill})^{\frac{\upsilon_{1}}{\sigma-1}-1}\iota_l^{-1}}{\sum_{jk}\tilde{A}_{ijk}(1-t_{ijk})^{\frac{\upsilon_{1}}{\sigma-1}-1}\iota_j^{-1}G_{i,jk}({\bf \tilde{A}}_{i},{\bf t})}.
\end{align}    
    Moreover, the fraction of shifted income generated by firms from $i$ that is produced in $l$ and reported in tax haven $h$ is given by: 
\begin{align}\label{gravityPSmodel}
	\frac{X_{ilh}}{\sum_{jk,j\neq k}X_{ijk}}=\frac{\tilde{A}_{ilh}^{\frac{\upsilon_{2}}{\upsilon_{1}}}(1-t_{ilh})^{\frac{\upsilon_{2}}{\sigma-1}-1}\iota_l^{-1}}{\sum_{jk,j\neq k}\tilde{A}_{ijk}^{\frac{\upsilon_{2}}{\upsilon_{1}}}(1-t_{ijk})^{\frac{\upsilon_{2}}{\sigma-1}-1}\iota_j^{-1}}. 
\end{align}
As a consequence, from (\ref{gravitynoPSmodel}), the partial  elasticity of the tax base in $l$ to $1-t_{ill}$ is $\tilde{\upsilon}_1:=\frac{\upsilon_{1}}{\sigma-1}-1$. 
Moreover, from (\ref{gravityPSmodel}), the partial elasticity of profits shifted from $l$ to $h$ w.r.t. $1-t_{ilh}$ is equal to $\tilde{\upsilon}_2:=\frac{\upsilon_{2}}{\sigma-1}-1$. 
\end{prop}
The proof is relegated to Appendix \ref{proofgravity}.
The model captures tax competition for paper profits across tax havens. 
Formally, the multilateral resistance terms in the denominator of (\ref{gravityPSmodel}) show that beyond the characteristics of tax haven $h$, those of the other tax havens also matter for bilateral profit shifting.
A decrease in a tax haven's tax rate $t_{ilh}$ triggers two main effects. 
First, it increases the total share of profits shifted from $l$ toward tax havens (see Equation \ref{probafrechet}). 
Second, it reshuffles these profits among tax havens (see Equation \ref{gravityPSmodel}).
Some affiliates in $l$ start shifting their profits to $h$ and some others producing in $l'\neq l$ move their production site to $l$ and engage in profit shifting. 
Moreover, some affiliates that were previously shifting their profits to $h'\neq h$ now switch to tax haven $h$.

This gravity-based profit-shifting enriches the reduced-form set-up \emph{à la} \citet{hines_fiscal_1994}, standard in the corporate tax avoidance literature, in which bilateral profit-shifting abstracts from other tax havens’ attributes and reallocation mechanisms across tax havens.\footnote{In these models, bilateral profit shifting between $l$ and $h$ is proportional to the difference in tax rates between $l$ and $h$. This implies that the elasticity of profit shifting is not constant. In section 4.2, we augment our model to allow for a varying profit-shifting elasticity.} 

\subsection{Tax Reforms and Welfare: a Discussion}\label{model_welfare}
While it is not feasible to derive the welfare effects of tax reforms in our model, we can identify and analyze the key mechanisms through which these reforms influence economic outcomes and, consequently, welfare.
\paragraph{Corporate Taxation.} 
The efficiency of production of the private consumption bundle depends on the number of varieties available for consumption and the allocation of consumption across goods with respect to their relative costs of production. 
Corporate tax policy impacts real income through both these channels: a tax rate hike in one jurisdiction may lower the number of active firms by decreasing post-tax profits but it also changes the spatial allocation of production across countries similarly to \citet{fajgelbaum_state_2019}. 
\paragraph{Minimum Taxation.}
The effect of corporate taxation can help us to understand better minimum taxation policies, where an increase in the minimum tax rate increases, on average, the effective tax rate of firms.
Starting from a world with positive tax rates, a decrease in the number of firms is expected to have a negative impact on the efficiency of production of the private consumption bundle. The intuition for this result can be traced back to closed-economy models like \citet{dixit_monopolistic_1977} and \citet{dhingra_monopolistic_2019} where the first-best number of firms is obtained without any taxes. In an open economy, the spatial reallocation of economic activity may benefit from a minimum tax rate to the extent that it reduces the dispersion in tax rates. Formally, if all countries had the same corporate tax rates, location probabilities would no longer depend on the level of taxes but on country fundamentals.
Turning to the production of the public good, a minimum tax is beneficial whenever it raises tax revenues - depending on the exact design of the tax, this effect may actually not only be at play for non-haven countries but also for tax havens as soon as they change their tax rate in response to the minimum tax.

\paragraph{Destination-based Cash Flow Tax.}
A DBCFT-like proposal combines a border-adjusted tax (BAT) with a (potentially large) reduction in the corporate tax rate. As such, this proposal will generally affect both real income and trade patterns. In addition, it should be noted that even if the reform was a pure BAT, it would not be neutral on trade either, be it for the presence of multinational firms under imperfect competition (\citealp{costinot2019lerner}) or the income effects arising from curbing profit shifting. A reduction in the corporate tax rate - to the extent that it dominates the increase in the ETR driven by the elimination of profit-shifting - will benefit real expenditure on the private-consumption bundle at the expense of the public good. 
We analyze the quantitative importance of these channels in Section 6.

\section{Profit Shifting and Profit-shifting Frictions}\label{secPSestimate}

To bring our model to the data and estimate policy counterfactuals, we need to estimate several parameters and calibrate endogenous variables at the current equilibrium.
This includes $X_{ln}$, the trade in goods and services between source country $l$ and destination country $n$, $X_{ill}$, the multinational production sales made by plants from country $i$ producing in country $l$, and $\mathbb{P}_{ilh}$, the probability for an affiliate of a firm from country $i$ to locate its production in $l$ and book its profits in $h$.
The calibration of the parameters, trade and multinational production is described in Section \ref{sec:empirics_results}. 
Here, we focus on estimating profit-shifting flows, which are key for the calibration procedure.
From profit shifting flows, we can back out $\mathbb{P}_{ilh}$ through a set of structural relations in the model. We describe our methodology below.\footnote{As typical in the literature, we assume that no profit is shifted out of tax havens ($\alpha_{lh}\to \infty$, when $h=l$). Therefore, we back out the profit-shifting shares for the residence $i$ and non-haven country $l$.}


\subsection{A New Approach to Estimating Profit Shifting} \label{section_calib}

We provide a new methodology to estimate profit shifting.
Our identification strategy rests on two pillars. 
The first is a decomposition implied by our model, which we formalize in Proposition \ref{Pilhdecomp}.
We start by noting that equation (\ref{probafrechet}) describes the probability for a firm from $i$ to select the pair $lh$ to locate its plant and book its post-tax profits. 
The firm can either report its profits in the source country ($h=l$) or shift profits from the source country to a tax haven ($h\neq l$).  
We denote by $\Pi_{ill}$ the total post-tax profits declared in $l$ by firms from $i$ producing in $l$ and by $PS_{ilh}$ post-tax profits shifted to $h$ by firms headquartered in $i$ and producing in $l$. 
Total profits - shifted or not - by firms from $i$ are denoted $\Pi_i\coloneqq \sum_l \Pi_{ill}+\sum_{lh} PS_{ilh}$, while $PS_{i}\coloneqq \sum_{lh}PS_{ilh}$ (resp. $PS_{ih}\coloneqq \sum_{l}PS_{ilh}$) represents total shifted profits by firms from $i$ (resp. from $i$ to $h$).
We use the separability of $\mathbb{P}_{ilh}$ across country pairs to derive accounting equations determining bilateral profit shifting. 
	\begin{prop}[Decomposition of $\mathbb P_{ilh}$]\label{Pilhdecomp}The following decomposition holds
		\begin{align}\label{eqpilhdecomp}
			\mathbb{P}_{ilh} = \mathcal P_i \times \zeta_{il} \times \chi_{lh} ~ \text{, for} ~ h \neq l,
	\end{align}
where $\mathcal P_i=\frac{PS_i}{\Pi_i}$ is the probability that plants of firms headquartered in $i$ shift profits, $\zeta_{il}$ is the probability that a tax-avoiding firm headquartered in $i$ locates production in $l$ and $\chi_{lh}$ is the probability that a tax-avoiding firm producing in $l$ books its profits in $h$. 
	\end{prop}
The proof is provided in Appendix \ref{app_triangle_ps}.
This proposition states that to infer $\mathbb{P}_{ilh}$, it is enough to observe three simpler probabilities: $\mathcal{P}_i,\,\zeta_{il}$ and $\chi_{lh}$.

Our strategy's second pillar helps us identify these probabilities. 
First, we show that $\zeta_{il}$ can be recovered as a function of multinational production flows, of aggregated profit shifting in residence countries $i$ and in source countries $l$, and of $\tilde{\upsilon_1}$ and $\tilde{\upsilon}_2$. 
Intuitively, for profits to be shifted from $l$, production must occur in $l$. 
However, because production and paper profits have different elasticities, the patterns of shifted profits are a distorted representation of real activity (captured by MP shares): our model implies that this distortion is shaped by $\frac{\tilde{\upsilon}_2+1}{\tilde{\upsilon}_1+1}$ (see in Appendix \ref{app_algo}). 
Second, to pin down $\chi_{lh}$, we use the following ``triangle identities'': the profits from firms with residence in $i$ that are booked in a tax haven $h$ must match the profits that they shift from any source country $l$ where they operate to a given tax haven $h$.  
Since our data allows us to compute ${PS_{ih}}$, we can thus recover the share of profits shifted from $l$ to any $h$, i.e., $\chi_{lh}$.

The triangle identities are illustrated in Figure \ref{ps_triangle} and formalized in Proposition \ref{triangleprop}:
\begin{prop}[Triangle of Profit Shifting]\label{triangleprop}
The following holds
\begin{align}
	\label{calib_ps_3}\frac{PS_{ih}}{PS_i} = \sum_{l\ne h} \zeta_{il} \times \chi_{lh}.
\end{align}
\end{prop}

The system shown in equation (\ref{calib_ps_3}) gives a set of $ (N-H) \times H$ equations, with $N-H$ the number of non-haven countries and $H$ the number of tax havens. It allows us to recover $\chi_{lh}$ in output from $\frac{PS_{ih}}{PS_i}$ and $\zeta_{il}$ as inputs.

In summary  , $\frac{PS_{ih}}{PS_i}$ is estimated as detailed in Section \ref{sub_ps_ih}. 
The function $\zeta_{il}$, which depends on $X_{ill}$, $PS_i$, $\frac{PS_l}{\sum_l PS_l}$, and $\frac{\tilde{\upsilon_1} + 1}{\tilde{\upsilon_2} + 1}$, is derived from the algorithm outlined in Appendix \ref{app_algo}. 
The calibration of $X_{ill}$ is given in Appendix \ref{app_data}.
Section \ref{sub_ps_ih} and Appendix \ref{app_algo} provide estimates for $PS_i$ and $\frac{PS_l}{\sum_l PS_l}$.
 $\frac{PS_l}{\sum_l PS_l}$ is quantified using country-level shares of worldwide value-added.
Additionally, the estimates for $\tilde{\upsilon_1}$ and $\tilde{\upsilon_2}$ are discussed in Section \ref{main_elasticities}. 
Finally, $\chi_{lh}$ is recovered through the application of triangle identities.
As formalized in Propositions \ref{Pilhdecomp} and \ref{triangleprop}, $\mathbb{P}_{ilh}$ is readily obtained from $\mathcal P_i$, $\zeta_{il}$ and $\chi_{lh}$.

\begin{figure}[htbp]
\begin{center}
	\begin{tikzpicture}[scale=1.8]
		\node[shape=circle,draw=black] (A) at (0,0) {i};
		\node[shape=circle,draw=black] (B) at (3,0) {l};
		\node[shape=circle,draw=black] (C) at (1.5, 2.5) {h};
		\path [->](A) edge node [below] {$\zeta_{il}$} (B);
		\path [->](B) edge node [right] {$\chi_{lh}$} (C);
		\path [->](C) edge node [left] {$\frac{PS_{ih}}{PS_i}$} (A);
	\end{tikzpicture}
	\\
	\small{$i$: headquarter \quad $l$: production \quad $h$: haven \\ 
 \begin{flushleft}
	    \footnotesize Note: $\frac{PS_{ih}} {PS_i}$ is estimated (section \ref{sub_ps_ih}), $\zeta_{il}$ is a function of $X_{ill},PS_i, \frac{PS_l}{\sum_l PS_l},\frac{\tilde{\upsilon_1}+1}{\tilde{\upsilon_2+1}}$ (see the algorithm in Appendix \ref{app_algo}). $MP_{il}$ is observed (see Appendix \ref{app_data}), $PS_i$ and $\frac{PS_l}{\sum_l PS_l}$ are estimated (see section \ref{sub_ps_ih} and Appendix \ref{app_algo}), $\tilde{\upsilon_1}$ and $\tilde{\upsilon_2}$ are estimated (section \ref{main_elasticities}). $\chi_{lh}$ is recovered using the triangle identities.
	\end{flushleft}}
\end{center}
\caption{Triangle of Profit Shifting}
\label{ps_triangle}
\end{figure}

\subsection{Profit-shifting Frictions} 

Bilateral profit-shifting frictions are an important novel ingredient of our framework. 
They govern how multinational firms decide whether and where to shift profits and produce. 
In this subsection, we back out these frictions, consistently with the observed flows of shifted profits by firms in residence $i$ to tax haven $h$ from source country $l$. 
We first detail the procedure and then explore the magnitude and determinants of these frictions in section \ref{main_frictions}.

\paragraph{Identifying Profit-shifting Frictions.}
We start by noting that, at the calibrated equilibrium, we know profit-shifting probabilities
$\mathbb{P}_{ilh}$, taxes $t_{ll}$
and $t_{lh}$, and our estimated elasticities $\tilde{\upsilon_{1}},\tilde{\upsilon_{2}}$. 
We formalize the identification result for profit-shifting frictions in the next Proposition. 

\begin{prop} [Identifying Profit-Shifting Frictions] {\label{prop_frictions}} At the calibrated equilibrium the following holds 
\begin{align} \label{frictioneq} 
\bar\theta\tilde\theta_i\alpha_{lh} \left(\frac{1-t_{ll}}{1-t_{ilh}}\right)^{\frac{1}{\sigma-1}}=\left(\frac{\mathbb{P}_{ilh}}{\mathbb{P}_{ill}}\right)^{-\frac{1}{\upsilon_1}}\left(\frac{\mathbb{P}_{ilh}}{\mathcal{P}_i}\right)^{\frac{1}{\upsilon_1}-\frac{1}{\upsilon_2}},\end{align} 
where $\bar{\theta}$ is a normalizing constant such that $\theta_i=\bar{\theta}\tilde{\theta}_i$. We specify $\bar\theta$ in Appendix \ref{appendix_ps_frictions}. \end{prop}

On the right-hand side of equation (\ref{frictioneq}), we have observable profit-shifting flows. On the left-hand side, we have the tax differential gain of firms from $i$ producing in $l$ choosing to shift to haven $h$ rather than booking profits domestically in $l$. Given our previous definitions, we know that  $\left(\frac{\mathbb{P}_{ilh}}{\mathcal{P}_i}\right)^{\frac{1}{\upsilon_1}-\frac{1}{\upsilon_2}}$<1. Typically, in the data, $\left(\frac{\mathbb{P}_{ilh}}{\mathbb{P}_{ill}}\right)^{-\frac{1}{\upsilon_1}}>1$ as less profits are shifted than booked domestically. We also know that tax gains are typically such that $\left(\frac{1-t_{ll}}{1-t_{ilh}}\right)^{\frac{1}{\sigma-1}}$<1 as $t_{ll}$ is typically larger than $t_{ilh}$. Our frictions $\bar\theta\tilde\theta_i \alpha_{lh}$ are what we need to explain the observed PS flows, given the observed tax differential and elasticities. Whenever $\mathbb{P}_{ilh}$ is very large relative to $\mathbb{P}_{ill}$, the model will call for small profit-shifting frictions to rationalize the data. Importantly, frictions can be both above and below 1, depending on whether the tax differential gain is sufficient to explain observed profit-shifting flows. 

We note that $\tilde\theta_i=\theta_i/{\bar \theta}$ and $\alpha_{lh}$ can be mapped into a marginal cost equivalent $Cost_{ilh}\coloneqq \tilde\theta_i\alpha_{lh}$. 
This is the marginal cost increment associated with profit
shifting from any $l$ to any $h$ by firms from $i$ \emph{if
all profit-shifting frictions were such that $\alpha_{l'h'}=\alpha_{lh}$}.
In contrast with trade or multinational
production frictions, the interaction of the tax base
and profit-shifting elasticities implies that bilateral profit-shifting
flows do not verify the irrelevance of independent alternatives. The
cost of shifting profits from $l$ to $h$ depends on the profit-shifting frictions between other $l'-h'$ pairs.

\section{Data and estimations}

This section presents the data and methods used to estimate profit shifting from headquarter countries to tax havens through source countries. Using bilateral FDI income, multinational production data, and firm-level pre-tax profits, we calibrate the key parameters of the model. The following subsections describe the data sources, address potential measurement issues, and outline the estimation methods and provide the main results. 

\label{sec:empirics_results}
\subsection{Data}\label{sec:data}

Our baseline sample consists of 40 countries from 2012-2017, which account for 90\% of the world GDP in 2017.
The sample includes seven major tax havens: Hong Kong, Ireland, Luxembourg, Netherlands, Singapore, Switzerland, and Offshore Financial Centers, an aggregate of smaller tax havens.

We use data on Foreign Direct Investment income and multinational production as building blocks to estimate profit shifting, elasticities, and frictions. 
We also use other data sources in the analysis (tax rates, tax havens' policies, trade, and other national accounts data). 
Details on the construction of the datasets and auxiliary sources of information are provided in the data Appendix \ref{app_data}.

We use bilateral FDI income as the primary source to estimate profit shifting.\footnote{See \citet{UNCTAD_world_2015}, \citet{Jansky2019}, and \citet{vicard_profit_2022} for studies studying FDI returns in tax havens.} 
The data are sourced from OECD balance of payments data and include foreign affiliate income returned to the residence country through dividends, interest, or reinvested earnings. 
We construct our FDI income series by summing reinvested earnings and dividends from tax havens (see \citealp{wright_exorbitant_2018}). To increase its coverage, we complete the database with an imputation procedure based in unilateral rates of returns. This procedure is described in Appendix \ref{app_fdiinc}. Note that this procedure is conservative as it tends to underestimate profits in tax havens (see Appendix Figure \ref{comparison_imputation}). FDI income data are subject to double counting and might misreport the location of the MNC's foreign earnings (\citealp{blouin_double_2021}, \citealp{damgaard_phantom_2019}). 
This occurs mostly because international statistics follow the immediate investor principle: each investor in an ownership chain reports the income from its immediate direct investment. 
As we move up the ownership chain, income is aggregated and thus appears double-counted. 
We propose two corrections to address double counting and misreporting.
First, investment flows originating from tax havens, such as the Netherlands and Ireland, are excluded from the construction of the sample. 
This restriction targets common profit-shifting arrangements, where ownership chains span several tax havens, thereby resulting in more accurate information that better aligns with true reporting.
It reduces the possibility of double counting. 
Second, we correct our FDI income series to limit double counting issues due to conduit FDI. 
Conduit FDI is realized through Special Purpose Entities (SPEs), which are entities without substantial activities, owned by non-residents, and established to obtain specific advantages, such as tax benefits (\citealp{bop6}). Some OECD countries are now reporting inward FDI income statistics in SPEs and in standard entities. In particular, the Netherlands, Ireland, Luxembourg, and the United Kingdom are among these countries. These countries are among the five most important for conduit investment according to \citet{garcia-bernardo_uncovering_2017}.
We exclude FDI income transiting through SPEs to limit the issue of double counting. For instance, 87\% of FDI income sent by Luxembourg in 2017 is excluded from our estimation, while only 14\% of that sent by Denmark is excluded. As shown below, applying this procedure deflates the estimated global profit shifting in our dataset from \$525 billion to \$357 billion (-32\%) in 2017, and from 40\% of total multinationals' profits in the sample to 33\%, close from the estimate of 310\$bn by \citet{blouin_double_2021}. Section \ref{app_fdiinc} of the Appendix, discusses in details the construction of the bilateral FDI income series. It also describes an additional robustness exercise where we follow \citet{damgaard_phantom_2019} methodology to impute conduit FDI income when it is not directly declared in raw statistics. Figure \ref{fig:comparison_fdi_income_variables} compares our baseline sample with the uncorrected sample and with the robustness sample and shows the extent of the aggregate correction. 

The adjusted FDI income series is the most comprehensive bilateral source of data about multinational profits, covering more country pairs and years than alternatives. 
We use it to estimate bilateral profit shifting between residence countries and tax havens.
We examine the sensitivity of our estimates using two alternative datasets, Country-by-Country Reporting (CbCR) and Orbis, with the latter serving as the basis for calibrating tax base and profit-shifting elasticities.
As mentioned by \citet{fuest2022global}, the CbCR dataset has the major advantage of reporting profits where they are actually booked, as opposed to where reporting takes place due to accounting conventions. 
However, while the coverage of the activities of MNCs in tax havens is important, CbCR has limitations: it lacks bilateral series of pre-tax profits as there is no disaggregated data for some pairs (e.g the United Kingdom and the Netherlands), and the data only covers the years 2016-2021. 
Moreover, CbCR series are also affected by double counting issues, that are harder to correct at the macro level than FDI income data.\footnote{The OECD includes an ``important disclaimer regarding the limitations of the Country-by-country report statistics" alongside its publication of CbCR data (available at \url{https://www.oecd.org/content/dam/oecd/en/topics/policy-sub-issues/corporate-taxation/anonymised-and-aggregated-cbcr-statistics-disclaimer.pdf}).} We also use Orbis firm-level data, following the procedure of \citet{delis2022global}, who construct a global database on MNCs activities using all ``vintages" of Bureau Van Dijk’s Orbis Historical database. 
 Their intensive data matching procedure results in a dataset that includes historical financial and ownership data, covering firms' locations, including tax havens, across the years of our analysis. 
 For the estimation of profit shifting, the data is aggregated at the country pair-year level.

Turning to multinational production (MP), we construct $X_{ill}$, the sales resulting from the production in the country $l$ by firms headquartered in the country $i$, using the Multinational Revenue, Employment, and Investment Database (MREID) created by \citet{Ahmad2023}. 
The dataset also sources information from Orbis Historic to provide harmonized data on cross-border multinational sales. 
Trade ($X_{iln}$), is taken from the International Trade and Production Database, also sourced from the US ITC gravity portal (\citealp{borchert2022}). 

In the following, we outline our methodologies and present estimates of profit shifting and tax elasticities. 
This procedure still requires the calibration of $\sigma$ and $\iota_l$.
We use administrative French firm-level data from the FARE dataset and follow the methodology provided by \citet{deloecker2012markups} to estimate firm-level markups. 
The results give a median markup equal to 17\%, which corresponds to $\sigma=6.88$.\footnote{This is in line with estimates found in the literature, e.g., \citet{tintelnot2017global}. 
Similarly, \citet{deloecker2020rise} find a median markup of around 20 percent using Compustat data.}
Given the estimated $\sigma$, $\iota_l$ is then calibrated using the wedge between gross output and profits. 
Overall, $\iota_l$ absorbs any non-labor cost that impacts profits but not sales.

\subsection{Estimation of Bilateral Profit Shifting}\label{sub_ps_ih}

Our goal is to determine the probability that a plant operating in source country $l$ and from residence country $i$, records profits in tax haven $h$, denoted as $\mathbb{P}_{ilh}$. We follow the procedure described in section \ref{secPSestimate}.
The first step in our analysis is to estimate the profit shifted from residence countries to tax havens, $PS_{ih}$. 
To estimate $PS_{ih}$, we 
use a gravitational model of FDI income to determine bilateral excess incomes booked by residence country in tax havens. 
These bilateral excesses are our measure of bilateral profit shifting.

\paragraph{Estimating $PS_{ih}$.}
What would be the level of profits recorded in country $h$ if this country was not a tax haven?
Estimating the amount of profit shifted requires defining a benchmark level of profit due to real activities (e.g., \citealp{hines_fiscal_1994}; \citealp{garcia2021profit}; \citealp{guvenenetal_2022}; \citealp{torslov_missing_2022}). 
Our model provides guidance for this benchmark. 
According to Proposition 1, the level of profits booked in a country $l$ by firms from country $i$ due to real activities follows a gravity structure (\ref{gravitynoPSmodel}). 
This gravity relationship helps us separate the profits booked by country $i$'s firms in tax haven $h$ ($\sum_l \Pi_{ilh}$) due to real activities in country $h$ ($\Pi_{ihh}$) from the profits shifted to country $h$ ($\sum_{l \neq h} \Pi_{ilh}$).
Because our model assumes that profit shifting occurs only in tax havens, our counterfactual exercise computes the amount of profits booked by country $i$'s firms in country $h$ if country $h$ were not a tax haven. 
We therefore regress bilateral FDI incomes or bilateral profits on tax haven dummy variables and a set of gravity factors, which explain real activity in country $h$.\footnote{In a previous version of the paper, we proposed an alternative strategy that applies the elasticity of profits to the effective tax rate to compute bilateral profit shifting. This methodology aligns with recent public finance literature (see \citealp{international_beer_2020} and \citealp{garcia2021profit}). However, this approach requires defining a counterfactual effective tax rate to estimate bilateral profit shifting. While both strategies yield similar aggregate values for profit shifting, determining a consensus on the effective tax rate under which firms shift profits is more challenging than simply muting the tax haven dummy variable.}



We estimate the following equation:
\begin{eqnarray} \label{eq:income}
Y_{ikt}&=& \exp(\beta_1 \text{Haven}_{k} + \theta^{'} \bm{X_{ikt}} + \mu_{it} + \mu_{r_{(k)}t} ) + u_{ikt}.
\end{eqnarray}
$Y_{ikt}$ represents the FDI income or pre-tax profit booked by residence country $i$ in country $k$ in year $t$.
$Haven_{k}$ is an indicator variable equal to 1 if country $k$ is a tax haven. 
We use gravity variables to parametrize $\bm{X_{ikt}}$, with $\theta$ being the vector of coefficients associated with these variables.
$\mu_{it}$ are residence country $\times$ year fixed effects, $\mu_{r_{(k)}t}$ are country $k$'s region $\times$ year fixed effects. 
$u_{ikt}$ are the residuals. In order to increase the accuracy of the predictions and to capture the geographical specialization of tax havens (\citealp{Laffitte_LTPS_2020}), we also include in the specification interactions of region fixed effects with the tax haven indicator variable.
We estimate equation (\ref{eq:income}) using the Poisson pseudo-maximum likelihood (PPML) estimator to take into account heteroskedasticity (\citealp{santos_silva_log_2006}) and to allow us to work with predictions in levels, avoiding the (log) OLS prediction's transformation issue (\citealp{duan_smearing_1983}).

\begin{table}[!h]	
\begin{center}
\caption{Estimation of $PS_{ih}$ } \label{tab:estimation_ps}
\setlength{\tabcolsep}{3.5pt}
\footnotesize{															
\begin{tabular} {l cc c c c c}			
\noalign{\smallskip}  \hline \noalign{\smallskip}												
\multicolumn{1}{l}{Dep. Variable } & \multicolumn{2}{c}{FDI Incomes} 		&&   \multicolumn{3}{c}{Pre-tax Profits} \\			\noalign{\smallskip} 
					\cline{2-3} \cline{5-7} 
\noalign{\smallskip}
						   &Unadjusted & Adjusted         &&   CbCR && ORBIS \\						
						&& ($no SPEs$) && && \\
					&	(1)	&	(2)	&&	(3)	&&	(4)	\\
\noalign{\smallskip} 
					\cline{2-3} \cline{5-5} \cline{7-7} 
\noalign{\smallskip}

$Haven_{k}$			&	2.524***	&	1.808***&&	1.535***	&&	1.470***	\\
					&	(0.281)	&	(0.271)	&&		(0.266)	&&(0.257)	\\
\noalign{\smallskip} 
					\cline{2-3} \cline{5-5} \cline{7-7} 
\noalign{\smallskip}
$\mu_{jt}$				&	Yes		&	Yes		&&	Yes		&&	Yes	\\	
$\mu_{r_{(k)}t}$			&	Yes		&	Yes		&&	Yes		&&	Yes	\\	
 Haven $\times$ Region dummies &	Yes		&	Yes		&&	Yes		&&	Yes \\
Controls & Yes		&	Yes		&&	Yes		&&	Yes	\\	
\noalign{\smallskip} 					
\cline{2-3} \cline{5-5} \cline{7-7} 
 \noalign{\smallskip}
\# Countries				&	146	&	146	&&	140	&&	57	\\	
 Observations				&	73,443	&	73,116	&&	4,354	&&	6,725	\\	
Pseudo $R^2$				&	0.82	&	0.81	&&	0.78	&&	0.79	\\	
\noalign{\smallskip} 					
\hline
\noalign{\smallskip}
\textbf{Proft Shifted (2017)}			&	\textbf{525303}	&	\textbf{357210}	&&	\textbf{540730}	&&	\textbf{77793}	\\
\textbf{Sample's profits (\%, 2017)}	&	\textbf{41}		&	\textbf{33}		&&	\textbf{30}		&&	\textbf{18}		\\	
\noalign{\smallskip} \hline \noalign{\smallskip}
\end{tabular}}														
\parbox{14cm}{\footnotesize Note: In column (1), we use unadjusted series of FDI incomes while in column (2) we exclude FDI income transiting through SPEs. In column (3), we use pre-tax profits from CbCR, and in column (4) we use series from Orbis Historic. The controls include the GDP and per capita GDP in logs of the destination country, log distance, contiguity, shared colonial ties, common colonizer, and common legal origin. Estimates of profit shifting are obtained from an estimation that also includes haven $\times $ region fixed effects. All reported estimates are obtained from PPML estimation. Standard errors are robust to clustering at the country level and reported in parentheses. $^{***}$, $^{**}$, and $^{*}$ indicate statistical significance at 1\%, 5\%, and 10\% confidence levels, respectively.}\\	
\end{center}
\end{table}

Profit shifting from residence $i$ to a tax haven $h$, $PS_{ih}$, is defined as the difference between the predicted and counterfactual income that is predicted by muting the tax haven premium: $ PS_{ih} = \reallywidehat{Y_{ih}} - \reallywidehat{Y_{ih}^0}$.
$\reallywidehat{Y_{ih}}$ are predicted values on the sample of all pairs $ih$ composed of non-haven countries $i$ reporting incomes in tax havens $h$. 
$\reallywidehat{Y_{ih}^0}$ is defined on the same sample and corresponds to the predicted income when the tax haven premium is set to 0 for all countries (i.e., $\beta_1=0$).

Table \ref{tab:estimation_ps} reports the tax haven semi-elasticity using bilateral series of FDI income and pre-tax profits as dependent variables.
The estimation is done across different estimation samples and data sources.
Our preferred estimate of profit shifting is based on the adjusted FDI income data which covers more pairs of countries.
We estimate profit shifting to be \$357bn, which corresponds to 33\% of all profit in the sample in 2017. 
This is consistent with \citet{wier_global_2022}, who report that profit shifting amounts to 36\% of global multinational profits in 2017.
We find estimates of profit shifting of \$541bn when we use the FDI income series that do not correct for conduit FDI, showing that our correction is quantitatively important. using the same methodology, we find that 30\% of the profits in the sample are shifted into tax havens using CbCR data. Finally, the Orbis sample appears limited, in particular in its aggregate coverage of tax havens. 
%
%

\begin{figure}[!h]
	\centering
    \includegraphics[width=0.65\linewidth]{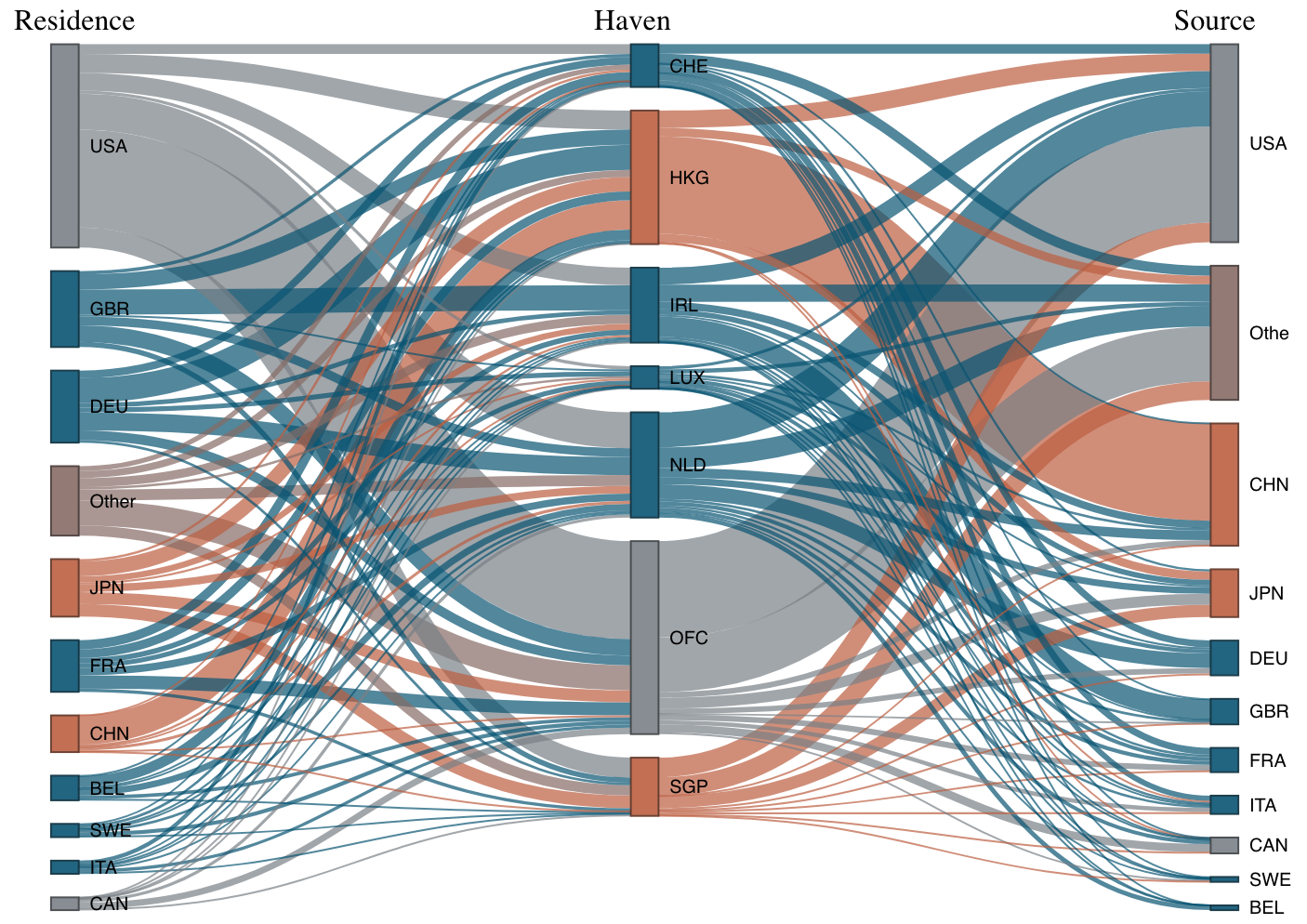} 
		\caption{Profit shifting shares from residence-country $i$ to $h$ ($PS_{ih}/PS_i$) and from source-country $l$ to $h$ ($\chi_{lh}$).}
	\label{fig:sankey_lh}
\end{figure}

\paragraph{Profit-shifting Flows.} 
We use Propositions (\ref{Pilhdecomp}) and (\ref{triangleprop}) to compute $PS_{ilh}$. 
Figure \ref{fig:sankey_lh} displays the estimated profit that has been shifted to tax havens (in the center) according to the residence country (on the left) and the source  country (on the right).
We display the top 10 countries and aggregate the bilateral shares for others. 

Figure \ref{fig:sankey_lh} shows the predominance of residence countries such as the U.S. and, to a lesser extent, the UK, and Germany, in shifting profits to tax havens. 
It also shows the relative importance of each tax haven in profit shifting, with a predominance of Offshore Financial Centers, mainly located in the Caribbean and attracting profits from the U.S.
The pattern displayed in Figure \ref{fig:sankey_lh} also confirms that gravitational frictions shape profit shifting. European tax havens prominently host profits from non-haven EU countries and the U.S., while China and Japan shift most of their profits to Hong Kong and Singapore. 

\paragraph{Comparisons.} 

Several papers provide estimates of profit shifting at the production country or tax haven level (\citealp{zucman_taxing_2014}, \citealp{clausing_effect_2016, clausing_2020}, \citealp{Jansky2019}, \citealp{garcia2021profit}, \citealp{delis2022global} and \citealp{torslov_missing_2022}).
Table \ref{correlations} reports Spearman's rank correlations of our vector of estimated profit shifting with different estimates from the literature. 
In Panel A, we aggregate our bilateral measure of profit shifting for each production country and display the correlations of this vector with unilateral profit-shifting measures constructed by TWZ, the Tax Justice Network (\citealp{tjn_2020}), and the European Commission using the CORTAX model (\citealp{cortax2016}). 
We find large positive rank correlations at the unilateral level, suggesting a stable relative position of each source country in profit shifting irrespective of the methodology used. 

\begin{table}[ht] 
	\centering
	\small
	\caption{Spearman's rank correlation}
	\label{correlations}
	\begin{tabular}{lcc} \toprule
		Source& Correlation &  Obs. \\ 
		\midrule
		A. Unilateral profit shifting ($PS_l$): \\
		\noalign{\smallskip}  \cline{1-1}    \noalign{\smallskip}
		\citet{torslov_missing_2022}  & 0.88& 33 \\
		\citet{tjn_2020} & 0.92 & 33\\
		\citet{cortax2016}  & 0.97 & 21\\ 
		\midrule
		B. Bilateral profit shifting ($PS_{lh}$): \\
		\noalign{\smallskip}  \cline{1-1}    \noalign{\smallskip}
		\citet{torslov_missing_2022}  & 0.68 & 109 \\
		\noalign{\smallskip}  \hline    \noalign{\smallskip}

	\end{tabular}
\end{table}

In Panel B, we compare our estimations with the bilateral estimates of \citet{torslov_missing_2022}, which is the first study to propose a measure of bilateral profit shifting across pairs of source and tax haven countries. 
Their methodology allocates bilaterally the global amount of profit shifting using an allocation key based on excess trade in high-risk services and excess cross-border flows of interest.\footnote{In Appendix \ref{app_ps_stateoftheart}, we review the methods used in the current literature, especially those by TWZ. We identify three additional sources of profit shifting beyond the ``high-risk" services exports and intra-firm interest payments that TWZ considers: profit shifting in goods, tax-haven deflated imports, and services that are not classified as ``high-risk". In Appendix \ref{PS_robustness}, we also evaluate the robustness of our calibration using inputs from TWZ.} 
We restrict our comparison to bilateral estimates for European tax havens as TWZ reports an aggregate for non-European tax havens.
We find a strong rank correlation of 0.68 between our bilateral measure and the one of TWZ despite using different data sources and estimation strategies.
In Appendix \ref{comparison}, we provide additional material that compares our profit-shifting estimates with other sources found in the literature. 

Finally, we propose different robustness exercises in Appendix \ref{PS_robustness}. 
We assess the correlation between our profit-shifting allocation and an allocation based on excess trade in services with tax havens only. 
The Spearman rank correlation coefficient of 0.6 indicates a relatively high correlation between both series but the $ PS_{lh}$ estimated in this paper is generally larger than the excess of services trade.
This result suggests that services trade is an important driver of profit shifting between source countries and tax havens but shall not be considered its unique determinant.
We also explore the role of the parameters $\tilde{\upsilon}_1$ and $\tilde{\upsilon}_2$ on the allocation of profit shifting.
We find that the estimated profit shifting is robust to different calibrations of these elasticities (see Appendix \ref{PS_robustness}). 
Finally, we show that our estimation of profit shifting is robust to an alternative calibration of $\frac{PS_l}{\sum_l PS_l}$ (Figure \ref{fig:comparison_sl_our_sl_twz_log}) and to the use of FDI series without our imputation procedure (Table \ref{estimation_ps_imputation}).

\subsection{Tax-base and Profit-shifting Elasticities \label{main_elasticities}}

Our model incorporates two tax elasticities: one for the tax base (governed by $\tilde{\upsilon}_1$) and one for profit shifting (governed by $\tilde{\upsilon}_2$). 
The model restrictions impose $\tilde{\upsilon}_2\geq \tilde{\upsilon}_1$, meaning that profit shifting is more elastic to taxes than real production (which is governed by both $\tilde{\upsilon}_1$ and $\tilde{\upsilon}_2$). 
Existing studies find that corporate tax rates have a larger impact on profits in low-tax jurisdictions compared to high-tax jurisdictions, holding real activity constant (e.g., \citealp{dowd_profit_2017}, \citealp{bratta_assessing_2021}, \citealp{fuest2021corporate}, \citealp{garcia2021profit}). However, these studies do not differentiate between profits arising from real production activities and those generated through profit shifting. Instead, they estimate an elasticity pooling data from tax havens, where a large share of profit are shifted, and high-tax countries, where real production dominates.
In contrast, our results explicitly show that the profit shifting elasticity is larger than the tax base elasticity, confirming that profit shifting is significantly more elastic to tax changes than real production.

Our model provides guidance on the functional form to be used to estimate elasticities. 
Proposition \ref{gravityprop} states that the fraction of shifted profits and the fraction of taxable profits exhibit a similar gravitational structure, depending on bilateral frictions, country wages, and market potential.
The use of firm-level variation in the Orbis Historical dataset allows us to account for these using controls and adequate sets of fixed effects. 
Each elasticity is identified based on different tax rates (see equations \ref{gravitynoPSmodel} and \ref{gravityPSmodel}): the fraction of shifted profits to tax havens is sensitive to changes in bilateral effective tax rates, while the fraction of taxable profits responds to changes in statutory tax rates.
In Appendix Table \ref{app:u2_rob_maintext}, we use our macro-estimate of profit shifting, $PS_{ilh}$, and data on profit shifting provided by TWZ to check the sensitivity of our firm-level reduced-form estimates. 
Using both a reduced-form and a structural approach, we find that the profit-shifting elasticity is approximately three times larger than the tax base elasticity.

\paragraph{Profit-shifting Elasticity.}
To calibrate $\tilde{\upsilon}_2$, we analyze the sample of firms reporting profits in tax havens.
Estimating bilateral profit shifting at the firm level requires defining a benchmark level of normal profits. 
As common in this literature, our ``excess profit" methodology benchmarks high profits in tax havens against firm-specific employment and fixed assets  (\citealp{fuest2022global}, \citealp{fuest2022corporate}, \citealp{guvenenetal_2022}).

Our model assesses the responsiveness of profit shifting to changes in effective tax rates across tax havens, also considering bilateral gravitational forces. 
The firm-level regressions include firm-year fixed effects and therefore identify $\tilde{\upsilon}_2$ using variations in excess profits across tax havens within firm. 
We estimate the following equation:
\begin{eqnarray} \label{gravityPSmodel_ppml}
PTI_{f_{i}ht} = \exp \Bigl( \delta_0 \ln(1-t_{f_{i}ht}) + \delta_1 \ln(Emp_{f_{i}ht}) + \delta_2 \ln(Assets_{f_{i}ht}) +\mu_{f_{i}t} + \mu_{iht} +  \mu_{ht}\Bigr) \times \epsilon_{f_{i}ht}
\end{eqnarray}
where $\delta_0= \tilde{\upsilon}_2 $ is our coefficient of interest. 
$PTI_{f_{i}ht}$ is the pre-tax profit, of firm $f$ from country $i$, reported in tax haven $h$ in year $t$. 
Using the Orbis Historical dataset, we compute the bilateral effective tax rates, $t_{f_{i}ht}$, that each firm from $i$ faces in each tax haven $h$ in which it has activity at time $t$. 
Our estimation includes firm-year fixed effects ($\mu_{f_{i}t}$), which control for time-varying characteristics of individual firms, such as changes in management practices, business strategies, or other firm-specific shocks that could affect firm profit shifting behavior independently of changes in effective tax rates.
Additionally, we include country pair and year fixed effects ($\mu_{iht}$) to control for time-varying bilateral factors that influence profit shifting between countries, such as historical ties or specific bilateral tax treaties that affect tax avoidance strategies.
The use of country pair and year fixed effects also mitigates the possibility that our results are influenced by country-specific accounting practices -- addressing a limitation highlighted by \citet{blouin_double_2021}.

\paragraph{Tax-base Elasticity.}
Similarly, firm-level variation in pre-tax profit across non-tax haven countries is used to identify $\tilde{\upsilon}_1$. 
The empirical model includes firm-year fixed effects. Identification is based on statutory tax rates, which vary only at the level of source countries and years. 
This restriction limits our use of fixed effects due to collinearity. 
The model includes controls for the production market's wage level and size, as well as for bilateral frictions between residence and source countries.
We estimate the following equation:
\begin{eqnarray} \label{gravitynoPSmodel_ppml}
PTI_{f_{i}lt} = \exp \Bigl(\kappa_0 \ln(1-t_{f_{i}lt}) + \kappa_1 \ln(GDP_{lt}) + \kappa_2 \ln(GDPPpc_{lt}) + \mu_{f_{i}t} + \mu_{il} +  \mu_{l} \Bigr) \times u_{f_{i}lt}
\end{eqnarray}
where $\kappa_0=\tilde{\upsilon}_1$ is our coefficient of interest.
$PTI_{f_{i}ht}$ is the pre-tax profit, of firm $f$ from country $i$, reported in non-tax haven $l$ in year $t$. 
Our estimation includes firm-year fixed effects ($\mu_{f_{i}t}$), which control for time-varying characteristics of firms.
The regression analysis includes GDP and GDP per capita (in logs) to control for the production market’s size and wage level.
Additionally, we incorporate country pair fixed effects ($\mu_{il}$) to control for bilateral frictions between residence and source countries.

$\epsilon_{f_{i}ht}$ and $u_{f_{i}lt}$ are the error terms. 
Given heteroscedasticity and nonlinearity in the profit data generating process, we estimate our empirical model using PPML estimation--but also reports the estimates using OLS.
Note that, unlike log-OLS estimation, PPML yields consistent parameter estimates even when the error term of the regression is heteroskedastic (\citealp{santos_silva_log_2006}; \citealp{fuest2022global}).
 
\begin{result} 
The paper profit elasticity is three times as large as the elasticity of the tax base.\end{result}

Table \ref{reg_u1u2} reports the estimated coefficients and the corresponding parameter elasticities $\tilde{\upsilon}_1$ and $\tilde{\upsilon}_2$.
In columns (1) and (2), we use the statutory tax rates ($t_{lt}$) as the corporate tax variable, while we use the bilateral effective tax rate ($t_{f_{i}ht}$) in columns (3) and (4).
We report the result using OLS in columns (1) and (3), and PPML in columns (2) and (4). 
\begin{table}[htbp]
	\centering
 	\caption{Estimation of elasticities $\tilde{\upsilon}_1$ and $\tilde{\upsilon}_2$\label{reg_u1u2}}
	\footnotesize{	
\begin{tabular}{l@{\extracolsep{0.05in}}cccc} \toprule
& \multicolumn{2}{c}{Estimation $\tilde{\upsilon}_1$} & \multicolumn{2}{c}{ Estimation $\tilde{\upsilon}_2$} \\ 
\noalign{\smallskip} \cline{2-3} \cline{4-5}\noalign{\smallskip} 
& (1) & (2) & (3) & (4) \\
\noalign{\smallskip} 
\noalign{\smallskip} \cline{1-3} \cline{4-5}\noalign{\smallskip} 
\noalign{\smallskip} 
$ln(1-t_{lt})$			&	1.085***	&	2.048**	&			&			\\
					&	(0.204)	&	(0.866)	&			&			\\
$ln(1-t_{f_{i}ht})$			&			&			&	3.844***	&	6.827***	\\
					&			&			&	(0.541)	&	(1.307)	\\

 \midrule
Observations			&216,397		&	216,397	& 	2,649	&	2,649 	\\
Adj.  R$^2$			& 0.524		&	0.896	&	0.602	&	0.979	\\
Estimator & OLS & PPML  & OLS & PPML\\ 
 \midrule
Controls								& 	Yes	&	Yes	&	Yes	&	Yes	\\
Firm $\times$ Year						&	Yes	&	Yes	&	Yes	&	Yes	\\
Origin $\times$ Destination				&	Yes	&	Yes	&	No	&	No	\\
Origin $\times$ Destination  $\times$ Year 	&	No	&	No	&	Yes	&	Yes	\\

 \bottomrule
\end{tabular}}
\parbox{12cm}{\footnotesize Note: Columns (1) and (2) include controls for GDP and GDP per capita, while columns (3) and (4) include controls for employment and fixed assets. All controls are logged. The reported estimates are derived from OLS and Poisson Pseudo-maximum Likelihood (PPML) estimations. Full results reported in Table \ref{app:reg_u1u2}. Standard errors, robust to clustering at the country level, are shown in parentheses. $^{***}$, $^{**}$, and $^{*}$ indicate statistical significance at 1\%, 5\%, and 10\% confidence levels, respectively.} \\
\end{table}

We find a profit-shifting elasticity, $\tilde{\upsilon}_2=6.8$, about three times the tax base elasticity, $\tilde{\upsilon}_1=2.0$. Our estimate suggests that multinational production - which is governed by both elasticities - is relatively mobile across countries.\footnote{For comparison, \citet{wang2020multinational} estimates an elasticity of MP sales to taxes between 1.8 and 2.1, in the same range as our estimates for $\upsilon_1$. 
We also compute the \textit{semi-elasticity} of the tax base to taxes using the same specification as in Table \ref{reg_u1u2} to compare our results with other estimates in the literature. We find a semi-elasticity of the tax base of 1.2 using the OLS estimator and 2.7 using PPML (see Appendix Table \ref{semi-elasticities}). In their meta-study, \citet{international_beer_2020} find that the average semi-elasticity
of profits to taxes estimated with micro-level data is between 1 (Table 2, column 3) and 1.2 (Table 2, column 2). Our semi-elasticity of the tax base to taxes lies in the upper bound of this range. In their estimation of a non-linear elasticity of profit to taxes using micro-level country-by-country reporting data, \citet{fuest2021corporate} find a semi-elasticity of profits to taxes that goes from -5 when the tax rate is 0.15 to -13 when the tax rate is close to zero, a situation which typically corresponds to tax haven affiliates. This result confirms large elasticities for profits (essentially paper profits) located in tax havens. 
}

\paragraph{Profit-shifting Elasticity: Robustness.}

So far, our estimation relies on estimation using firm-level data. 
To assess the sensitivity of our implied elasticities, we use the macro-estimate of profit shifting, $PS_{ilh}$, implied by our model and data on profit shifting provided by TWZ.
Table \ref{u2_rob_maintext} reports a summary of our results. 

\begin{table}[htbp]		
\begin{center}	
	\caption{Alternative identification of $\tilde{\upsilon}_2$}
	\label{u2_rob_maintext}
 \begin{tabular} {lcc}			
	\noalign{\smallskip}  \hline \noalign{\smallskip}			

	Data Source		&	$PS_{ilh}$	&	TWZ	 \\
					& (1)			 & (2) 	 \\ 
 	\noalign{\smallskip}  \hline \noalign{\smallskip}												

Implied $\tilde{\upsilon}_2$ 	&	5.205***	&	6.617***	\\
                                &	(2.000)	    &	(1.641)	\\
	\noalign{\smallskip}  \hline \noalign{\smallskip}												
Gravity Controls			&	Yes	&	Yes		\\
$i \times l$ fixed effects	&	Yes	&	Yes		\\
\noalign{\smallskip} \hline \noalign{\smallskip}
\end{tabular}											
\parbox{14.7cm}{\footnotesize Note: This table reports $\tilde{\upsilon}_2$ using alternative quantification of bilateral profit shifting. As in our baseline estimate, both columns use a PPML estimator. Full results reported in Table \ref{app:u2_rob_maintext}. In column (1), we use the macro-estimate of profit shifting, $PS_{ilh}$, implied by our model. In column (2), we directly use profit-shifting data from \citet{wier_global_2022} (WZ). }\\
\end{center}											
\end{table}

The results summarized in Table \ref{u2_rob_maintext} indicate that the implied profit-shiting elasticity derived from different data sources, is consistent with the reduced form estimation using firm-level data. 
Column (1) employs the macro-estimate of profit shifting, $PS_{lh}$, derived from our model, yielding an elasticity of 5.3 with a high level of statistical significance. Column (2) uses profit-shifting data from TWZ, resulting in an elasticity of 6.6, also highly significant.

\paragraph{Extension: a Variable Profit-shifting Elasticity.}\label{nonlinear_elasticity_main}

Our calibration of $\tilde{\upsilon}_2$ rests on the assumption that the share of profits shifted to tax havens is a constant elasticity function of the corporate tax rate. 
While this assumption is reasonable for small changes in corporate tax rates, policies like a minimum taxation reform could generate large variations in effective tax rates and tax rate differentials. 
We refine our parametrization of the profit-shifting elasticity and allow for an additional variable profit-shifting elasticity.   
We augment our profit-shifting friction $\alpha_{lh}$ with $\left(t_l-t_{lh}\right)^k$ where $k$ is a shape parameter.
The partial elasticity of profit shifting then becomes $\tilde{\upsilon_{2}}-k\left(\tilde{\upsilon_{2}}+1\right)\left(\sigma-1\right)\frac{1-t_{h}}{t_{l}-t_{h}}$. In this calibration, the shape parameter $k$ and the elasticity $\tilde{\upsilon_2}$ are determined by matching moments of the data. In particular, we calibrate the non-linear elasticity so that it equals the estimated linear elasticity when $t_{l}$ and $t_{h}$ are at their average value in the sample, and so that it stays larger than $\tilde{\upsilon_1}$, when the tax differential between non-haven countries and tax havens goes towards 1. $\tilde{\upsilon_1}$ is seen here as a natural upper bound for the elasticity of profit shifting when tax rates are very different. The details of the exercise are provided in Appendix \ref{app_nonlin}. 
In this setting, the non-linear elasticity will be above the linear elasticity for small tax rates differentials, as demonstrated by Figure \ref{nonlin} in the Appendix. 
This property will have implications for the implementation of the minimum tax rate. We implement this varying profit-shifting elasticity to simulate minimum taxation policy scenarios.

\subsection{Magnitude and Determinants of Profit-shifting Frictions \label{main_frictions}} 

The model allows us to back out the bilateral profit shifting frictions following Proposition \ref{prop_frictions}. 
In the following, we explore the magnitude and determinants of profit-shifting frictions.

\begin{figure}[htbp]
\begin{subfigure}[b]{0.499\textwidth}
\centering
     \includegraphics[width=1\linewidth]{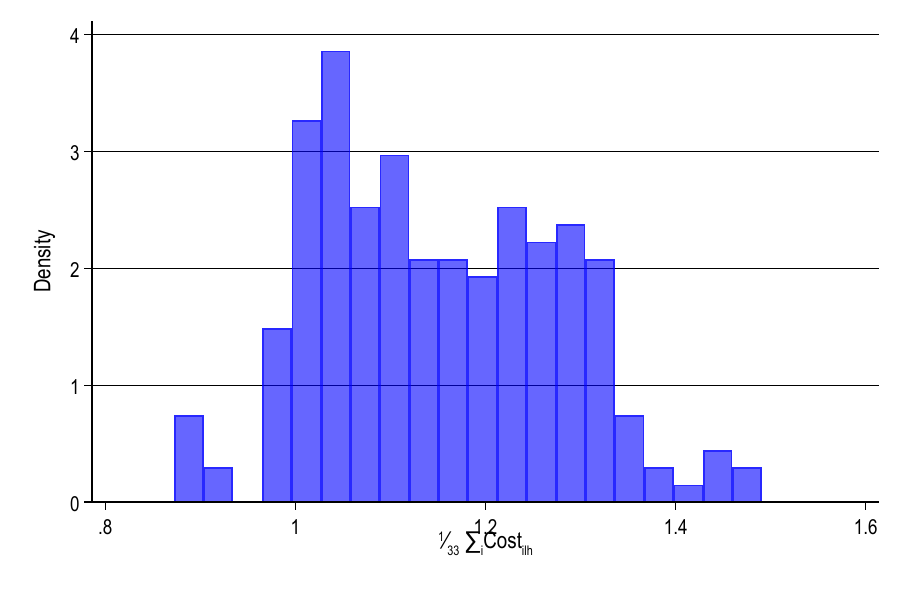} 
		\caption{Average cost of profit shifting ($Cost_{lh}$)} \label{fig:costps}
\end{subfigure}
\begin{subfigure}[b]{0.499\textwidth}
		\includegraphics[width=1\linewidth]{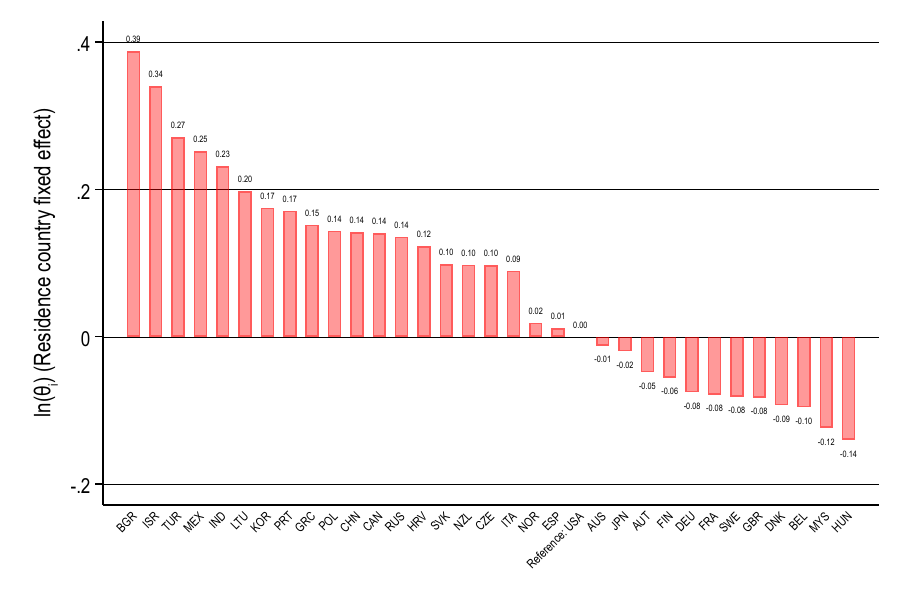} 
		\caption{Distribution of $ln(\theta_i)$} \label{fig:theta_i}
\centering
\end{subfigure}
\caption{Profit-shifting frictions}
\caption*{\footnotesize Note: Profit shifting frictions are estimated following Proposition \ref{prop_frictions}. Panel (a) show the distribution of $Cost_{lh}=\frac{1}{33} \sum_i Cost_{ilh}$. Panel (b) shows the distribution of $\theta_i$, the tax aggressiveness of the residence country. To obtain it, $Cost_{ilh}$ is separated into its residence-specific component ($\theta_i$) and its source-haven component ($\alpha_{lh}$) using fixed effects: $ln(Cost_{ilh})=ln(\tilde{\theta_i})+ln(\alpha_{lh})$.}
\label{fig:ps_frictions}
\end{figure}

\paragraph{Average Profit-shifting Costs.}
We start by describing the distribution of average $Cost_{ilh}:=\tilde{\theta }_i \alpha_{lh}$ between $l$ and $h$ in the panel (a) of Figure \ref{fig:ps_frictions}. 
We plot the distribution of the profit-shifting cost averaged over (non-haven) $i$ countries: $Cost_{lh}=\frac{1}{33}\sum_i Cost_{ilh}$.

Conditional on observing profit shifting, the median value of profit-shifting costs calculated in our sample is 1.12. 
A profit-shifting cost of 1.12 means that shifting from a residence country $i$ to a tax haven $h$ through a production affiliate $l$ generates an increase in the cost of production of 12\%, all other things being equal. 
The friction can be compared to the variable friction $\gamma_{il}$, which represents the costs of separating the location of production from headquarters. 
We find a median value of $\gamma_{il}$ of 1.5, somewhat larger than the multinational production costs of 1.31 provided by \citet{head2019brands} for the car industry. 

Given the magnitude of our estimated frictions, one should wonder why we observe any profit shifting since, for most triplets, the cost of shifting is higher than the tax advantage. For a given pair, frictions larger than the tax advantage reveal that the median firm does not shift profits. The firms who do book profits in tax havens are the selected ones for which the payoff of booking profits in tax havens is both the tax advantage $(1-t_{ilh})/(1-t_{ll})$ and their relative technology draw $\varphi_{lh}/\varphi_{ll}$. Whenever the technology advantage is large, firms might find it profitable to shift even if the frictions are large. In this sense, our model replicates salient empirical findings on firms' selection into profit shifting (\citealp{davies_knocking_2018}, \citealp{bilicka_comparing_2019}, \citealp{wier_2020jpube}, \citealp{wier_dominant_2023}).

\paragraph{Components of Profit-shifting Costs: $\tilde{\theta}_{i}$ and $\alpha_{lh}$.}
The profit-shifting cost has two components: the tax aggressiveness of the residence country $\tilde{\theta}_{i}$ and the bilateral friction $\alpha_{lh}$.
Our model suggests that the costs are separable via a fixed effect for $i$ and one for $lh$ pairs: $ln(Cost_{ilh})=ln(\tilde{\theta_i})+ln(\alpha_{lh})$. 
The residence country fixed effects correspond to the log of $\tilde\theta_{i}$. 
The source and tax haven dyadic fixed effects ($lh$) capture the bilateral profit-shifting frictions $\alpha_{lh}$.
About 39\% of the variation in profit-shifting costs is explained by the (log) bilateral frictions, $\alpha_{lh}$.\footnote{Note that the different abilities of each residence country to reduce the costs of shifting profits should be interpreted as deviations from the tax aggressiveness of one reference country that we choose to be the U.S.}

In Figure \ref{fig:theta_i}, we plot the cross-country distribution of $\ln(\theta_{i})$. 
In our sample, U.S. multinational corporations appear to be among the most aggressive in their profit shifting behavior alongside those from Western European countries such as Germany, the United Kingdom, and France. The MNCs from the majority of countries, except some Western European ones, incur cost penalties compared to U.S. MNCs.
This finding is similar to the results of \citet{klassen2012us} and \citet{delis2022global}. 
We find a relatively large dispersion in profit-shifting costs.
The differences in tax aggressiveness across residence countries in panel (b) of Figure \ref{fig:ps_frictions} show the key role of headquarters in firms' profit-shifting practices. 


Finally, we study the determinants of $\alpha_{lh}$ in Appendix Table \ref{table_costPS}. We show that gravity variables play an important role in explaining its variation. The distance between the source countries and tax havens is found to have a positive impact on the profit shifting cost while a former colonial domination by a source country on a tax haven reduces the cost of shifting profit there. In other words, profit-shifting frictions shape the geographic incidence of international tax reforms.


\section{Policy Simulations}\label{sec_counter}

This section analyzes policy simulations of international tax reforms across countries, examining their effects on tax revenues, GDP, profit shifting, real income, and welfare. 
It begins with a description of the methodological assumptions and an exploration of simple tax changes to highlight the model's core mechanisms. 
Subsequently, it quantifies scenarios of minimum taxation, evaluating these policies' short- and long-term impacts and their optimality. 
Finally, it assesses the introduction of a border-adjustment tax following the DBCFT proposal. In the following results, we generate counterfactual outcomes for 40 countries using aggregate data from 2015-2017 and focus primarily on outcomes for the U.S.
Our calibration procedure is summarized in Table \ref{calibration_data}. We assess the external validity of the calibration in Figure \ref{validation_tax_gdp}. 

\subsection{Methodological Assumptions and Model Mechanisms}

To simulate counterfactual tax reforms, we use the exact hat algebra, as developed by \citet{Dekleetal2007} and \citet{CostinotRC2014tradenumbers}. 
This method expresses the new equilibrium in proportional changes relative to the baseline (see Appendix \ref{app_eha} for details). 
We assume constant technological parameters and frictions throughout the analysis.

Importantly, our welfare metric cannot be inferred from observables only, as it relies on a preference parameter for the public good. We compute this parameter by assuming that the observed data as a Nash equilibrium. This allows us to pin down country-specific motives to obtain tax revenues. 
Formally, we use a revealed preferences approach and back out the vector of $\beta_n$ such that at the initial equilibrium, under territorial taxation, countries would not have an incentive to change their statutory rate. Namely, $ U_n$ must be maximized around the initial tax rates with $\beta^\star_n: \frac{\partial U_n(\beta^*_n)}{\partial t_n}=0,\,\forall n$. 
Then, log changes in welfare are $\frac{d  U_n }{ U_n}=\frac{dY_n/P_n}{Y_n/P_n}+\beta^\star_n \frac{dB_n/P_n}{B_n/P_n}$, namely a combination of private and public goods consumption. As demonstrated by Appendix Figure \ref{beta_and_taxrate}, the preference for public goods correlates well with the tax rate. Throughout our counterfactuals, we hold the vector of $\beta_n^\star$ fixed.

To illustrate the model's mechanisms and benchmark our results, Appendix Table \ref{scenario_appendix} simulates the impact on the U.S. economy of three simple reforms. First, a 5\% reduction in the U.S. tax rate highlights a key trade-off in corporate taxation: lower tax rates boost real income (+0.44\%) but reduce tax revenues (-4.21\%). 
The decline in tax revenues is less than mechanically expected due to decreased profit shifting (-6.36\%) and a rise in GDP (+0.34\%). 
By definition of our welfare function, any deviation from the pre-reform statutory rate reduces welfare (-0.04\%). 
Second, turning Singapore into a non-haven country has minor effects on profit shifting (-3.99\%) and tax revenues (+0.29\%) as profit shifting reallocates, notably to Hong-Kong. 
Tax revenues rise more in countries with lower profit-shifting frictions with Singapore, like Malaysia and Japan (Figure \ref{fig:close_sgp}), highlighting the importance of geography and bilateral profit-shifting frictions in determining the winners and losers from such reform. 
Lastly, eliminating profit shifting altogether increases U.S. corporate tax revenues by 7.24\% but decreases real income by 0.52\%. 
Overall welfare improves (+0.25\%) due to the U.S. preference for tax revenues over private consumption.

\subsection{Minimum Taxation}
\paragraph{Policy Context.} The principle of minimum taxation ensures that no foreign affiliate can avoid a minimum tax rate $t^{min}$ by declaring operations in low-tax jurisdictions. 
However, the implementation of minimum taxation raises significant challenges, particularly in relation to the allocation of taxing rights. 
A key issue lies in determining which jurisdiction should have the priority to collect the minimum tax.
This decision depends on whether value creation is primarily seen as coming from the location of the headquarters, the research and development activities, or the production of physical output (see \citealp{devereux_taxing_2021}). 
Consequently, taxing rights on profits $\pi_{ilh}$ can be assigned to the residence country ($i$), the source country ($l$), or the location where the profits are recorded ($h$).

Against this backdrop, countries negotiating the global minimum tax initially proposed granting priority to residence countries for collecting these taxes. 
If residence countries choose not to collect the tax, source countries can then collect it based on their share of the group's activities.\footnote{This approach is known as the \textit{Income Inclusion Rule} (IIR) for residence countries and the \textit{Undertaxed Profits Rule} (UTPR) for source countries.} 
Later, the \textit{Qualified Domestic Minimum Top-Up Tax} (QDMTT) was introduced, prioritizing the jurisdiction where profits are booked, potentially tax havens, over residence countries. 
This motivates us to study the impact of these different taxing rights allocations under the global minimum corporate tax rate of 15$\%$ established by the OECD/G20 Inclusive Framework on BEPS.

Finally, a common concern regarding the introduction of a minimum effective tax rate is the risk of corporations relocating their headquarters to countries that do not enforce this rate. 
We study this potential endogenous response by using our model to distinguish between the short-run effects (with a fixed number and location of headquarters) and the long-run effects (where both the number and location of headquarters adjust endogenously). 
In the short-run scenario, corporate inversions are not permitted by design, although multinational firms can still shift production across countries in both scenarios. 

We study all these counterfactuals under the assumption that real activities are fully deductible from the base of the minimum tax. 
Consequently, the minimum tax only applies to shifted profits, $PS_{ilh}$. 
This assumption is motivated by the OECD/G20 Inclusive Framework agenda's intended goal, to tackle the erosion of the tax base through profit shifting and not through tax competition for real resources. This assumption mimicks the substance-based carve-out that aims at deducting profits from real activity from the minimum tax base (see \citealp{schjelderup_economics_2024} for a theoretical analysis). 
Table \ref{MinTaxUS} presents the results for these scenarios, showing the outcomes for unilateral and multilateral implementation of the reform in the short-run (Panel A) and long-run (Panel B).
 

 \begin{table}[htbp]
	\centering
	\caption{Impact of a 15\% minimum tax rate for the U.S.}\label{MinTaxUS}
\scalebox{.85}{	\begin{tabular}{lc ccccc} \toprule
		&&\multicolumn{5}{c}{\% change in ...} \\ 
		\textbf{Minimum Taxation}   &		&	 Tax         & Profit 	&	Real 	&	Consumer 	& Welfare\\
	                               	&		&	 revenues   &  Shifting	&		Production        &	 Real Income	&\\
 \midrule
                                     & \multicolumn{6}{c}{\textbf{A. Short Run}} \\
\noalign{\smallskip}   
        Unilateral  \\
 \noalign{\smallskip}  \cline{1-2}   \noalign{\smallskip}
    	-- Residence	    &	&	4.1	&	-35.6	&	0.03	&	-0.04	&	0.40	\\
-- Source	        &   &	4.21	&	-38.0	&	-0.02	&	-0.07	&	0.38	\\
\noalign{\smallskip}    \noalign{\smallskip}
Multilateral \\
\noalign{\smallskip}  \cline{1-2}   \noalign{\smallskip}

-- Residence 	    &	&   4.23	&	-37.3	&	0.03	&	-0.04	&	0.42	\\
-- Source 	        &   &	4.22	&	-37.3	&	0.03	&	-0.04	&	0.42	\\
\midrule
\noalign{\smallskip}   
& \multicolumn{6}{c}{\textbf{B. Long Run}} \\
Unilateral  \\
\noalign{\smallskip}  \cline{1-2}   \noalign{\smallskip}
-- Residence	    &	&	3.89	&	-35.4	&	-0.14	&	-0.26	&	0.17	\\
-- Source	        &   &	3.97	&	-37.1	&	-0.18	&	-0.05	&	0.38	\\
\noalign{\smallskip}    \noalign{\smallskip}
Multilateral \\
\noalign{\smallskip}  \cline{1-2}   \noalign{\smallskip}
-- Residence 	    &	&  4.05	&	-37.5	&	-0.15	&	-0.23	&	0.21	\\
-- Source 	        &   &	4.04	&	-37.5	&	-0.15	&	-0.23	&	0.21	\\
-- Tax havens' adjustment &   &   2.71	&	-37.5	&	-0.15	&	-0.23	&	0.07	\\
  \midrule
\bottomrule
	\end{tabular}}
\end{table}
   \paragraph{Unilateral Minimum Taxation.}  
Under a residence-based minimum tax rate $t^{min}$, the U.S. can tax U.S. MNCs that continue to shift profits to tax havens at a rate that is equal to the difference between the minimum rate and their effective tax rate ($t^{min}-t_{ilh}$), regardless of the source countries where they operate. Additionally, the reform also endogenously increases the U.S. tax base as some U.S. firms operating in the U.S. no longer find it profitable to engage in profit shifting.
As a result, corporate tax revenues in the U.S. increase (+4.1\% in the short run) because of both the reduction in profit shifting (-35.6\%) and the implementation of the minimum tax.
Ex-ante, the impact of residence-based minimum taxation on production is ambiguous. 
Under the minimum tax regime, U.S. firms give more weight to U.S. fundamentals ($A_{U.S.}$) and less weight to the effective tax rate when deciding where to book profits and allocate production. We find that the minimum tax positively affects production (+0.03\%), but negatively impacts real income (-0.04\%). The effect on welfare is positive and sizeable (+0.4\%), primarily driven by increased tax revenues.\footnote{Recall that our status quo that define the preference for public-good consumption is a Nash equilibrium where the instruments are the unilateral statutory rates. Instead, the minimum tax is a different instrument that induces changes to the tax rates applied to shifted profits $t_{ilh}$ for $h\neq l$, not to the statutory tax rates. Hence deviations from the status quo with minimum taxation can be welfare improving.} 

Our findings show that the effects of implementing a \textit{unilateral source-based minimum tax} would differ from those described above. Under this scenario, the effective tax rate of all profit-shifting firms operating in the U.S. increases, resulting in a decrease in production by 0.02\%. The overall real income effect is negative with a decrease of 0.07\%. The impact on welfare is smaller than in the residence-based scenario, with an increase of 0.38\%. 

In the long run, the exit of headquarters weakens the positive impact on welfare by decreasing private consumption. 
This negative effect on real income is especially significant when a residence-based minimum tax targets all U.S.-headquartered firms, compared to a source-based minimum tax that targets firms operating in the U.S. 
In fact, a unilateral source-based minimum tax proves similarly beneficial for welfare in the long run, yielding a 0.38\% increase.

\paragraph{A Global Minimum Tax.}
Implementing a multilateral minimum tax reduces the dispersion of effective tax rates across countries and increases them for all tax-avoiding firms, regardless of their headquarters. 
The distribution of corporate tax rates remains the same in both the residence and source scenarios since firms face the same minimum tax, whether levied by the countries of operation or headquarters. 
As a result, the direct effects on profit shifting and production are identical. However, in general equilibrium, these effects may differ as the two reforms allocate tax revenues differently across countries. 
For the U.S., residence- and source-based minimum taxation lead to a similar increase in tax revenues.

From a global efficiency perspective, the long-run effect of the reform is mixed. It raises the effective tax rate and reduces its dispersion across countries, impacting welfare in opposing ways. 
Higher tax rates boost public good provision, increasing welfare, but also lead to headquarters relocation, which reduces private consumption -- a long-run effect only. 
Additionally, the lower dispersion in tax rates decreases the influence of tax factors on firms' decisions, leading them to base choices more on countries' economic fundamentals, which improves the spatial allocation of activity and, therefore, welfare.

In the short run, the reform increases public good provision without affecting product variety. 
In the long run, however, the exit of firms has a greater negative impact on real income, as shown in Panel B of Table \ref{MinTaxUS}, where this effect outweighs the efficiency gains from smaller tax dispersion. 
Nevertheless, welfare increases in all scenarios, as the rise in public consumption compensates for the reduction in private consumption.

\begin{figure}[htbp]
\centering
\begin{subfigure}[b]{0.9\textwidth}
    \centering
    \label{fig:mintax_1}
    \centering\hspace{-1cm}\includegraphics[width=.8\linewidth]{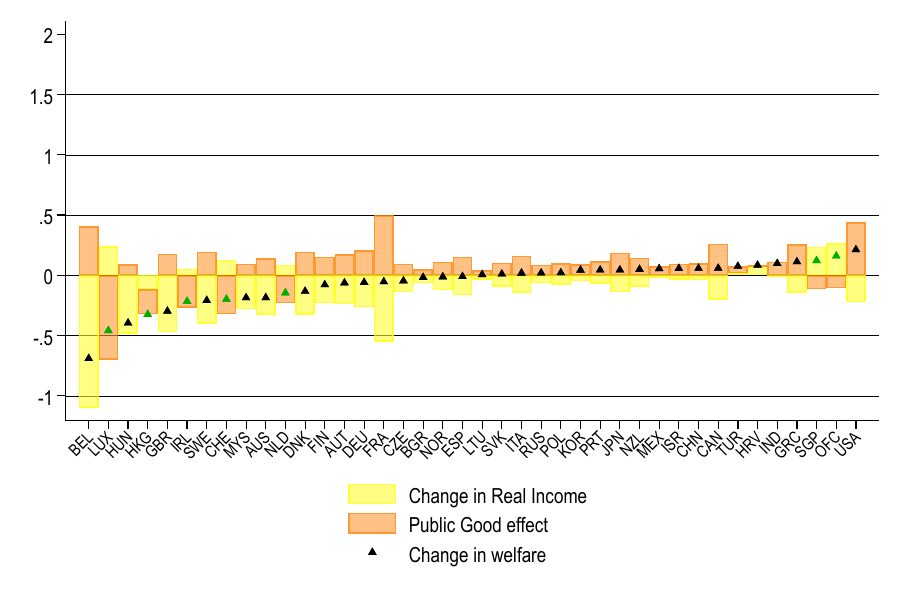}
        \caption{Multilateral Residence-based Minimum Tax}
\end{subfigure}
\begin{subfigure}[b]{0.9\textwidth}
    \label{fig:mintax_2}
    \centering\hspace{-1cm}\includegraphics[width=.8\linewidth]{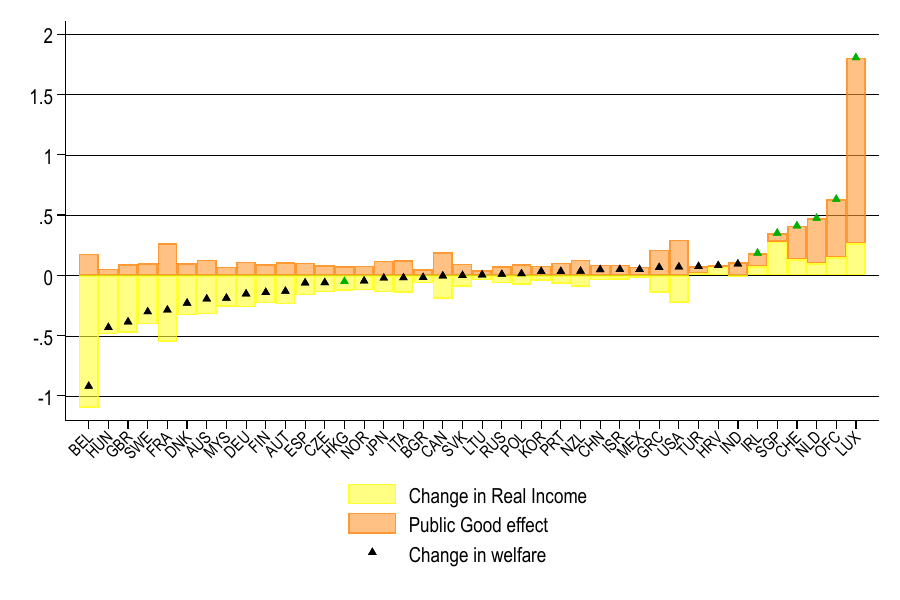}
                \caption{Multilateral Residence-based Minimum Tax - Tax Havens Response}
\end{subfigure}
 \caption{Effect of a 15\% multilateral residence-based minimum tax.}
	 \caption*{\footnotesize Note: Real income of country $n$ is defined as its real income ${Y_n}/{P_n}$. Welfare of country $n$ is defined as $\tilde U_n=(B_n/P_n)^{\beta_n} {Y_n}/{P_n}$. The public good effect denotes the change in $(B_n/P_n)^{\beta_n}$. The change in welfare corresponds to the sum of the change in real income and the public good effect. Bars are stacked. See Section \ref{model_welfare} for details on the calibration of $\beta_n$. Green triangles denote tax havens. Panel (a) shows the long-run residence-based mintax scenario while Panel (b) shows the scenario in which tax havens adjust their tax rates.}
	\label{pubgood_allcountries}
\end{figure}

Panel (a) in Figure \ref{pubgood_allcountries} illustrates the welfare changes from a 15\% residence-based \textit{multilateral} minimum tax. 
Most countries, particularly non-havens, see a net welfare gain, while only a few non-haven countries experience significant real income losses that are not offset by public good provision. 
Tax havens generally lose from the reform as multinational enterprises reduce profit shifting. 
However, some havens, such as OFCs and Singapore, compensate for this loss through gains in real income due to the reallocation of economic activities, as firms react by relocating real activity to these havens where the effective tax rate remains unchanged.

Overall, these findings suggest that a residence-based multilateral minimum tax can generate welfare gains for most countries.
\paragraph{Tax Havens' Response to the Minimum Tax.}
Implementing minimum taxation must account for tax havens' incentives to adjust their corporate tax regimes\footnote{For theoretical analyses, see, e.g., \citealp{janeba_global_2022}, \citealp{johannesen_global_2022}, and \citealp{hebous_pareto_2022}.}. 
In our model, tax havens respond to the reform by setting their tax rate to the minimum threshold $t_{lh}=t^{min}$ to capture tax revenues that would otherwise go to source or residence countries, as permitted by the OECD QDMTT mechanism.\footnote{This adjustment does not necessarily mean that tax havens raise their statutory tax rates; see the example of Ireland in Footnote 3.} 
At the firm level, the effective tax rate remains unaffected by the allocation of taxing rights, leading to the same impact on profit shifting, real production, and real income regardless of which jurisdiction collects the tax. 
However, at the country level, tax revenues shift from non-havens to tax havens. 
Despite this shift, the gains from a global minimum tax persist, as the overall reduction in profit shifting broadens the tax base of non-haven countries. For the U.S., tax revenues increase by 2.71\%, leading to a welfare gain of 0.07\%.

Appendix \ref{app_equilibrium_effects} further decomposes the effects of the reform into mechanical impacts and firm responses in general equilibrium. 
Introducing a residence-based multilateral minimum tax initially raises tax revenues mechanically by 2.2\%, but in general equilibrium, the total increase in tax revenues is larger (+4.05\%). 
This increase is primarily driven by reduced profit shifting, which broadens the CIT tax base (+2.84\%, representing 129\% of the mechanical effect) while slightly reducing the minimum tax base (-0.86\%). 
Additionally, the relocation of production slightly decreases tax revenues (-0.12\%). 
When tax havens adapt, the minimum tax base becomes negligible, and the reduction in profit shifting remains the dominant effect. 
These results underscore the importance of accounting for firms' endogenous profit-shifting behavior and general equilibrium effects in evaluating tax policy reforms.

Panel (b) in Figure \ref{pubgood_allcountries} shows the distribution of welfare effects when tax havens respond to the mintax reform. This response only shifts the allocation of tax revenues across space. Since the increase in public good consumption was a key driver of welfare gains for non-havens and welfare losses for havens, this reallocation significantly changes the winners and losers from the reform. In particular, tax havens rip large benefits from the minimum tax both in terms of public and private good consumption. The intuition behind this result is instructive. Our model features competition among tax havens. This competitive force implies that while havens may want to increase their tax rate to obtain higher tax revenues, they cannot do so as profit-shifting incomes are very elastic to tax rates. The introduction of the global minimum tax reduces the competitive pressure, operating like a coordination device for tax havens. As a consequence, low-tax jurisdictions can obtain larger tax revenues without the fear of being undercut by competitors.

\paragraph{The Optimal Minimum Tax Rate.}
Is the 15\% minimum tax rate chosen by OECD countries optimal? 
Our framework is designed to address this question through various scenarios of taxing rights allocation. The objective function is global welfare, defined as the sum of country-level welfare. It formally writes:
$\max_t \bar{\mathcal{W}({t})} =\sum_{i}\tilde{U}_{i}(t)$.\footnote{Note that this is equivalent to using per capita utility and population weights to aggregate across countries. }

\begin{result}
The distribution of taxing rights critically shapes the welfare impact of the global minimum tax. Source-based and residence-based scenarios result in superior outcomes, with optimal welfare achieved at a 23\% tax rate. Tax havens' response through Qualified Minimum Domestic Top-up Tax (QMDTT) and non-haven response through Corporate Income Tax (CIT) adjustments result in lower welfare gains.
\end{result}

\begin{figure}[htb]
	\centering
	\includegraphics[width=0.7\linewidth]{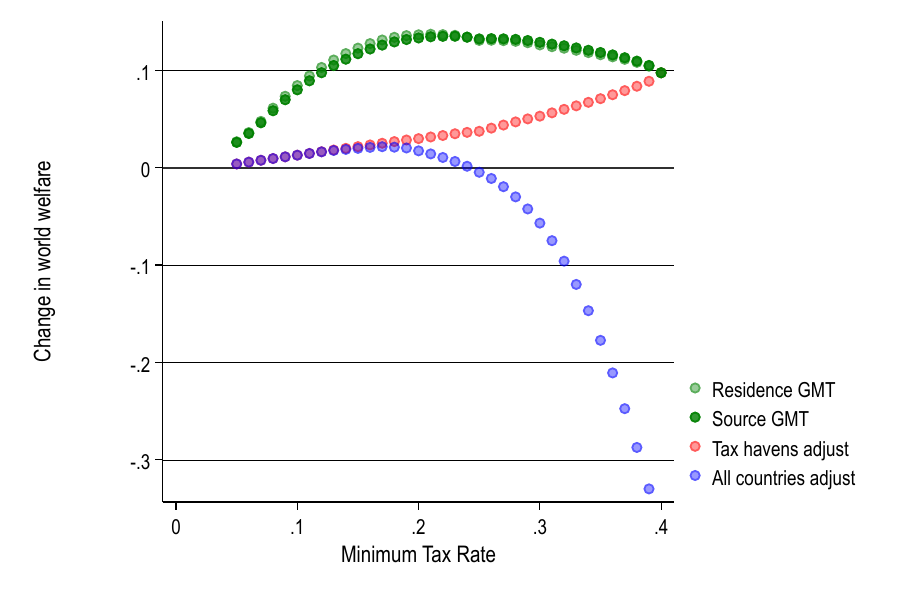}
	\caption{Optimal minimum tax rate.}
 \caption*{\footnotesize Note: This figure plots changes in world welfare for different minimum taxation scenarios, as described in the text. The residence GMT has an optimum at 21\%. The source GMT has an optimum of 23\%. The scenario where tax havens adjust reaches a maximum for a rate of 40\%, where all profit shifting is eliminated. When all countries adjust their tax rates, the optimum is reached for a rate of 17\%. At the optimum rates, profit shifting respectively declines by 66\%, 72\%, 100\%, and 52\%.}
	\label{optimal_min_tax}
\end{figure}

The results, shown in Figure \ref{optimal_min_tax}, suggest that the optimal global cooperative minimum tax rate lies between 17\% and 40\%. 
Across all scenarios, the global minimum tax generates positive welfare gains at rates between 0\% and 24\%. 
Below, we analyze the effects of these different tax designs on welfare in detail.

Early debates focused on two main scenarios: allocating minimum taxing rights to residence countries or to source countries. 
Our model indicates that, from a global welfare perspective, the distinction is minor. 
This outcome arises because the aggregate sizes and preferences for tax revenues of countries affected by profit shifting and those engaged in it are similar. 
Importantly, the optimal tax does not entirely eliminate profit shifting but still yields positive welfare gains. In the cases when havens do not adjust, the optimal rate is 21\% for residence-based taxing and 23\% for source-based taxing, both increasing global welfare by about 0.14\% (green lines in Figure \ref{optimal_min_tax}). These scenarios respectively reduce world profit shifting by 66\% and 72\%. 

However, these policies are likely to prompt tax havens to adjust their tax systems through mechanisms like the QDMTT. Fixing the minimum tax rate, any such adjustment induces only a reallocation of tax revenues across countries, with tax havens capturing the reform's gains. 
In this case (red line), global welfare is lower compared to the scenarios prioritizing residence or source countries, as the benefits accrue mainly to tax havens, which have a lower preference for tax revenues. As the minimum tax rate increases, however, profit shifting is reduced. As a consequence, at high rates, the gains shift to non-haven countries. At the rate of 40\%, profit shifting is completely eliminated. Hence, the three scenarios coincide.

All these scenarios consider reforms in which countries can effectively distinguish between profit shifting and real economic activity -- a rationale behind the ``substance-based carve-out''. 
Without this carve-out or if discrimination between real profits and profit shifting fails (blue curve), the minimum tax applies broadly, effectively harmonizing statutory tax rates. 
In this situation, welfare gains peak at a 17\% tax rate but decline beyond this point as countries raise taxes beyond their optimal preferences. Welfare effects remain positive up to a 24\% tax rate. The implementation of a 17\% rate leads to a 52\% reduction in global profit shifting.

%

\paragraph{Effects of Minimum Taxation on Tax Competition.}\label{unilateral_Nash}
The minimum tax reform, by changing the global allocation of taxing rights and tax rates, could influence countries' statutory corporate tax rates.
To study this issue, we start from the Nash equilibrium induced by the vector of $\beta_n^*$, implement the minimum tax reform with a rate of 15\% when tax havens adjust, and let countries unilaterally change their statutory rate at the margin (0.1 percentage point increase). To identify which countries benefit from such deviation, we compute the change in welfare $\left(\frac{\partial  U_n}{\partial t_n}\right)$.\footnote{We leave the resolution of the new Nash equilibrium for future work. Although the general approach of \citet{ossa_trade_2014} and \citet{wang2020multinational} could be adapted to our framework, a policy-relevant study should thoroughly consider all available instruments, from subsidies (e.g., those announced by Switzerland following the global tax deal) to domestic minimum taxes (e.g., the U.K.).}
\begin{figure}[htbp]
	\centering
	\includegraphics[width=0.5\linewidth]{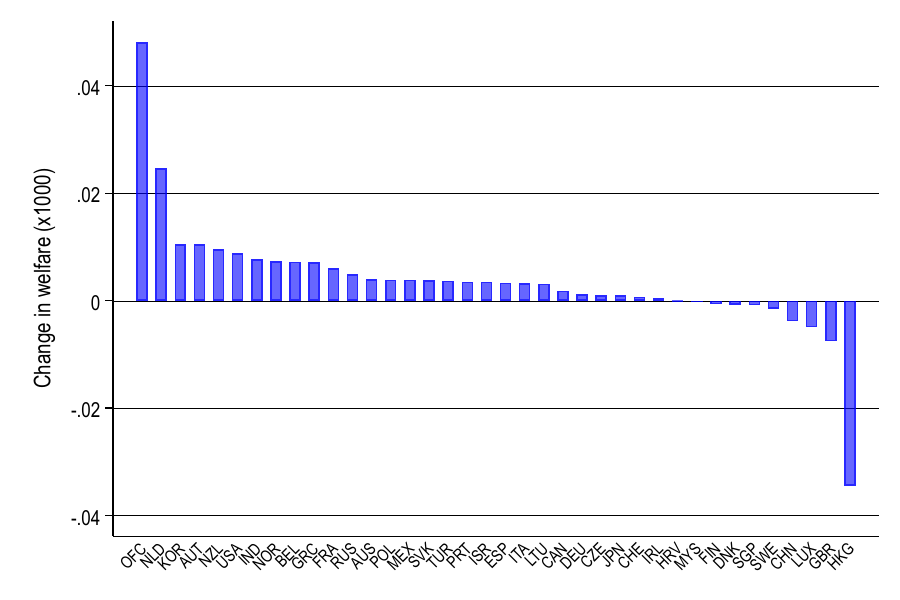}
	\caption{Marginal welfare effect of the statutory tax rate after a 15\% global minimum tax.}
	\label{unilateral_deviations}
\end{figure}

\begin{result}
A 15\% minimum tax reduces tax competition incentives for most countries.
\end{result}

Figure \ref{unilateral_deviations} plots the change in welfare implied by a marginal increase in the statutory rate. 
Most countries would benefit from a unilateral increase in their statutory rate. Intuitively, they would trade-off a loss in real income with an increase in real tax revenues. 
The introduction of the global minimum tax reduces the cost of increasing the statutory rate as it limits the erosion of the tax base through profit shifting. This, in turn, implies that countries are incentivized to increase their statutory tax rate and suggests that the global minimum tax can have beneficial effects on tax competition and on the race to the bottom.

 \subsection{Destination-Based Cash Flow Tax (DBCFT)}
So far, we have studied the impact of minimum tax reforms. A 15\% minimum tax could reduce U.S. profit shifting by at most 30-40\% and increase U.S. welfare by 0.07\% in the most likely scenario.
This raises the question: is it possible to find an alternative design that eliminates profit shifting without compromising efficiency? 
We explore a reform akin to the destination-based cash flow tax (DBCFT). This policy proposal has several distinctive features. First, it eliminates opportunities for profit shifting that exist in the current system (\citealp{auerbach_international_2017}). The intuition is simple: when taxation occurs at the location of sales, tax revenues accrue in the country of consumption rather than where profits are recorded. Second, it can be implemented unilaterally. Third, it relates to Lerner symmetry, which provides a benchmark for efficiency.

Table \ref{tab:dbcft} presents our results for different values of the CIT-equivalent DBCFT rate in the United States, building on our theoretical discussion of this tax in section \ref{overview}. For comparison, we also include simulation results for the adoption of a BAT in Japan. Table \ref{dbcft_breakdown} in the Appendix further breaks down government revenues after the reform between revenues coming from domestic sales and revenues coming from the border adjustment. It also decomposes the change in GDP between the activity of domestic firms and the activity of foreign firms in the domestic market.

To build intuition, consider the introduction of DBCFT in the U.S., such that the profits from domestic sales of domestic firms are taxed at 5\%. 
This is equivalent to an 87.5\% reduction of the corporate tax rate, from 40\% to 5\%. 
The government revenues from domestic firms fall by 78\%. This can be decomposed into -87\% real revenues from domestic sales of U.S. firms and +9\% from border adjustment revenues. These accrue as the U.S. is running a trade deficit and, therefore, obtains larger revenues from imports than it spends to subsidize exports. 
Firm entry in the U.S. boosts the demand for labor, which, in turn, generates increases in wages and income. However, these gains are approximately halved by the full pass-through of the border tax to CPI (+5\%), in line with \citet{barbiero2019macroeconomics}.  
The net gain in real income is undone by the large drop in tax revenues, which translates into an overall large negative welfare effect of -10\%. 

The insights of this relatively small policy carry through to higher rates. 
Higher DBCFT rates generate smaller revenue drops from domestic firms but reduce the contribution of the border adjustment and, importantly, generate a strong appreciation of the terms of trade. For most rates, we find a trade-off between the change in revenues and in household consumption. For example, at a CIT equivalent rate of 40\%, which leaves the effective tax rate on domestic sales of domestic firms unchanged, we estimate gains in terms of tax revenues but losses in terms of consumption. For smaller rates, around 30\%, we find that there is no trade-off. The U.S. experiences gains both in terms of real tax revenues and consumption. The gains in tax revenues come about as the negative effect of lowering the effective CIT is dominated by the BAT contribution due to the trade deficit. Similarly, real activity increases as foreign firms move to the U.S. and more than compensate for the decline in domestic firm production. We find that the unilaterally optimal DBCFT for the U.S. corresponds to a CIT equivalent rate of 33\%. This policy is associated with a 2.1\% increase in welfare. 
Finally, we study the adoption of a Border Adjustment Tax (BAT) in Japan, a large economy with trade surpluses. The BAT implementation in this context of large trade surpluses leads to a significant decline in tax revenues (-30\%), resulting in welfare losses (-2.4\%). Japan's optimal BAT rate is 25\%, which results in a welfare decline of 2.2\%, indicating that maintaining the status quo is preferable. This analysis reveals that the welfare effects of a Destination-Based Cash Flow Tax (DBCFT) are primarily determined by a country's trade balance, leading to substantial fluctuations in tax revenues and prices. These fluctuations are significantly larger than those observed under cooperative minimum taxation.

\begin{result}
    Unilaterally replacing the CIT with a DBCFT induces welfare variations that are an order or magnitude higher than the minimum tax. While DBCFT also entails a private vs. public good consumption trade-off, the net impact is highly sensitive to trade imbalances.
\end{result}

Returning to our comparison with BAT based on Lerner symmetry, we can quantify the importance of each deviation for our non-neutrality result. Considering the 40\% rate in the U.S., which represents a pure BAT, we find that real income ($Y_n+B_n$) declines by 2.3\%. In a world with Lerner symmetry, this should be exactly zero. We find that the presence of MP imbalances alone would have implied a drop of 1.67\%, while windfall from eliminating profit shifting induces a 0.11\% gain. 
\begin{table}[htb]
  \centering
    \caption{Implementation of DBCFT}
    \label{tab:dbcft}
\scalebox{0.92}{ \hspace{-.7cm} \begin{tabular}{@{\extracolsep{.2em}}lccccccccc}
    \toprule
    & \multicolumn{8}{c}{\% change in ... } & \\ \noalign{\smallskip} \cline{2-9}  \noalign{\smallskip}
    & \multicolumn{3}{c}{Real} & \multicolumn{3}{c}{Nominal} & && \\ \noalign{\smallskip} \cline{2-4} \cline{5-7}  \noalign{\smallskip}
    CIT-equiv. rate  	& Tax Rev.	&	GDP	& Income &Tax Rev.	&	GDP	& Income &	P &	Welfare & $\frac{NX'}{GDP'}$ \\\midrule
    \textbf{USA} & & & & & & & & & \\
5\%	&	-78.4	&	4.5	&	6.3	&	-77.3	&	4.2	&	11.6	&	5.0	&	-10.3	&	-2.4	\\
10\%&	-57.9	&	4.1	&	5.2	&	-53.4	&	3.5	&	16.3	&	10.6	&	-4.4	&	-2.2	\\
20\%&	-19.6	&	3.0	&	2.9	&	-0.4	&	2.0	&	27.5	&	23.9	&	0.5	&	-1.9	\\
30\%&	14.8	&	1.5	&	0.5	&	62.0	&	0.2	&	41.7	&	41.1	&	2.0	&	-1.6	\\
33.3\%$^\star$&	25.4	&	1.0	&	-0.4	&	85.5	&	-0.5	&	47.3	&	47.9	&	2.1	&	-1.5	\\
40\% (BAT)	&	44.6	&	-0.2	&	-2.3	&	137.0	&	-2.0	&	60.2	&	63.9	&	1.8	&	-1.2	\\ \midrule
    \textbf{Japan} &&&&&&&&& \\
    25\%$^\star$&	-39	&	1.6	&	1.2&	-19.3	&	0.7&	33.8	&	32.3	&	-2.2&	3.5	\\
    31\% (BAT)	&	-30&	1.2	&	-0.02	&	0.3	&	0.1	&	43.2	&	43.2	&	-2.4	&	4	\\ \bottomrule

    \end{tabular}}
    \caption*{\footnotesize Note: This table shows the impact of a unilateral adoption of DBCFT by the United States and Japan. The CIT equivalent rate corresponds to $1-\frac{1}{1+tr_{US}}$ with $tr_{US}$ being the DBCFT rate. The $^\star$ marks the optimal unilateral CIT-equivalent tax rate, and ``BAT'' identifies that for a CIT-equivalent tax rate equal to the statutory tax rate pre-reform, DBCFT corresponds to a pure border-adjusted tax.}
    \end{table}

Overall, we find that the U.S.\ would experience substantial welfare gains from unilaterally adopting a DBCFT. These gains stem primarily from its trade deficit, which generates increased tariff revenues. While the policy leads to decreases in real income and real GDP, these are outweighed by the large increase in tax revenues, resulting in net welfare gains. As a consequence, if a government prioritizes only private consumption (formally, $\beta_n=0$), then abolishing corporate taxes and implementing a DBCFT at around 5\% can generate sizeable gains. However, when households value public good consumption, optimal DBCFT rates are substantially higher.


\section{Conclusion}\label{conclusions}

The current international corporate tax system is outdated because it is not robust to a variety of tax avoidance strategies used by firms to shift their profits to tax havens.
The ongoing reform of international taxation discussed in the OECD/G20 Inclusive Framework is meant to crowd out profit shifting by implementing a multilateral residence-based minimum taxation. 
This paper examines this tax policy proposal as well as alternative reforms such as Destination-Based Cash-Flow Taxation (DBCFT).  

We use a general equilibrium model of multinational production augmented with corporate taxation and profit shifting to assess the short- and long-run consequences of different scenarios of domestic and international corporate taxation reforms. 
Our focus is on real activity and welfare, in addition to tax revenues and profit shifting. 
The model delivers a set of simple equations to recover the distribution of profits shifted across source-haven country pairs. 
Exploiting our theoretical framework, we derive profit-shifting frictions and the tax-base and profit-shifting elasticities, which are key determinants of how changes in the tax environment affect entry, production, and profit-shifting decisions.
We highlight the importance of profit-shifting frictions and the role of geography in shaping profit shifting and production locations.  

Our findings show that a global minimum tax improves welfare in most countries. According to the allocation of taxing rights under minimum taxation, the optimal multilateral tax rate is between 17\% and 40\%. In contrast with some critics of the proposal, we find little support for a ``race to the minimum tax''. Instead, we find that a global minimum tax reduces the cost for countries to increase their corporate tax rate. We benchmark this multilateral reform against a unilateral border-adjustment tax that eliminates profit shifting. We find that border-adjusted taxes generate large variations in prices and tax revenues, driven by trade balances. The United States would experience large welfare gains due to its trade deficit, while Japan, a country maintaining trade surpluses, would experience large welfare losses.


	\phantomsection
	\addcontentsline{toc}{section}{References}
\printbibliography[heading=subbibintoc]	
\end{refsection}	
\begin{refsection}
	\clearpage
		\appendix
\counterwithin{figure}{section}
\counterwithin{table}{section}
\renewcommand\thefigure{\thesection\arabic{figure}}
\renewcommand\thetable{\thesection\arabic{table}}
	\setcounter{footnote}{0} 
\setcounter{page}{1} 

\addcontentsline{toc}{section}{Appendix} 

\renewcommand \thepart{}
\renewcommand \partname{}

\begin{center}
    \part{Online Appendix}
\end{center}
\vspace{-1cm}
\singlespacing 
\parttoc 

\onehalfspacing 

\section{Model\label{appendix_theory}}

\subsection{Representation of the model}

\begin{figure}[ht]
\hspace{-1cm}	\centering
	\includegraphics[width=1\linewidth]{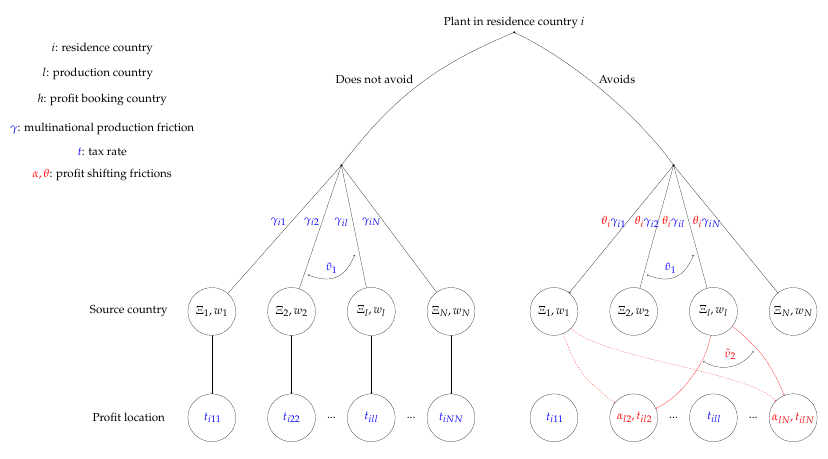} \\
      \tiny{Note: The \color{red} red color \color{black} refers to the proﬁt shifting activity of the ﬁrms and the \color{blue}blue color \color{black} to their real activity. Countries \color{red}2 \color{black} and \color{red}N \color{black} are tax havens.}
    	\caption{Structure of the theoretical framework }\label{graphmodel}
\end{figure}

Figure \ref{graphmodel} shows a schematic representation of the model under a territorial taxation regime. 
For non-tax avoiders, all taxes are levied where production takes place, in country $l$.
The location choice depends on corporate tax rates $t_{ill}$, market size and geography embedded in $\Xi_{l}$, and wages, $w_l$.
For tax avoiders, multinationals producing in non-haven countries can transfer their profits to a tax haven (countries 2 and N) upon paying the cost $\alpha_{lh}$.

\subsection{Proof of proposition \ref{gravityprop}} \label{proofgravity}

Taking equations (\ref{probafrechet}) and (\ref{profitGEV}) together, we have:

	$$ \frac{X_{ilh}}{X_{i}}=\frac{\tilde{A}_{ilh}(1-t_{ilh})^{\frac{\upsilon_{1}}{\sigma-1}-1}\iota_l^{-1}G_{i,lh}({\bf \tilde{A}}_{i},{\bf t})}{\sum_{jk}\tilde{A}_{ijk}(1-t_{ijk})^{\frac{\upsilon_{1}}{\sigma-1}-1}\iota_j^{-1}G_{i,jk}({\bf \tilde{A}}_{i},{\bf t})} $$

Thus, we can deduct easily:
\begin{eqnarray}
		\frac{X_{ilh}}{\sum_{l,h,h\neq l}X_{ilh}} =&\frac{\tilde{A}_{ilh}^{\frac{\upsilon_{2}}{\upsilon_{1}}}(1-t_{lh})^{\frac{\upsilon_{2}}{\sigma-1}-1}}{\sum_{l,h,h\neq l}\tilde{A}_{ilh}^{\frac{\upsilon_{2}}{\upsilon_{1}}}(1-t_{lh})^{\frac{\upsilon_{2}}{\sigma-1}-1}} \nonumber
	\end{eqnarray}


\clearpage

\section{Estimation of profit shifting: Theory} \label{app_estimation_PS}

\subsection{Proof of Proposition \ref{Pilhdecomp} \label{app_triangle_ps}}
\begin{proof} Using equation (\ref{probafrechet}) and the specific $G(\cdot)$ function, the statement follows by defining $\zeta_{il}$ and $\chi_{lh}$ that are given are given by
\begin{align*}\zeta_{il}=\frac{\sum_{h,h\neq l}\tilde{A}_{ilh}^{\frac{\upsilon_{2}}{\upsilon_{1}}}\left(1-t_{lh}\right)^{\frac{\upsilon_{2}}{\sigma-1}}}{\sum_{l,h,h\neq l}\tilde{A}_{ilh}^{\frac{\upsilon_{2}}{\upsilon_{1}}}\left(1-t_{lh}\right)^{\frac{\upsilon_{2}}{\sigma-1}}} \quad \mbox{ \text{and} }\quad  \chi_{lh}=\frac{A_{ll}^{\frac{\upsilon_{2}}{\upsilon_{1}}}\left(\alpha_{lh}\left(\left(1-t_{lh}\right)\iota_{l}\right)^{\frac{1}{1-\sigma}}\right)^{-\upsilon_{2}}}{\sum_{h,h\neq l}A_{ll}^{\frac{\upsilon_{2}}{\upsilon_{1}}}\left(\alpha_{lh}\left(\left(1-t_{lh}\right)\iota_{l}\right)^{\frac{1}{1-\sigma}}\right)^{-\upsilon_{2}}}. \end{align*}
\end{proof}

\subsection{Computing $\zeta_{il}$}\label{app_algo}

We describe how $\zeta_{il}$ can be backed-out from the data using the model's equations. 
We proceed in three steps.

\textbf{1.} Denote 0 as a reference country, such that $i_0$ and $l_0$ denote the reference country for the location of the HQ and as a source country respectively. We write $\Gamma_{il}=\left( \frac{\gamma_{il}/\gamma_{il_{0}}}{\gamma_{i_{0}l}/ \gamma_{i_{0}l_{0}}}\right)^{\frac{\upsilon_2}{\upsilon_1}}$ the propensity of country $i$ to shift profits out of source country $l$, relative to the reference country. Then
\begin{eqnarray}
	\label{calib_ps_2} \zeta_{il} &=& \frac{\Gamma_{il} \zeta_{i_{0}l}}{\sum_k \Gamma_{ik} \zeta_{i_{0}k}}.
\end{eqnarray}
%
A higher elasticity of paper profits relative to the tax base implies that differences in attractiveness for multinational production (governed by $\gamma_{il}$) are magnified when attracting tax avoiders, as shown by $\Gamma_{il}$. 
From Equation (\ref{calib_ps_2}), we can recover all $\zeta_{il}$ from the reference country $\zeta_{i0l}$ and the frictions $\gamma_{il}$.

\textbf{2.} We use an accounting identity to back out $\zeta_{i_0 l}$.
Profits shifted by multinational firms from source country $l$ to tax havens, $PS_l$, are equal to the sum of profits shifted from headquarters countries, $PS_i$, times $\zeta_{il}$.
 \begin{align}
	\label{calib_ps_1} PS_l= \sum_i PS_i \overbrace{\frac{\Gamma_{il} \zeta_{i_{0}l}}{\sum_l \Gamma_{il} \zeta_{i_{0}l}}}^{\zeta_{il}}.
\end{align}
Conditional on observing $PS_i$ and $PS_l$, there are 33 equations and 33 unknowns ($\zeta_{i0l}$).
Consequently, the system described in equation (\ref{calib_ps_1}) is perfectly identified. 

\textbf{3.}  Solving equation (\ref{calib_ps_1}) implies to observe $PS_i$ and $PS_l$. 
$PS_i$ is recovered in section \ref{secPSestimate} using our estimation of $PS_{ih}$: $PS_i=\sum_h PS_{ih}$
To proxy the share of world profits that are shifted from shifted from $l$, i.e., $s_l=PS_l/\sum_{i,h} PS_{ih}$, we use the share of world value-added produced in country $l$. 

\begin{figure}[ht]
\centering
	\includegraphics[width=.6\linewidth]{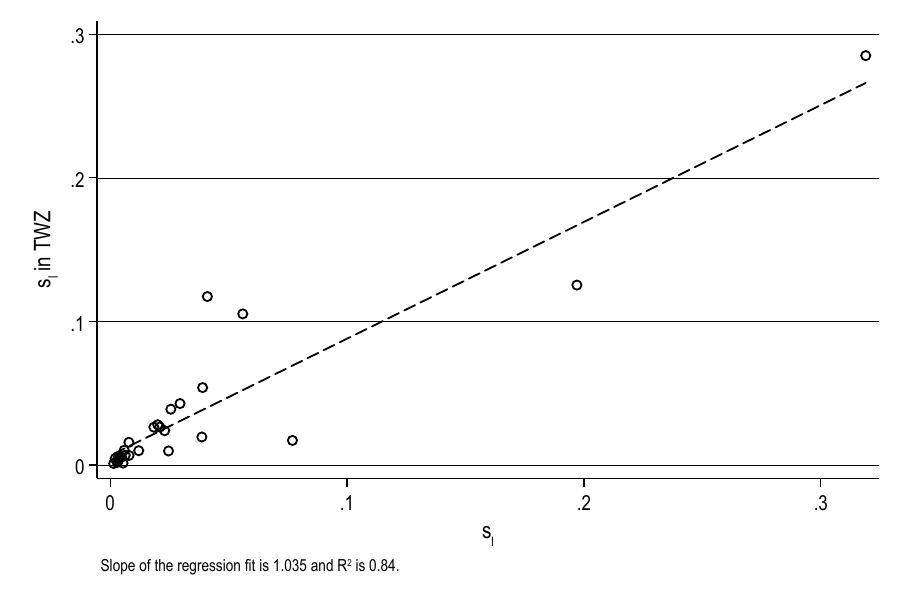}
    	\caption{Comparison of $s_l$ with TWZ estimations \label{s_l_vs_twz}}
     \caption*{\footnotesize Note: $s_l$ corresponds to the share of world profit shifting done from country $l$ as calibrated in our algorithm to retrieve $PS_{ilh}$.\textit{$s_l$ in TWZ} corresponds to the share of world profit shifting made from production countries to tax haven as estimated by TWZ. It relies on the estimates provided in Appendix Table C4 of their paper.}
\end{figure}

In Figure \ref{s_l_vs_twz}, we compare the calibrated values of $s_l$ to the estimates from TWZ. For this comparison, we sum over $h$ the TWZ estimates of $PS_{lh}$ (Appendix Table C4 in their paper).\footnote{These estimates from TWZ cannot be directly used for calibration because they do not cover all the countries in our sample.} This estimation is in line with TWZ estimations using independent estimation techniques. The graph displays an almost one-to-one relationship and a good correlation as highlighted by the $R^2$ of 0.84.

This procedure allows us to obtain $\zeta_{i_0 l}$ from $PS_l$ and $PS_i$ and thereby $\zeta_{il}$.

\clearpage

 \section{Data} \label{app_data}
	\subsection{FDI Income\label{app_fdiinc}}

We collect information on bilateral FDI income from 2012 to 2019 using the bilateral balance of payments data from the OECD. 
This data is completed with data from the Hong-Kong national accounts that provide information on aggregated inward and outward FDI income for the 10 largest immediate recipients and investors in Hong-Kong. 

FDI income has three components: reinvested earnings, dividends, and interest payments. 
Ideally, we want to use only information about reinvested earnings and dividends to construct the FDI income data.
We want to exclude interest payments because in a typical tax avoidance scheme where the non-haven is indebted towards a tax haven affiliate, interests would be paid from the parent company to the tax haven affiliates (\citealp{wright_exorbitant_2018}). 

\paragraph{Taking conduit FDI into account} Foreign Direct Investment (FDI) income data, produced on an immediate investor basis under the BPM6 methodology \citep{bop6}, can be influenced by conduit FDI. Conduit FDI refers to investment that passes through a country solely to take advantage of regulatory benefits. Conduit FDI is FDI that transits through a country exclusively to benefit from regulatory advantages. Since the recommendations of the Benchmark Definition $3^{rd}$ edition in 1996, some OECD countries are producing inward FDI statistics that separate standard FDI from FDI in Special Purpose Entities (SPEs).\footnote{See \cite{oecd2008benchmark}.} These SPEs are the instruments of conduit FDI and are characterized by \citet{imf2014bop} as following: ``their owners are not residents of the territory
of incorporation, main parts of their balance sheets
are claims on or liabilities to nonresidents, they are
companies with little or no physical presence in their
host economy, little or no employment, little or no
significant production, and few (if any) nonfinancial assets, and many SPEs have bank accounts in the 
host economy (although they may be of a temporary
nature).''

These statistics are not necessarily available for all the components of FDI income (dividends, reinvested earnings and interest). To address this issue, we compute the ratio of FDI through SPEs to total FDI for the most aggregated category (Total FDI income) and apply this ratio to each of the components of FDI income. This approach allows us to adjust the FDI income series for SPEs in the economies that report such data. Additionally, we exclude transactions where the immediate investor is a tax haven, which accounted for 29\% of total conduit FDI income in 2015.

In a robustness sample, we apply the methodology of \citet{damgaard_global_2017} to impute conduit FDI income based on the relationship between the ratio of conduit FDI income to total FDI income and the ratio of total FDI income to GDP. Using the estimated correlation between these two variables, we can extrapolate the amount of conduit FDI income passing through an economy. Assuming that the shares of FDI income and conduit FDI income are proportionally the same, we can estimate the bilateral amount of conduit FDI income for countries not covered by the SPE statistics. This methodology is applied only to countries that are not classified as ``\textit{sink} FDI'' economies as defined by \citet{garcia-bernardo_uncovering_2017}.\footnote{These are countries that ``attract and retain foreign capital while conduit-OFCs are attractive intermediate destinations in the routing of international investments and enable the transfer of capital without taxation.'' (\citealp{garcia-bernardo_uncovering_2017}).}

\begin{figure}
	\centering
	\includegraphics[width=0.7\linewidth]{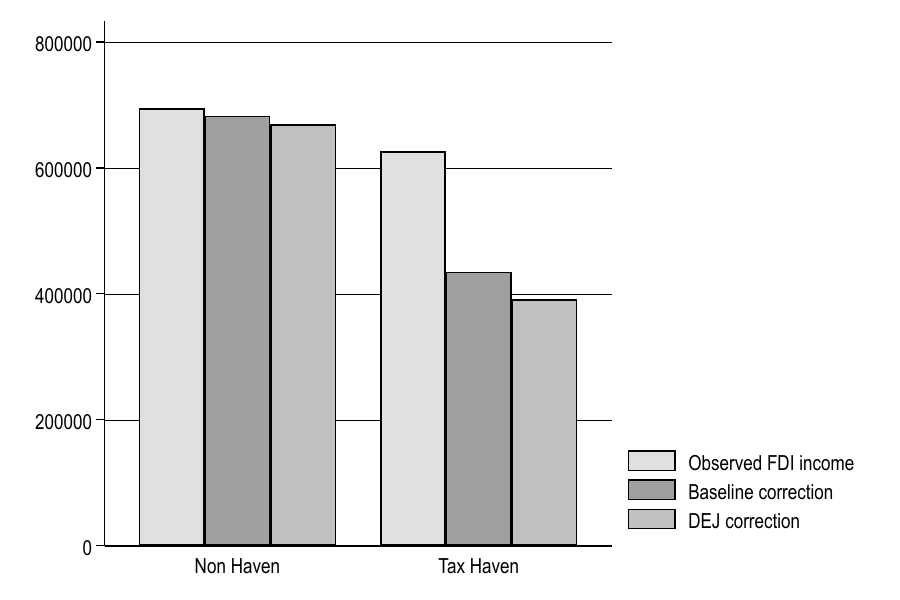}
	\caption{Total FDI income by receiving country type}
	\caption*{\footnotesize Note: This figure compares different FDI income variables according to the type of the economy that receives the investment in 2017. It shows the differences in aggregate amount between the observed FDI income, the variable constructed using available information on SPEs (\textit{Baseline correction}) and the variable constructed using a imputation inspired by DEJ. It excludes FDI income from investment originating from tax havens. }
	\label{fig:comparison_fdi_income_variables}
\end{figure}

Figure \ref{fig:comparison_fdi_income_variables} compares different FDI income variables by the type of economy receiving the investment in 2017. 
It shows the aggregate differences between observed FDI income, the variable adjusted for available information on SPEs (our baseline correction), and the variable adjusted using an imputation method inspired by DEJ. 
For non-tax havens, the correction is small and aligns with expectations. For tax havens, the uncorrected FDI income is almost as large as that for non-havens. Once corrected, it decreases significantly, by nearly one third. 
A further correction using the DEJ method slightly reduces the observed FDI income in tax havens.

\paragraph{Imputation procedure} To improve the coverage of our bilateral FDI income series, we impute some missing values. The imputed flows are obtained in two steps. First, we use the unilateral balance of payments data from the IMF, which informs on inward FDI income, inward FDI stock, outward FDI income and outward FDI stock. This dataset helps us compute the unilateral rates of return on inward and outward investments. 
Second, we apply the unilateral rates of returns on bilateral FDI stock data from the IMF CDIS. 
We use the outward rates of return only in the case of missing information on the inward rate. 
The correlation between imputed bilateral rates of return and observed rates of return in our dataset is 0.9. 

\begin{figure}
    \centering
    \includegraphics[width=0.7\linewidth]{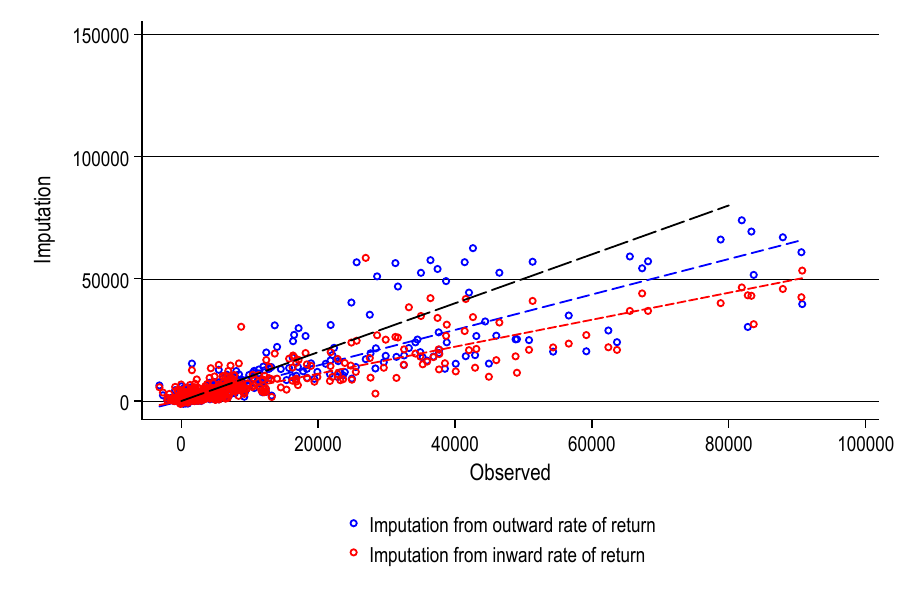}
    \caption{Comparison of observed and imputed flow FDI income.}
    \caption*{\footnotesize Note: This figure compares, for cases where the receiving country of the investment is a tax haven, the observed bilateral FDI income to the imputed one for the flows for which we observe both of them (in which case we keep the observed flow).}
    \label{comparison_imputation}
\end{figure}

It is important to note that this strategy tends to be conservative as it assigns the average rate of return to observed bilateral FDI flows in tax havens while the literature suggests that tax havens generally have a rate of return larger than the average (\citealp{vicard_profit_2022}).
In Figure \ref{comparison_imputation}, we show that for country pairs involving a destination tax haven, where both the actual flow and the imputed flow are observed, the imputed flow tends to underestimate the actual flow. 
This suggests that the imputation procedure underestimates profits in tax havens.

\paragraph{Final dataset construction}
The final data on bilateral FDI income is constructed following this procedure:

\begin{enumerate} 
\item We use the participation revenue values directly provided by the OECD (60\% of the aggregate value). 
\item If missing, we sum dividend income and reinvested earnings (0.001\% of the aggregate value). 
\item If missing, we use the available value between dividend income and reinvested earnings (representing 1\% and 12\% of the aggregate value, respectively). 
\item If missing, we set the value of FDI income from dividends and reinvested earnings as the difference between total FDI income and income from debt instruments (0.03\% of the aggregate value). 
\item If any component information is missing, we use total FDI income to estimate dividends and reinvested earnings (16\% of the aggregate). 
\item Steps 1 to 5 are applied separately for flows reported under the inward directional principle and the outward directional principle. For the estimation, we select the maximum value between the inward and outward flows. 
\item We follow the imputation procedure described below for any remaining missing information (10\% of the aggregate). 
\end{enumerate}

The final dataset includes 170 investing (non-haven) countries and 146 destination countries (34 non-haven countries plus Hong Kong, Ireland, Luxembourg, Netherlands, Singapore, Switzerland, and 28 countries later aggregated to form the Offshore Financial Centers, a composite tax haven). 

\subsection{Country-by-country reporting (CbCR)} We use the OECD CbCR dataset to obtain an alternative measure of aggregated bilateral profits and to compute effective tax rates. This dataset consists in the aggregation of mandatory firm-level country-by-country reports at the residence country $\times$ source country level. Only large firms with a turnover larger than EUR 750 million have to fill the reports. This restriction allows us to concentrate on the firms that are the most likely to engage in profit shifting activities. The aggregation distinguishes profit-making from loss-making firms. We focus on profit-making firms to avoid an aggregation bias when computing the effective tax rates. The requirement to report CbCR for these large firms begins in 2016. We use the reports from the year 2016 and 2017 that are filled by firms from 25 different residence countries.

\subsection{Orbis' Firm-level pre-tax profits.} 

We follow closely to the methodology outlined by \citet{delis2022global}, which constructs a global database on MNE activities using all available ``vintages" of Bureau Van Dijk's Orbis Historical database. As noted by \citet{delis2022global}, Orbis Historical offers several advantages over the online version and other datasets. First, it captures dynamic ownership changes, crucial for avoiding misclassification of affiliates under different GUOs. Second, it covers a period longer than the standard ten years typically available online, addressing reporting lags identified by \citet{kalemli2024construct}. The detailed procedure for building the micro-level dataset is provided in \citet{delis2022global}. The dataset includes firms with non-negative pre-tax income and total assets. The closing date variable is used to determine the fiscal year. As in \citet{delis2022global}, the average statutory tax rate for both firm and GUO countries in our dataset is 0.25. This average closely aligns with the global average statutory corporate tax rate of 0.24, as reported by \citet{torslov_missing_2022}.
We keep data on pre-tax income, cash tax paid, fixed assets, and employment for the foreign affiliates of Global Ultimate Owners along with their locations by country. The cash effective tax rate (ETR) is calculated as the ratio of cash tax paid to pre-tax income and is winsorized for outliers at 1\% and 90\%. 
We aggregate the data at the GUO level and report profits by country. The estimation sample includes 13,331 GUOs across 60 countries for the period 2010-2017. 

\subsection{Trade \label{app_trade}}
Trade data comes from the International Trade and Production Database for Estimation (ITPD-E) from \citet{borchert_international_2021, borchert2022}. This database provides consistent trade data for international and domestic flows using administrative data and avoiding any estimation of missing flows. It also includes trade in services making it a comprehensive and consistent data source for our purposes. 
 
\subsection{Multinational Production Sales \label{app_mp}} \label{app_mp_sales}
	
Multinational production (MP) sales corresponds to the sales made in the production country $l$ by firms headquartered in the country $i$ and reported in $l$ (country $l$ may be identical to country $i$). 
It corresponds to $X_{ill}$ in model's notations. 
We build a  $40 \times 40$ matrix of MP sales that covers the period 2015-2017. 
We use the Multinational Revenue, Employment, and Investment Database (MREID) developed by \citet{Ahmad2023}. This database provides bilateral sector-level data on multinational sales, employment, and investment. The dataset is compiled by aggregating firm-level data from Orbis and does not rely on imputed data from gravity estimation, unlike other similar databases such as the OECD's Analytical AMNE database (\citealp{AMNE_Cadestin_2018}), which are more suited for estimation purposes.
We then compute intra-national MP sales. 
It corresponds to the domestic sales made by domestic firms: $X_{lll}$.
They are obtained by summing the exports of country $l$ and its intra-national trade ($\sum_{i,n} X_{iln}$) and subtracting the MP sales made in $l$ by other countries $i$, with $i\neq l$ ($\sum_{i,i\neq l} X_{ill}$).

\subsection{Tax rates} \label{taxr}
	
\paragraph{Statutory tax rates.}
We use the KPMG Corporate Tax Rate Table accessed through the Tax Foundation's ``Corporate Tax Rates Around the World'' database (\citealp{tax_foundation_corporate_2022}).

\paragraph{Tax havens' tax rates.}
	
The model needs the tax rate available to tax-avoiding firms in tax havens ($t_{lh}$), which is not directly observable. 
Tax havens offer legal provisions that can make the effective tax rate differ greatly from the statutory tax rate. 
We use the OECD CbCR dataset to calculate effective tax rates based on taxes paid and profits. 
This data have been used in other studies that evaluate multinational firms' tax avoidance (\citealp{garcia2021profit} at the macro level, \citealp{delpeuch_laffitte_2019}, \citealp{bratta_assessing_2021} or \citealp{fuest2021corporate} at the micro-level). 

We calculate effective tax rates (ETR) as tax paid divided by pre-tax profits, and remove negative and outlier values.
For each tax haven in our sample, we observe the ETR paid by firms from each headquarter country reporting activity in the tax haven. 
It corresponds to 19 origin countries for Switzerland, 20 for Hong Kong, 19 for Ireland, 19 for Luxembourg, 23 for the Netherlands, 24 for OFCs, and 21 for Singapore. 

We define $t_{lh}$ as the median effective tax rate observed in each tax haven. 
Therefore, $t_{lh}$ does not vary with country $l$ for $l\neq h$.

Notice that \citet{torslov_missing_2022} provides data on the effective tax rate for many countries. However, this would measure $t_{lh}$ with a bias induced by firms having a real activity in tax havens and then paying a different tax rate than tax-avoiding firms. 
This is especially the case in large tax havens. 
	
\subsection{Profits} \label{app_data_profits}

The calibration of the model requires information on profits in each country of the sample. 
Profits are composed of three components. It is computed as gross operating surplus minus depreciation less net interest paid. 
The main data source is the UN National Accounts Table 14 (\citealp{UN}) accessed in January 2024. 
The data is complemented with data gathered from national authorities for Malaysia, Hong-Kong and Singapore.
When missing, we impute the profits' component using the ratio of the component to the Gross Operating Surplus of other countries in the sample. 
The information is missing for Croatia and OFCs.
We impute their profits by estimating them through a regression of profits on GNI, achieving an adjusted $R^2$ of 83\%.

\subsection{Tax haven policies}  \label{app_tjn}

We proxy tax havens' tax avoidance ``technologies'' using the TJN's Corporate Tax Haven Index (\citealp{CTHI}) for 2019 (the first available year). 
The index aggregates 20 \textit{de jure} and \textit{de facto} indicators from 5 categories of policies: Lowest available corporate income tax, Loopholes and gaps, Transparency, Anti-avoidance, and Double tax treaty aggressiveness. 
Out of the 20, we select 13 indicators that inform on the profit-shifting technology and take their average for each tax haven in our database (Foreign investment income treatment, Loss utilization, Capital gains taxation, Sectoral exemptions, Tax holidays and Economic zones, Fictional interest deduction, Public company accounts, Tax court secrecy, Interest deduction, Royalties deduction, Service payment deduction, CFC rules, and Tax treaties).

\clearpage

\section{Estimation of profit shifting: Empirics}

\subsection{Bilateral profit-shifting flows 
\label{app_ps_stateoftheart}}

\paragraph{Unilateral profit-shifting flows} 

The measurement of aggregate profit shifting at the country level is challenging.
Most of the literature follows, in spirit, the approach pioneered by \citet{hines_fiscal_1994}, which delivers estimated amounts of unilateral profit shifting. 
The premise of their methodology is that the observed pre-tax profits of a firm correspond to the sum of \textit{normal} profits and \textit{shifted} profits.
The combination of inputs and technology in production countries determines normal profits.
Shifted profits are generated thanks to the fiscal environment and the incentives offered to foreign firms to shift profits out of production countries. 
Profit shifting is then estimated as the difference between total profits and estimated normal profits. 
When the countries of interest are tax havens, these are ``excess profits''; when the countries of interest are non-havens these are ``missing profits''. 
Papers based on macro-level data estimate the amount of profit shifted to tax havens for the U.S. or at the global level (\citealp{zucman_taxing_2014}, \citealp{clausing_effect_2016, clausing_2020}, \citealp{Jansky2019}, \citealp{garcia2021profit}, or \citealp{torslov_missing_2022}).

\paragraph{The methodology from \citet{torslov_missing_2022}} 
Unilateral profit-shifting estimates may be allocated to bilateral pairs using an allocation key. TWZ are the first to propose a bilateral allocation of profit shifting across pairs of source countries and tax havens and pairs of residence countries and tax havens.

To estimate profit shifting, TWZ collect data on the geography of profits by local and foreign companies.
They proceed in two independent steps. 
They first compute a benchmark level of \textit{normal} profitability level from national account data.
This benchmark is defined as the ratio of pre-tax profits to wages of domestic-controlled firms.
The methodology assumes that, in the absence of profit shifting, the average ratio of pre-tax profits to wages of foreign-controlled firms is the same as that of domestic-controlled firms. 
They show that the ratio of foreign-owned firms in tax havens is an order of magnitude larger than the one of local firms.
In tax havens, profits that are above the benchmark level of profitability are considered as ``excessive''.	
The difference between the excessive level of profits and the benchmark level is the amount of shifted profits. 
TWZ provide estimates of profit shifting to each tax haven and then aggregate it to obtain a worldwide estimate of \$616bn in 2015. 
The estimation is extended to subsequent years in \citet{wier_global_2022}. 

In the second step, the profits shifted to tax havens are allocated across non-haven origin countries. 
Their methodology relies on the assumption that multinational corporations in high-tax countries use intra-firm interest payments and services imports to shift profits. 
Following \citet{hebous_2021}, TWZ identify ``high-risk'' services categories such as royalties and headquarter services (information and communication technologies, insurance, financial and management). 
TWZ define as a benchmark level of trade in ``high-risk'' services and intra-firm interest payment  the average share of high-risk services exports and intra-firm interest received in the GNI of non-haven EU countries. 
These shares are then computed for each tax haven and their difference with respect to the benchmark informs on excessive flows going to tax havens.

\paragraph{Profit shifting and ``high-risk'' services exports.} 

The approach of TWZ has many advantages, one of which is that it relies on available trade in services data, arguably having a broader coverage than FDI income data. 
Nevertheless, our approach, developed in section \ref{secPSestimate} of the paper for our baseline calibration, is agnostic about the sources of profit shifting.
We do not rely on specific information about the methods used to shift profits to tax havens. 

One important advantage of our methodology is that it does not require intra-firm transactions and prices. 
Take trade in ``high-risk" services, for instance. 
Profit shifting is due to the manipulation or mispricing of high-risk services transactions between entities of the multinational firm. 
Quantifying profit shifting at the aggregate level requires information on intra-firm services transactions. 
These flows could be approximated by service trade if they constituted a non-negligible share of it. 
\citet{hebous_2021} note that less than half of ``high-risk'' services imports from tax havens in Germany are intra-firm. 
A back-of-the-envelope calculation implies that around \$26bn of ``high-risk'' services are imported intra-firm by German firms from tax havens in 2015 (50\% of \$51.5bn, as reported in TWZ replication guide, Table C1). 
In comparison, TWZ find \$44bn in excess services imported by German firms from tax havens (replication guide, Table C2). 

To illustrate this point from a different angle, we compare in appendix \ref{PS_robustness} the bilateral excess exports of ``high-risk'' services by tax havens (computed using a gravity equation) with our estimated distribution of bilateral profit shifting. 
The figure shows a good correlation (in line with TWZ assumptions) but also that the implied estimates of profit shifting are generally larger than those from  excess trade in ``high-risk'' services only.

The gap between these two suggests that while high-risk services are an important channel for profit shifting, they may not fully account for profit-shifting practices.

\paragraph{Missing profit shifting.} 

We see at least three possible explanations for this gap: i) profit-shifting estimates through trade in goods are admittedly small in the academic literature. Yet, it is backed out by a lot of anecdotal evidence and even dispute settlements with large fines that go beyond the rather conservative econometric approaches; ii) while profits can be shifted by inflating firms' exports from tax havens, it is also possible for firms to symmetrically deflate their imports; iii) other services, not considered as high-risk, can account for an important share of profit shifting.

The case study of Caterpillar provided by the U.S. Subcommittee on investigations (\citealp{levin_caterpillars_2014}) illustrates ii) and iii). 
The tax avoidance strategy of Caterpillar allowed them to shift more than \$8bn to Switzerland between 2000 and 2012. A part of this strategy was based on the fact that Caterpillar's Swiss affiliate entered into tolling agreements that require the French and Belgian affiliates to provide manufacturing services at a reduced margin of 7\% (see \citealp{levin_caterpillars_2014}, page 51).
This strategy, which relies on an under-priced import of a manufacturing service, allowed Caterpillar to shift its profits from France and Belgium to Switzerland. 
The case of Procter and Gamble (\citealp{bensoussan_bras_2019}) provides a similar narrative. Procter and Gamble's Swiss affiliate contracts with French affiliates to provide a manufacturing service. 
Once the production is done, the goods are owned by the Swiss affiliate against the payment of a margin to the manufacturing affiliate. 
Procter \& Gamble has been accused of shifting its profits to Switzerland by under-pricing this margin compared to similar production activities that would have been conducted with a non-related entity. 
Both case studies highlight that the under-evaluation of imports of ``manufacturing services'' (that are not considered as ``high-risk'' services) by firms located in tax havens is not an uncommon tax avoidance practice.

\subsection[Comparing PS to other estimations]{Comparing $PS_{lh}$ to other estimations} \label{comparison}
	
	\paragraph{Comparison with TWZ.}
To our knowledge \citet{torslov_missing_2022} (TWZ) is the only other paper in the literature that proposes a bilateral measure of profit shifting. 
We compare our measure of bilateral profit shifting to the one of TWZ. 
We also compare our estimates of profit shifting aggregated at the country level with other estimates from the literature. 

\begin{figure}[!h]
    \centering
    \begin{subfigure}[b]{0.48\linewidth}
        \centering
        \includegraphics[width=\linewidth]{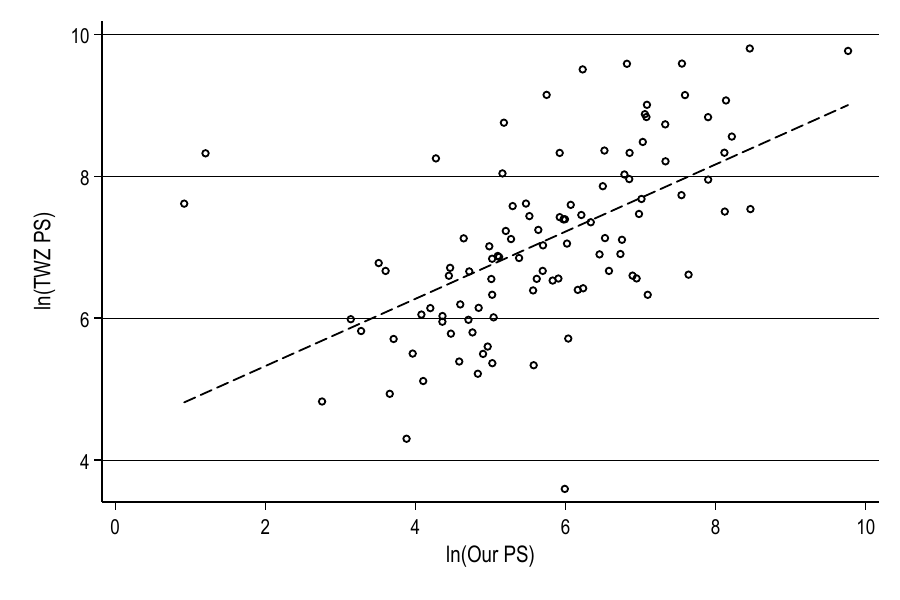}
        \caption*{Panel A: European tax havens.}
        
    \end{subfigure}
    \hfill
    \begin{subfigure}[b]{0.48\linewidth}
        \centering
        \includegraphics[width=\linewidth]{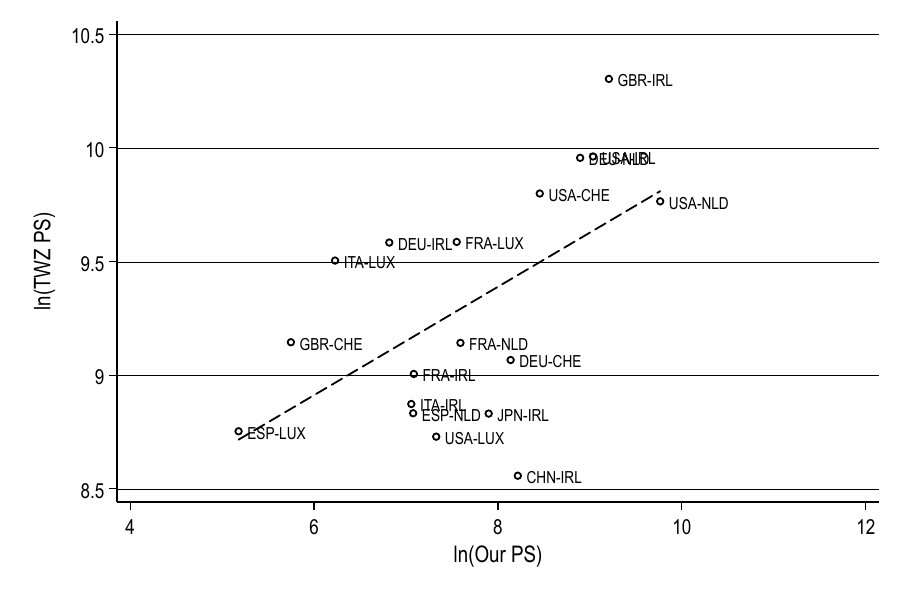}
        \caption*{Panel B: PS flows greater than \$5bn.}
    \end{subfigure}
    \caption{Comparison between \citet{torslov_missing_2022} estimation of PS and ours.}
    \caption*{\footnotesize Note: This figure compares the logarithms of the bilateral profit shifting from source countries $l$ to tax havens $h$ in this paper and in \citet{torslov_missing_2022}. Panel A shows the comparison for European tax havens, while Panel B focuses on large values of bilateral profit shifting.}
    \label{fig:twz_vs_our_ps}
\end{figure}

In Panel A of Figure \ref{fig:twz_vs_our_ps}, we show for European tax havens the correlation between TWZ estimation of profit shifting and ours (in naperian logarithm).\footnote{Due to aggregation of OFC, Hong-Kong and Singapore in TWZ files, we are not able to display a similar graph that separately includes these countries.} 
There is a positive relationship between the two variables. 
The Pearson correlation is 0.62, and the Spearman rank correlation is 0.68.

In Panel B, we focus on large profit-shifting flows (those greater than \$5bn). 
We observe larger differences for higher values of profit-shifting flows. 
While the correlation remains high, most of the profit-shifting flows estimated by TWZ are larger than our estimates, reflecting higher aggregate profit shifting in their estimation.
	
\paragraph{Comparison with unilateral estimations.}
	
\begin{figure}[!h]
		\centering
		\includegraphics[width=0.75\linewidth]{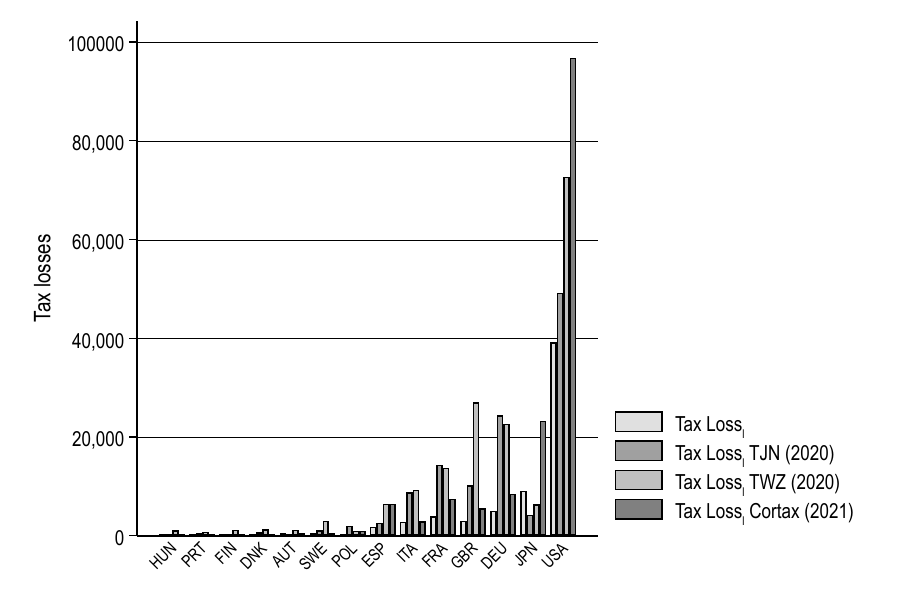}
		\caption{Comparison with other estimations.}
		\caption*{\footnotesize Note: This figure compares the (unilateral) tax losses from profit shifting with \citet{tjn_2020}, \citet{torslov_missing_2022} and \citet{cortax2016}. Tax losses are obtained by multiplying profit shifting out of source countries $l$ by their statutory tax rate.}
		\label{fig:all}
	\end{figure}

We now compare our estimates aggregated at the source-country level with other estimates in the literature. 
These estimates are taken from TWZ, the Tax Justice Network report (\citealp{tjn_2020}) and CORTAX, the model of the European Commission (\citealp{cortax2016}). 
To match with CORTAX data, we transform estimates of profit shifting into tax losses by multiplying them by the statutory tax rate. 
	Figure \ref{fig:all} displays tax losses in selected source countries based on the available data in the CORTAX estimations-- the study with the smallest sample of countries. 

This graph first reveals that the estimates of profit shifting are sensitive to methodologies and data. 
However, these studies find a similar order of magnitude for many countries. 
The CORTAX estimation is particularly high for the U.S while our estimation, despite being lower than others, is close to the ones of the TJN.
Overall, our quantification is in the range of the other studies but tend to display lower aggregate amount of profit shifting. 
	
\subsection{Robustness of profit-shifting estimates} \label{PS_robustness}
\begin{center}
\begin{figure}[!h] 
	\centering
	\includegraphics[width=0.64\linewidth]{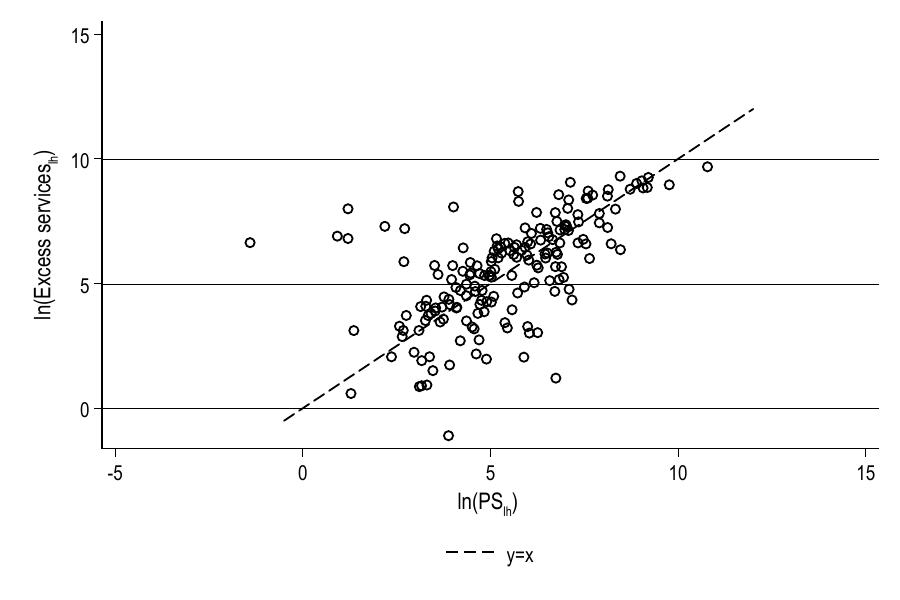}
	\caption{Excessive high-risk services and our measure of bilateral profit shifting}
	\caption*{\footnotesize Note: This figure compares our series of profit shifting between production countries $l$ to tax havens $h$, to the excess of high-risk services exported by tax havens. High-risk services are defined following \citet{torslov_missing_2022} as insurance and pension services, financial services, charges for the use of intellectual property, telecommunications, computer, and information services, and other business services. }
	\label{excess_services}
\end{figure}
\end{center}

This section explores the robustness of our bilateral profit-shifting estimates. 
\paragraph{Comparing $PS_{lh}$ with excess trade in services in tax havens.}
In Figure \ref{excess_services}, we assess the correlation between our profit-shifting allocation and an allocation based on excess imports of services from tax havens. 
We use a reduced-form methodology to directly approximate $PS_{lh}$ from the observations of bilateral services flows. 
For each pair of countries $l$ and $h$, we estimate the amount of bilateral profit shifting as excessive ``high-risk'' services computed from a gravity equation. 

Using the OECD-WTO's BATIS database, we regress the trade values in services  exported from country $k$ to the country $n$ for the service category $s$ at date $t$ on a dummy equal to one when a ``high-risk'' service $s$ is exported by a tax haven $k$. 
``High-risk'' services are defined following \citet{torslov_missing_2022} as insurance and pension services, financial services, charges for using intellectual property, telecommunications, computer and information services, and other business services. 
The methodology that is used to estimate excesses follows the one used to estimate profit shifting in Section \ref{secPSestimate} of the paper. 
An advantage in the context of service data is that we can include exporting country $\times$ year fixed effects. 
Therefore, the estimation of excesses is based on the excess exports of high-risk services compared to standard services in tax havens compared to this excess in non-tax-haven countries. 
We estimate $Service_{knst} =
\beta_1 \left(High-Risk_s \times Haven_k\right) + \mu_{nst} + \mu_{kt} + \mu_{kn} + \mu_s + \epsilon_{knst}$. 
We compute the excess high-risk services exported by tax havens as the difference between the prediction of this equation and its prediction assuming that $\beta_1=0$.

Figure \ref{excess_services} shows a positive and significant correlation between excessive high-risk services and the theoretically consistent measure of bilateral profit shifting. 

The Spearman rank correlation coefficient of 0.6 indicates a relatively high correlation between both series. 
This result suggests that services trade is an important driver of profit shifting between source countries and tax havens but it is not the unique determinant.

\begin{table}[htbp]
    \centering
    \begin{tabular}{lcc} \toprule
     in \$bn  & $PS_{lh}$ & $Excess_{lh}$  \\ \midrule
    Total  & 312 & 265\\
    Mean  & 1.43 &  1.21 \\
    Median & 0.30 & 0.29 \\ \bottomrule
    \end{tabular}
    \caption{Comparing estimated profit shifting and excess high-risk services.}
    \label{table_ps_vs_serv}
\end{table}

In particular, the excess services are sometimes too low to account for the estimated level of $PS_{lh}$. This is evident from the comparison of the aggregate, mean, and median values of both variables in Table \ref{table_ps_vs_serv}. These findings suggest that services alone are insufficient to explain the total amounts of bilateral profit shifting.

\paragraph{Sensitivity to $\tilde{\upsilon}_1$ and $\tilde{\upsilon}_2$.} 

\begin{center}
\begin{figure}[htbp]
    \centering
	\includegraphics[width=0.65\linewidth]{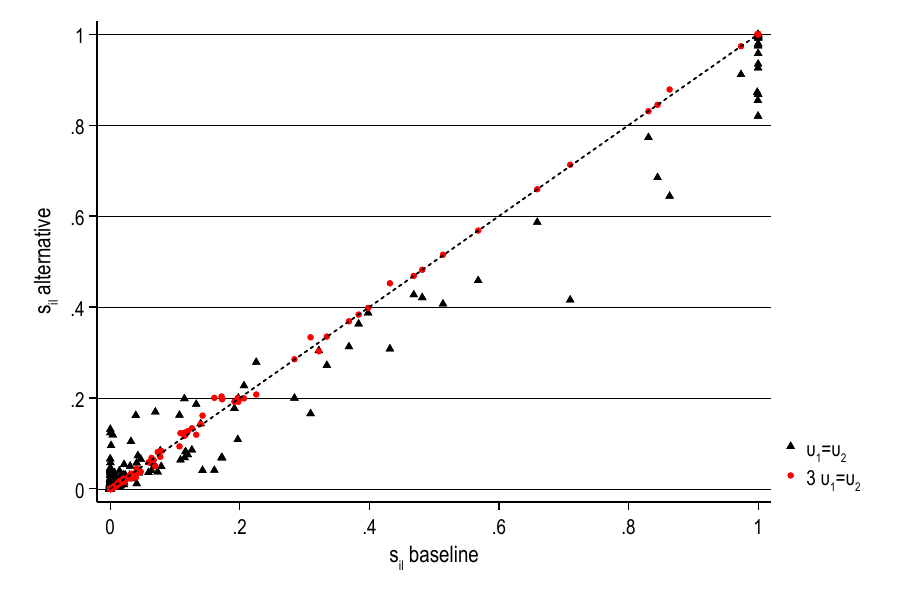}
 \caption{Profit shifting: sensitivity to elasticities calibration}
	\caption*{\footnotesize Note: This figure plots $\zeta_{il}$ as obtained in the baseline exercise (horizontal axis) and compares it to alternative $\zeta_{il}$ obtained with a different calibration of the ratio $\frac{\upsilon_2}{\upsilon_1}$.}
	\label{fig:ps_il_robustness_upsilon}
\end{figure}
\end{center}

In our methodology to estimate profit shifting, the value of $\zeta_{il}$ depends on $\Gamma_{il}=\left( \frac{\gamma_{il}/\gamma_{il_{0}}}{\gamma_{i_{0}l}/ \gamma_{i_{0}l_{0}}}\right)^{\frac{\upsilon_2}{\upsilon_1}}$, that itself depends on the elasticities $\upsilon_1$ and $\upsilon_2$. 
We explore the sensitivity of our estimates to the values of these elasticities. 
Note that only the ratio of these elasticities, not their level, matters for estimating profit shifting. 
In Figure \ref{fig:ps_il_robustness_upsilon}, we plot the baseline estimation of $\zeta_{il}$ and alternative allocations obtained by i) setting $\upsilon_1$ equal to $\upsilon_2$, and ii) increasing the ratio $\frac{\upsilon_2}{\upsilon_1}$ to 3. 
In both cases, the allocation of $\zeta_{il}$ is similar to the baseline allocation and displays a Spearman correlation coefficient larger than 0.97.

\paragraph{Sensitivity to $PS_l$ calibration.}

\begin{figure}[htbp]
\centering
 	\includegraphics[width=0.6\linewidth]{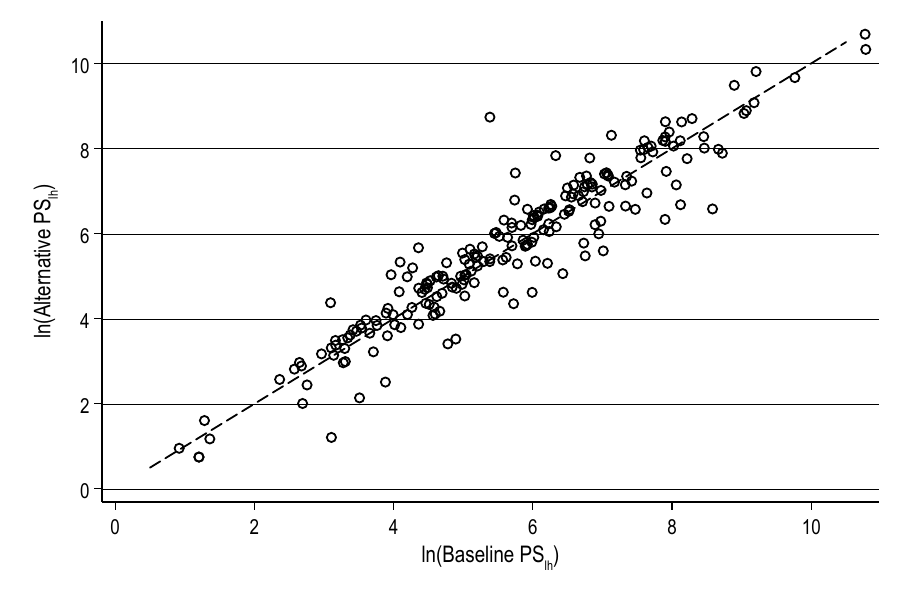}
	\caption{Profit shifting: sensitivity to $s_l$ calibration}
\caption*{\footnotesize Note: This figure plots the log value of $PS_{lh}$ obtained in the baseline exercise and the log value of $PS_{lh}$ obtained when we calibrate $PS_l$ using TWZ data. }
	\label{fig:comparison_sl_our_sl_twz_log}
\end{figure}

As detailed in section \ref{secPSestimate} of the paper, the share of world profits shifted from production countries $l$ needs to be calibrated to recover $\zeta_{il}$. 
We use the share provided in \citet{torslov_missing_2022} data to assess the sensitivity of our estimates to this assumption. 
In figure \ref{fig:comparison_sl_our_sl_twz_log}, we observe a large correlation between both $PS_{lh}$ measures, showing the robustness of our estimates to the calibration of $PS_l$.  

\paragraph{Sensitivity to imputation procedure.} 
We estimate profit shifting using the procedure described in Section \ref{sub_ps_ih} without relying on imputed FDI income series. We find a coefficient on the tax haven dummy similar to those estimated in table \ref{tab:estimation_ps}. In addition, the estimated profit shifting represents 30\% of the total profits in the sample in line with the estimate of 33\% in the sample that includes imputation. 

\begin{table}[htbp]
	\centering
 	\caption{Estimation of profit shifting robustness: sample without imputation   \label{estimation_ps_imputation}}

	\footnotesize{	
\begin{tabular}{lc} \toprule
& (1) \\ 
	&	$FDI\textrm{ }Income_{ij}$ (no imputation) \\ \midrule
$Tax\textrm{ }Haven_j$	&	1.916***	\\
	&	(0.273)	\\
 \midrule
 $\mu_{jt}$	& Yes \\
 $\mu_{r_{(k)}t}$			&	Yes	 \\
 Haven $\times$ Region dummies & Yes \\
 Controls & Yes	\\ \midrule
Number of destination countries	&	144	\\
Observations	&	67,741	\\
Pseudo R2	&	0.831	\\
Controls & Yes \\
Headquarter-Year FE	&	Yes	\\
Region-Year FE	&	Yes	\\
Region $\times$ Haven dummies	&	Yes	\\

 \midrule
\textbf{Profit Shifted (2017)}			&	\textbf{299}	\\
\textbf{Sample's profits (\%, 2017)}	&	\textbf{30}		\\	
 \bottomrule
\end{tabular}}
\caption*{\footnotesize Note: The table replicates column (2) of Table \ref{tab:estimation_ps} without using imputed values. The controls include the GDP and per capita GDP in logs of the destination country, log distance, contiguity, shared colonial ties, common colonizer, and common legal origin. Estimates of profit shifting are obtained from an estimation that also includes haven $\times $ region fixed effects. PPML estimation. Standard errors are robust to clustering at the country level and reported in parentheses. $^{***}$, $^{**}$, and $^{*}$ indicate statistical significance at 1\%, 5\%, and 10\% confidence levels, respectively.}
\end{table}

\clearpage 

\section{Estimation of elasticities and robustness}\label{app:elast}

\begin{table}[htbp]
	\centering
 	\caption{Estimation of elasticities $\tilde{\upsilon}_1$ and $\tilde{\upsilon}_2$\label{app:reg_u1u2}}
	\footnotesize{	
\begin{tabular}{l@{\extracolsep{0.05in}}cccc} \toprule
& \multicolumn{2}{c}{Estimation $\tilde{\upsilon}_1$} & \multicolumn{2}{c}{ Estimation $\tilde{\upsilon}_2$} \\ 
\noalign{\smallskip} \cline{2-3} \cline{4-5}\noalign{\smallskip} 
& (1) & (2) & (3) & (4) \\
\noalign{\smallskip} 
\noalign{\smallskip} \cline{1-3} \cline{4-5}\noalign{\smallskip} 
\noalign{\smallskip} 
$ln(1-t_{lt})$		&	1.085***	 &	2.048**	&			  &			    \\
					&	(0.204)	     &	(0.866)	&			  &			    \\
$ln(1-t_{iht})$		&			     &			&	3.844***  &	6.827***	\\
					&			     &			&	(0.541)	  &	(1.307)	    \\
Employment (log)	&		         &		    &	0.290***  &	0.428***	\\
	                &		         &		    &	(0.056)	  &	(0.077)	\\
Asset (log)	        &		         &		    &	0.345***   &	0.551***	\\
	                  &		           &		  &	(0.029)	    &	(0.051)	\\
GDP (log)	        &	-1.271***	 &	-1.892	&		     &		\\
	                &	(0.428)	     &	(2.612)	&		     &		\\
Per-Capita GDP (log)	&	1.598***	&	2.590	&		&		\\
	                &	(0.431)	&	(2.669)	&		&		\\
 \midrule
Observations			&216,397		&	216,397	& 	2,649	&	2,649 	\\
Adj.  R$^2$			& 0.524		&	0.896	&	0.602	&	0.979	\\
Estimator & OLS & PPML  & OLS & PPML\\ 
 \midrule
Firm $\times$ Year						&	Yes	&	Yes	&	Yes	&	Yes	\\
Origin $\times$ Destination				&	Yes	&	Yes	&	No	&	No	\\
Origin $\times$ Destination  $\times$ Year 	&	No	&	No	&	Yes	&	Yes	\\

 \bottomrule
\end{tabular}}
\parbox{12cm}{\scriptsize Note: Columns (1) and (2) include controls for GDP and GDP per capita, while columns (3) and (4) include controls for employment and fixed assets. All controls are logged. The reported estimates are derived from OLS and Poisson Pseudo-maximum Likelihood (PPML) estimation. Standard errors, robust to clustering at the country level, are shown in parentheses. $^{***}$, $^{**}$, and $^{*}$ indicate statistical significance at 1\%, 5\%, and 10\% confidence levels, respectively.} \\
\end{table}

\begin{table}[htbp]		
\begin{center}	
	\caption{Alternative identification of $\tilde{\upsilon}_2$}
	\label{app:u2_rob_maintext}
 \begin{tabular} {lcc}			
	\noalign{\smallskip}  \hline \noalign{\smallskip}			

	Data Source		&	$PS_{ilh}$	&	TWZ	 \\
					& (1)			 & (2) 	 \\ 
 	\noalign{\smallskip}  \hline \noalign{\smallskip}												

Implied $\tilde{\upsilon}_2$ 	&	5.205***	&	6.617***	\\
                                &	(2.000)	    &	(1.641)	\\

Distance 	(log)&	-0.190*	&	-0.582***	\\
	&	(0.104)	&	(0.122)	\\
Contiguity	&	2.155***	&	-0.0382	\\
	&	(0.477)	&	(0.154)	\\
Common Legal Origin	&	-0.591*	&	0.0392	\\
	&	(0.310)	&	(0.172)	\\
Common Language	&	0.00113	&	-0.263	\\
	&	(0.399)	&	(0.165)	\\
Corporate Tax Haven Index	&	0.0694***	&	0.0540***	\\
	&	(0.00879)	&	(0.0208)	\\
                                
	\noalign{\smallskip}  \hline \noalign{\smallskip}									Observations &	2,742	&	589	\\	
 Estimator  &	PPML	&	PPML	\\	
 \noalign{\smallskip}  \hline \noalign{\smallskip}												
			
$i \times l$ fixed effects	&	Yes	&	Yes		\\
\noalign{\smallskip} \hline \noalign{\smallskip}
\end{tabular}											
\parbox{14.7cm}{\footnotesize Note: This table reports $\tilde{\upsilon}_2$ using alternative quantification of bilateral profit shifting. As in our baseline estimate, both columns use a PPML estimator. In column (1), we use the macro-estimate of profit shifting, $PS_{ilh}$, implied by our model. In column (2), we directly use profit-shifting data from \citet{wier_global_2022} (WZ). }\\
\end{center}											
\end{table}

\begin{table}[htbp]
	\centering
 	\caption{Estimation of semi-elasticities of the tax base and profit shifting to taxes \label{semi-elasticities}}
\footnotesize{	
\begin{tabular}{l@{\extracolsep{0.05in}}cccc} \toprule
& \multicolumn{2}{c}{Semi-elasticity of the tax base} & \multicolumn{2}{c}{ Semi-elasticity of profit shifting} \\ 
\noalign{\smallskip} \cline{2-3} \cline{4-5}\noalign{\smallskip} 
& (1) & (2) & (3) & (4) \\
\noalign{\smallskip} 
\noalign{\smallskip} \cline{1-3} \cline{4-5}\noalign{\smallskip} 
\noalign{\smallskip} 
$ t_{lt}$	&	-1.248***	&	-2.723**	&		&		\\
	&	(0.275)	&	(1.219)	&		&		\\
$t_{f_{i}ht}$	&		&		&	-4.518***	&	-9.210***	\\
	&		&		&	(0.783)	&	(1.634)	\\
    
 \midrule
Observations			&216,397		&	216,397	& 	2,649	&	2,649 	\\
Adj. R$^2$	&	0.524	&	0.896	&	0.593	&	0.979	\\
Estimator & OLS & PPML  & OLS & PPML\\ 
 \midrule
Controls								& 	Yes	&	Yes	&	Yes	&	Yes	\\
Firm $\times$ Year						&	Yes	&	Yes	&	Yes	&	Yes	\\
Origin $\times$ Destination				&	Yes	&	Yes	&	No	&	No	\\
Origin $\times$ Destination  $\times$ Year 	&	No	&	No	&	Yes	&	Yes	\\

 \bottomrule
\end{tabular}}
\parbox{15cm}{\footnotesize Note: Columns (1) includes controls for GDP and GDP per capita, while columns (2) includes controls for employment and fixed assets. All controls are logged. The reported estimates are derived from OLS estimations. Standard errors, robust to clustering at the country level, are shown in parentheses. $^{***}$, $^{**}$, and $^{*}$ indicate statistical significance at 1\%, 5\%, and 10\% confidence levels, respectively.} \\
\end{table}

\clearpage

\section{A variable profit-shifting elasticity \label{app_nonlin}}

In our baseline model, the elasticity of profit-shifting with respect to the keep rate is constant and given by $\tilde{\upsilon}_2$. We introduce a variable elasticity of profit shifting by augmenting $\alpha_{lh}$ with $\left(t_{l} - t_{h}\right)^{k}$, so that the cost of shifting profits depends on the tax differential. 
In this setup, the elasticity of profit shifting to the net-of-tax rate is $\tilde{\upsilon}_2 - k\upsilon_{2}\frac{1-t_{h}}{t_{l}-t_{h}}$. 
There are two parameters to calibrate: $\tilde{\upsilon_{2}}$ and $k$. 
We write a system of two equations with two unknowns to calibrate these parameters, targeting two different moments in our data.

\begin{figure}[htbp]
\centering
 	\includegraphics[width=0.8\linewidth]{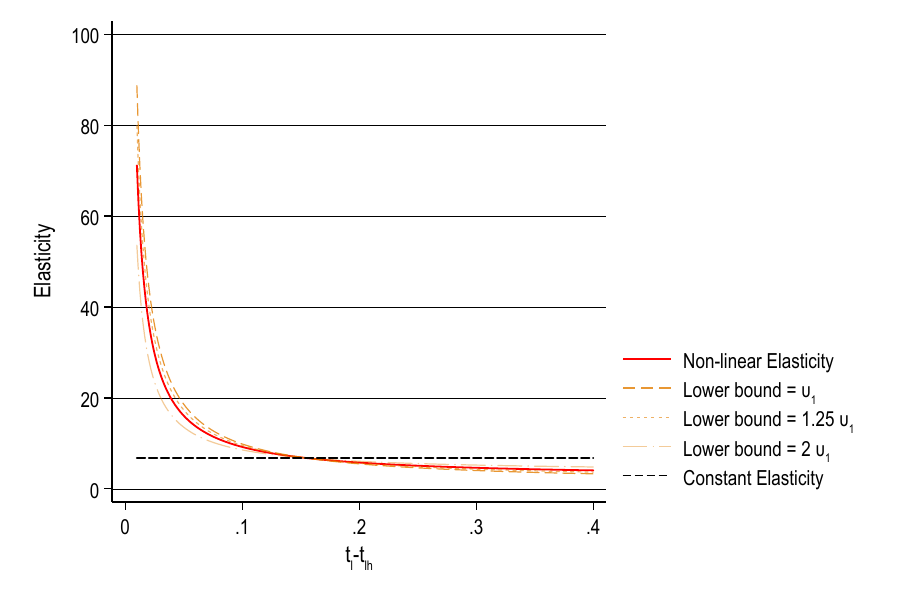}
	\caption{Elasticity of profit shifting w.r.t. net-of-tax rate, as a function of tax differentials.} \label{nonlin}
\caption*{\footnotesize Note: This figure plots the value of the non-linear elasticity of profit shifting to taxes according to the tax differential between the source country $l$ and the tax haven $h$. It is calibrated such as to match different data moments as detailed in section \ref{app_nonlin}. The graph is plotted for a given tax haven tax rate of 5\%. The black dashed line corresponds to the baseline constant elasticity $\tilde{\upsilon_2}$ while the plain red line displays the main calibration of the non-linear elasticity where the lower bound of the non-linear elasticity is set to $1.25\tilde{\upsilon_{1}}$. 
Thin orange lines correspond to alternative shapes of the non-linear elasticity, according to the chosen lower bound.}
\end{figure}

First, we calibrate the non-linear elasticity so that it equals the estimated constant elasticity $\tilde{\upsilon_2}$ when $t_{l}$ and $t_{h}$ are at their average value, respectively, in the sample. 
Second, we target the elasticity to be larger than $\tilde{\upsilon_{1}}$ when $\frac{1-t_{h}}{t_{l}-t_{h}} \to 1$, meaning $t_{l} \to 1$ and $t_{h} \to 0$. 
When the tax differential is large, we expect the elasticity to be small, but bounded by $\tilde{\upsilon_{1}}$. 
To ensure the non-linear $\tilde{\upsilon_{2}}$ remains larger than $\tilde{\upsilon_{1}}$, we set a lower bound of $1.5\tilde{\upsilon_{1}}$. 
As shown in Figure \ref{nonlin}, the choice of the lower bound does not significantly affect the shape of the non-linear function. 
Formally, we solve the following system:
\[
\begin{cases}
\begin{array}{c}
\frac{\upsilon_{2}}{\sigma-1}-1-k\upsilon_{2}\frac{1-\bar{t_{h}}}{\bar{t_{l}}-\bar{t_{h}}}=\upsilon_{2}^{linear}\\
\frac{\upsilon_{2}}{\sigma-1}-1-k\upsilon_{2}=\frac{3}{2}\upsilon_{1}
\end{array}\end{cases}.
\]
Solving the system, we find $k=0.66$ and $\hat{\upsilon_2}=2.35$. Figure \ref{nonlin} shows the value of the profit-shifting elasticity based on the tax differential between the production affiliate and the tax haven, with $t_h = 0.05$. 
The black dashed line represents the baseline constant elasticity, while the solid red line illustrates the main calibration of the non-linear elasticity. 
Thin orange lines depict alternative shapes of the non-linear elasticity, depending on the chosen lower bound. The choice of lower bound does not significantly impact the shape of the non-linearity.

\clearpage

\section{Profit shifting frictions} \label{appendix_ps_frictions}

\subsection{Derivation of profit shifting frictions.}

We start from the following definitions of $\mathbb{P}_{ilh}$and
$\mathbb{P}_{ill}$, assuming that, for all $l$, $\alpha_{ll}=1$:
\[
\mathbb{P}_{ilh}=\frac{\left(A_{lh}\right)^{\frac{\upsilon_{2}}{\upsilon_{1}}}\left(\gamma_{il}\alpha_{lh}\tilde{t}_{ilh}\iota_{l}^{\frac{1}{1-\sigma}}w_{l}\Xi_{l}\right)^{-\upsilon_{2}}\times\theta_{i}^{-\upsilon_{1}}\left(\sum_{j,k,k\neq j}\left(A_{jk}\right)^{\frac{\upsilon_{2}}{\upsilon_{1}}}\left(\gamma_{ij}\alpha_{jk}\tilde{t}_{ijk}\iota_{j}^{\frac{1}{1-\sigma}}w_{j}\Xi_{j}\right)^{-\upsilon_{2}}\right)^{\frac{\upsilon_{1}}{\upsilon_{2}}-1}}{\sum_{j}A_{jj}\left(\gamma_{ij}\tilde{t}_{ijj}\iota_{j}^{\frac{1}{1-\sigma}}w_{j}^{p}\Xi_{j}\right)^{-\upsilon_{1}}+\theta_{i}^{-\upsilon_{1}}\left(\sum_{j,k,k\neq j}\left(A_{jk}\right)^{\frac{\upsilon_{2}}{\upsilon_{1}}}\left(\gamma_{ij}\alpha_{jk}\tilde{t}_{ijk}\iota_{j}^{\frac{1}{1-\sigma}}w_{j}\Xi_{j}\right)^{-\upsilon_{2}}\right)^{\frac{\upsilon_{1}}{\upsilon_{2}}}};
\]
\[
\mathbb{P}_{ill}=\frac{A_{ll}\left(\gamma_{il}\tilde{t}_{l}\iota_{l}^{\frac{1}{1-\sigma}}w_{l}\Xi_{l}\right)^{-\upsilon_{1}}}{\sum_{j}A_{jj}\left(\gamma_{ij}\tilde{t}_{ijj}\iota_{j}^{\frac{1}{1-\sigma}}w_{j}^{p}\Xi_{j}\right)^{-\upsilon_{1}}+\theta_{i}^{-\upsilon_{1}}\left(\sum_{j,k,k\neq j}\left(A_{jk}\right)^{\frac{\upsilon_{2}}{\upsilon_{1}}}\left(\gamma_{ij}\alpha_{jk}\tilde{t}_{ijk}\iota_{j}^{\frac{1}{1-\sigma}}w_{j}\Xi_{j}\right)^{-\upsilon_{2}}\right)^{\frac{\upsilon_{1}}{\upsilon_{2}}}}.
\]
We replace $\theta_{i}$ with $\tilde{\theta_{i}}\bar{\theta}$ to
allow for the following normalization. We impose that, given the same
fundamentals and absent profit shifting frictions ($\alpha_{lh}=1,\,\forall l,h$)
firms are indifferent between shifting and not shifting. Formally,
we impose $\sum_{l,h,l\neq h}\mathbb{P}_{ilh}=\sum_{l}\mathbb{P}_{ill}$.
We obtain: 
\[
\frac{\sum_{l,h,l\neq h}\left(A_{lh}\right)^{\frac{\upsilon_{2}}{\upsilon_{1}}}\left(\gamma_{il}\alpha_{lh}\tilde{t}_{ilh}\iota_{l}^{\frac{1}{1-\sigma}}w_{l}\Xi_{l}\right)^{-\upsilon_{2}}\times\left(\sum_{j,k,k\neq j}\left(A_{jk}\right)^{\frac{\upsilon_{2}}{\upsilon_{1}}}\left(\gamma_{ij}\alpha_{jk}\tilde{t}_{ijk}\iota_{j}^{\frac{1}{1-\sigma}}w_{j}\Xi_{j}\right)^{-\upsilon_{2}}\right)^{\frac{\upsilon_{1}}{\upsilon_{2}}-1}}{\sum_{l}A_{ll}\left(\gamma_{il}\tilde{t}_{l}\iota_{l}^{\frac{1}{1-\sigma}}w_{l}\Xi_{l}\right)^{-\upsilon_{1}}}=\left(\tilde{\theta_{i}}\bar{\theta}\right)^{\upsilon_{1}}.
\]
We now set all fundamentals identical across countries and all the
$\tilde{\theta}_{i}=1$. This gives us $\frac{\left(\sum_{j,k,k\neq j}A_{jk}^{\frac{\upsilon_{2}}{\upsilon_{1}}}\right)^{\frac{1}{\upsilon_{2}}}}{\left(\sum_{l}A_{ll}\right)^{\frac{1}{\upsilon_{1}}}}=\bar{\theta}$.
Since we cannot separately identify $A_{lh}$ and $\alpha_{lh}$,
we load all bilateral variation on $\alpha_{lh}$: $A_{lh}=A_{ll},\forall\;l,h$.
As a consequence, we can write:
\begin{align*}
\frac{\left(\sum_{j,k,k\neq j}A_{jj}^{\frac{\upsilon_{2}}{\upsilon_{1}}}\right)^{\frac{1}{\upsilon_{2}}}}{\left(\sum_{l}A_{ll}\right)^{\frac{1}{\upsilon_{1}}}} & =\bar{\theta},
\end{align*}
which simplifies to
\[
\bar{\theta}=\frac{\left(H\sum_{j}A_{jj}^{\frac{\upsilon_{2}}{\upsilon_{1}}}\right)^{\frac{1}{\upsilon_{2}}}}{\left(\sum_{l}A_{ll}\right)^{\frac{1}{\upsilon_{1}}}}.
\]
Given this normalization, we start by comparing $\mathbb{P}_{ill}$to
an alternative $\mathbb{P}_{imm}$. We obtain the following vector
of relative productivities:
\[
\frac{A_{ll}}{A_{mm}}=\frac{\mathbb{P}_{ill}}{\mathbb{P}_{imm}}\frac{\left(\gamma_{il}\tilde{t}_{l}\iota_{l}^{\frac{1}{1-\sigma}}w_{l}\Xi_{l}\right)^{\upsilon_{1}}}{\left(\gamma_{im}\tilde{t}_{m}\iota_{m}^{\frac{1}{1-\sigma}}w_{m}\Xi_{m}\right)^{\upsilon_{1}}}.
\]
We set $A_{USUS}=1$ as a normalization to recover the vector of productivities
in levels. This implies that, for all countries $l$: 
\[
A_{ll}=\frac{\mathbb{P}_{ill}}{\mathbb{P}_{iUSUS}}\frac{\left(\gamma_{il}\tilde{t}_{l}\iota_{l}^{\frac{1}{1-\sigma}}w_{l}\Xi_{l}\right)^{\upsilon_{1}}}{\left(\gamma_{iUS}\tilde{t}_{US}\iota_{US}^{\frac{1}{1-\sigma}}w_{US}\Xi_{US}\right)^{\upsilon_{1}}}.
\]
Using a similar rationale on $\mathbb{P}_{ilh}$ and $\mathbb{P}_{ill}$we
obtain:
\begin{align*}
 \frac{\tilde{\theta_{i}}^{^{-\upsilon_{1}}}A_{lh}^{\frac{\upsilon_{2}}{\upsilon_{1}}}\alpha_{lh}^{-\upsilon_{2}}}{A_{ll}}=\frac{\mathbb{P}_{ilh}}{\mathbb{P}_{ill}}\frac{\left(\gamma_{il}\tilde{t}_{ilh}\iota_{l}^{\frac{1}{1-\sigma}}w_{l}\Xi_{l}\right)^{\upsilon_{2}}\times\bar{\theta}^{\upsilon_{1}}\left(\sum_{j,k,k\neq j}\left(A_{jk}\right)^{\frac{\upsilon_{2}}{\upsilon_{1}}}\left(\gamma_{ij}\alpha_{jk}\tilde{t}_{ijk}\iota_{j}^{\frac{1}{1-\sigma}}w_{j}\Xi_{j}\right)^{-\upsilon_{2}}\right)^{1-\frac{\upsilon_{1}}{\upsilon_{2}}}}{\left(\gamma_{il}\tilde{t}_{l}\iota_{l}^{\frac{1}{1-\sigma}}w_{l}\Xi_{l}\right)^{\upsilon_{1}}}.
\end{align*}
Using again $A_{lh}=A_{ll},\forall\;l,\:h$ allows us to write:
\begin{equation}
\tilde{\theta_{i}}^{-\upsilon_{1}}A_{lh}^{\frac{\upsilon_{2}}{\upsilon_{1}}-1}\alpha_{lh}^{-\upsilon_{2}}=\frac{\mathbb{P}_{ilh}}{\mathbb{P}_{ill}}\frac{\left(\gamma_{il}\tilde{t}_{ilh}\iota_{l}^{\frac{1}{1-\sigma}}w_{l}\Xi_{l}\right)^{\upsilon_{2}}\times\bar{\theta}^{^{\upsilon_{1}}}\left(\sum_{j,k,k\neq j}\left(A_{jk}\right)^{\frac{\upsilon_{2}}{\upsilon_{1}}}\left(\gamma_{ij}\alpha_{jk}\tilde{t}_{ijk}\iota_{j}^{\frac{1}{1-\sigma}}w_{j}\Xi_{j}\right)^{-\upsilon_{2}}\right)^{1-\frac{\upsilon_{1}}{\upsilon_{2}}}}{\left(\gamma_{il}\tilde{t}_{l}\iota_{l}^{\frac{1}{1-\sigma}}w_{l}\Xi_{l}\right)^{\upsilon_{1}}}. \label{eq:frictions1}
\end{equation}
We now consider the ratio of $\mathbb{P}_{ilh}$ and an alternative
triplet $\mathbb{P}_{irs}$ and find:
\begin{equation}
\left(\frac{A_{lh}}{A_{rs}}\right)^{\frac{\upsilon_{2}}{\upsilon_{1}}}\left(\frac{\alpha_{lh}}{\alpha_{rs}}\right)^{-\upsilon_{2}}=\frac{\mathbb{P}_{ilh}}{\mathbb{P}_{irs}}\frac{\left(\gamma_{ir}\tilde{t}_{ilh}\iota_{l}^{\frac{1}{1-\sigma}}w_{l}\Xi_{l}\right)^{\upsilon_{2}}}{\left(\gamma_{ir}\tilde{t}_{irs}\iota_{r}^{\frac{1}{1-\sigma}}w_{r}\Xi_{r}\right)^{\upsilon_{2}}}.\label{eq:rel_A}
\end{equation}
With this result we can go back to \ref{eq:frictions1} and divide
by $A_{lh}^{\left(1-\frac{\upsilon_{1}}{\upsilon_{2}}\right)\left(-\frac{\upsilon_{2}}{\upsilon_{1}}\right)}\alpha_{lh}^{\left(1-\frac{\upsilon_{1}}{\upsilon_{2}}\right)\upsilon_{2}}$:
\[
\tilde{\theta_{i}}^{-\upsilon_{1}}\alpha_{lh}^{-\upsilon_{1}}=\frac{\mathbb{P}_{ilh}}{\mathbb{P}_{ill}}\frac{\left(\gamma_{il}\tilde{t}_{ilh}\iota_{l}^{\frac{1}{1-\sigma}}w_{l}\Xi_{l}\right)^{\upsilon_{2}}\times\bar{\theta}^{\upsilon_{1}}\left(\sum_{j,k,k\neq j}\left(\frac{A_{jk}}{A_{lh}}\right)^{\frac{\upsilon_{2}}{\upsilon_{1}}}\left(\frac{\alpha_{jk}}{\alpha_{lh}}\right)^{-\upsilon_{2}}\left(\gamma_{ij}\tilde{t}_{ijk}\iota_{j}^{\frac{1}{1-\sigma}}w_{j}\Xi_{j}\right)^{-\upsilon_{2}}\right)^{1-\frac{\upsilon_{1}}{\upsilon_{2}}}}{\left(\gamma_{il}\tilde{t}_{l}\iota_{l}^{\frac{1}{1-\sigma}}w_{l}\Xi_{l}\right)^{\upsilon_{1}}}.
\]
We can now use eq. \ref{eq:rel_A} to obtain:
\begin{align*}
\tilde{\theta_{i}}^{-\upsilon_{1}}\alpha_{lh}^{-\upsilon_{1}}=\frac{\mathbb{P}_{ilh}^{\frac{\upsilon_{1}}{\upsilon_{2}}}}{\mathbb{P}_{ill}}\bar{\theta}^{\upsilon_{1}}\left(\sum_{j,k,k\neq j}\mathbb{P}_{ijk}\right)^{1-\frac{\upsilon_{1}}{\upsilon_{2}}}\left(\frac{\tilde{t}_{ilh}}{\tilde{t}_{l}}\right)^{\upsilon_{1}}.
\end{align*}
The right-hand side of the equation is fully observable. After manipulations,
we obtain the formula of proposition \label{prop_frictions}: 
\[
\bar{\theta}\tilde{\theta_{i}}\alpha_{lh}\left(\frac{1-t_{ll}}{1-t_{ilh}}\right)^{\frac{1}{\sigma-1}}=\left(\frac{\mathbb{P}_{ilh}}{\mathbb{P}_{ill}}\right)^{-\frac{1}{\upsilon_{1}}}\left(\frac{\mathbb{P}_{ilh}}{P_{i}}\right)^{\frac{1}{\upsilon_{1}}-\frac{1}{\upsilon_{2}}}.
\]

\subsection{Determinants of profit shifting costs $\alpha_{lh}$.} 

In Table \ref{table_costPS}, we show the results of estimations of the log of $\alpha_{lh}$ on gravitational variables. We find that gravitational variables correlate well with the profit-shifting costs. In particular, increase in geographical distance increase bilateral profit shifting costs. This relationship is concave suggesting that distance is an important determinant for small distances and matters less when distance is more bigger. Bilateral historical relationships also matters, as illustrated by the importance of colonial links.

\begin{table}[htbp]
	\centering
	\caption{Gravitational determinants of profit-shifting frictions} \label{table_costPS}
	\begin{threeparttable}
		\footnotesize
		\begin{tabular}{lccc} \toprule
			& (1) & (2) & (3) \\
			&  \multicolumn{3}{c}{$ ln(\alpha_{lh}) $} \\ \midrule
$ ln(distance_{lh}) $	&	0.00855*	&	0.190***	&	0.110***	\\
	&	(0.00466)	&	(0.0328)	&	(0.0388)	\\
$ ln(distance_{lh})^2 $	&		&	-0.0114***	&	-0.00672***	\\
	&		&	(0.00217)	&	(0.00240)	\\
Ever colony $ _{lh} $	&		&		&	-0.0322*	\\
	&		&		&	(0.0188)	\\
Common legal origin $ _{lh} $	&		&		&	0.00511	\\
	&		&		&	(0.0113)	\\
Contiguity $ _{lh} $	&		&		&	-0.0310**	\\
	&		&		&	(0.0129)	\\  \midrule
Observations	&	218	&	218	&	218	\\
R-squared	&	0.952	&	0.958	&	0.962	\\
Source Fixed Effects	&	Yes	&	Yes	&	Yes	\\
Haven Fixed Effects	&	Yes	&	Yes	&	Yes	\\ \bottomrule
		\end{tabular}
		\begin{tablenotes}
			\small
			\item Robust standard errors clustered at the $l$ level in parentheses. *** p$<$0.01, ** p$<$0.05, * p$<$0.1
		\end{tablenotes}
	\end{threeparttable}
\end{table}
	

\clearpage

\section{Exact hat algebra \label{app_eha}} 

This section describes the Exact Hat Algebra algorithm used in the paper.

\subsection[Relative changes in probabilities ]{Relative changes in probabilities $\hat{\mathbb{P}}_{ilh}$ }

\paragraph{Non-haven residence countries $i\protect\notin\mathcal{H}$.}




We introduce $N_{ill}$ and $N_{ilh}$ to denote the numerator of
$\mathbb{P}_{ill}$ and $\mathbb{P}_{ilh}$ respectively 
and\\ {$\mathcal{D}_{i}={\sum_{l}N_{ill}+\left(\sum_{l\notin\mathcal{H},h,h\neq l}N_{ilh}\right)^{\frac{\upsilon_{1}}{\upsilon_{2}}}}$} their denominator so that:

$h\neq l\Rightarrow \mathbb{P}_{ilh}=\frac{N_{ilh}\left(\sum_{l\notin\mathcal{H},h,h\neq l}N_{ilh}\right)^{\frac{\upsilon_{1}}{\upsilon_{2}}-1}}{\mathcal{D}_{i}}\mbox{ and } h=l\Rightarrow \mathbb{P}_{ill}=\frac{N_{ill}}{\mathcal{D}_{i}}.$



Relative changes in $\mathbb{P}_{ill}$ and {$\mathbb{P}_{ilh}$}
are given by
\[
\widehat{\mathbb{P}}_{ill}\equiv\frac{\hat{N}_{ill}}{\sum_{l}\hat{N}_{ill}\mathbb{P}_{ill}+\left(1-\sum_{l}\mathbb{P}_{ill}\right)^{1-\frac{\upsilon_{1}}{\upsilon_{2}}}\left(\sum_{l\notin\mathcal{H},h,h\neq l}\hat{N}_{ilh}\mathbb{P}_{ilh}\right)^{\frac{\upsilon_{1}}{\upsilon_{2}}}}
\]
and
\[
\widehat{\mathbb{P}}_{ilh}\equiv\frac{\hat{N}_{ilh}\left(1-\sum_{l}\mathbb{P}_{ill}\right)^{1-\frac{\upsilon_{1}}{\upsilon_{2}}}\left(\sum_{l\notin\mathcal{H},h,h\neq l}\hat{N}_{ilh}\mathbb{P}_{ilh}\right)^{\frac{\upsilon_{1}}{\upsilon_{2}}-1}}{\sum_{l}\hat{N}_{ill}\mathbb{P}_{ill}+\left(1-\sum_{l}\mathbb{P}_{ill}\right)^{1-\frac{\upsilon_{1}}{\upsilon_{2}}}\left(\sum_{l\notin\mathcal{H},h,h\neq l}\hat{N}_{ilh}\mathbb{P}_{ilh}\right)^{\frac{\upsilon_{1}}{\upsilon_{2}}}}
\]
where 

\noindent 
\[
\hat{N}_{ill}=\widehat{w_{l}\Xi_{l}\tilde{t}_{ill}}^{-\upsilon_{1}}\quad\hat{N}_{ilh}=\widehat{w_{l}\Xi_{l}\tilde{t}_{ilh}}^{-\upsilon_{2}}.
\]

\paragraph{Haven-residence countries $i\in\mathcal{H}$.}

\noindent
Relative changes in the probability to locate in $l$ are given by
$
\widehat{\mathbb{P}}_{ill}=\frac{\hat{N}_{ill}}{\sum_{l}\mathbb{P}_{ill}\hat{N}_{ill}}
$.

\subsection{Computing counterfactual equilibria}

Notations: we introduce the share of sales by firms from $i$, sourcing
in $l$, booking their profits in $h$: $\eta_{ilh}=\frac{X_{ilh}}{\sum_{l,h}X_{ilh}}$.
From equation (10), we obtain

$$\eta_{ilh}=\frac{\mathbb{P}_{ilh}/\left((1-t_{ilh})\iota_{l}\right)}{\sum\mathbb{P}_{ilh}/\left((1-t_{ilh})\iota_{l}\right)}.$$

We denote by $\mu_{ln}$ the share of sales to country $n$ by firms
producing in $l$. This share does not depend on firm's residence:

$$\mu_{ln}=\frac{\tau_{ln}^{1-\sigma}Y_{n}P_{n}^{\sigma-1}}{\sum_{n}\tau_{ln}^{1-\sigma}Y_{n}P_{n}^{\sigma-1}}\equiv\left(\frac{\Xi_{ln}}{\Xi_{l}}\right)^{1-\sigma}.$$

The sales of firms from $i$ producing in $l$ is denoted by $X_{il}=\sum_{h=l;h\in\mathcal{H}}X_{ilh}$
and their sales in market $n$ by $X_{iln}=\mu_{ln}X_{il}$.

Endogenous variables $z$ are denoted $z$, and $z'$, respectively
the initial and the new equilibrium so that $\hat{z}=z'/z$. Following
\citet{Dekleetal2007}, we look for a fixed point in changes $\hat{{\bf w}}=(\hat{w}_{l})_{l\in[[1,N]]},{\bf \hat{Y}}=(\hat{Y}_{n})_{n\in[[1,N]]},{\bf \hat{P}}=(\hat{P}_{n})_{n\in[[1,N]]},{\bf \hat{N}}=(\hat{\mathcal{N}}_{i})_{i\in[[1,N]]}$
. Given $\hat{{\bf w}},{\bf \hat{Y}},{\bf \hat{{N}}},\hat{{\bf P}}$
and the change in policy, we can compute the implied change in market
potential $\hat{\Xi}_{l}$ . This pins down the change in $\hat{\mathbb{P}}_{ilh}$
(see next subsection) and thereby the changes $\hat{\eta}_{ilh}$ and $\hat{\mu}_{ln}$.
The output in $l$ produced by $l$ firms is then obtained as 
\[
X_{il}'=\frac{\mathcal{N}_{i}'}{T_{i}^{1-\sigma}}\left(\frac{\sigma}{\sigma-1}\right)^{-\sigma}\sum_{h}\left(\mathbb{P}'_{ilh}\iota_{l}^{-1}(1-t_{lh})^{-1}\right)\mathcal{D}_{i}^{\prime \frac{\sigma-1}{\upsilon_{1}}}\Gamma\left(1-\frac{\sigma-1}{\upsilon_{1}}\right)
.\]
We thus get $X_{iln}'=\mu_{ln}'X_{il}'$ and $X'_{ilh}=\eta'_{ilh}\left(\sum_{n}X'_{iln}\right)$.
A fixed point in changes is obtained when:

\noindent - wages satisfy the labor-market clearing
\[
w'_{k}=\frac{1}{\sigma L_{k}}\sum_{l,h,n}\eta'_{klh}\left(1-{t}'_{klh}\right)\iota_{l}X'_{kln}+\frac{\sigma-1}{\sigma L_{k}}\sum_{i}X'_{ik};
\]
 - total expenditures are equal to labor income, tax revenues, adjusted
for the friction $\iota_{l}$ and imbalances
\[
Y'_{k}=w'_{k}L_{k}+\frac{1}{\sigma}\left(\sum_{i,n}t'_{k}\eta'_{ikk}\iota_{k}X'_{ikn}+\sum_{i,l,n,l\neq k}t'_{ilk}\eta'_{ilk}\iota_{l}X'_{iln}\right)+\frac{1}{\sigma}\sum_{i,n}(1-\iota_{k})X'_{ikn}+\Delta_{k};
\]
 - price indices for all countries but the numeraire verify
\[
P_{n}^{\prime 1-\sigma}=\sum_{l}\tau_{ln}^{1-\sigma}\Xi_{l}^{\prime \sigma-1}\sum_{i}X'_{il};
\]
 - and the number of firms satisfies the free-entry condition
\[
\mathcal{N}'_{i}=\frac{\frac{1}{\sigma}\sum_{l,h,n}\left(1-t'_{ilh}\right)\iota_{l}X'_{ilh}}{w'_{i}f_{E}}.
\]

\clearpage

\section{Supplements to section \ref{sec_counter}} \label{app_figures}

\subsection{Calibration overview and validation}
\begin{table}[h]
	\centering
	\caption{Calibration overview}
	\label{calibration_data}
	\resizebox{.84\linewidth}{!}{\begin{tabular}{p{2cm}p{10cm}p{2.5cm}} 
			\textbf{Variables}	&	\textbf{Definition/Source/Methodology/Reference}	&	\textbf{Section}	\\
			\midrule
			\noalign{\smallskip}\noalign{\smallskip} 		
			\textit{Endogenous variables} & & \\ \cline{1-1}
			\noalign{\smallskip}\noalign{\smallskip}
			\multicolumn{1}{c}{$X_{ln}$} & Trade in goods and services and own trade from ITPD-E & Appendix \ref{app_trade}  \\

			\multicolumn{1}{c}{$X_{ill}$} & Multinational Production Sales from MREID & Appendix \ref{app_mp} \\ 

			\multicolumn{1}{c}{$X_{ilh}$} & Profit shifting. Estimated using accounting models' equations and using data from OECD bilateral balance of payments, IMF Balance of payments data. & Section \ref{sec:data}, Appendix \ref{app_fdiinc}\\
			\noalign{\smallskip}\noalign{\smallskip}
			\noalign{\smallskip}\noalign{\smallskip}	
			\textit{Parameters} & & \\ \cline{1-1}
			
			\multicolumn{1}{c}{$t_{l}$} & Statutory tax rate. KPMG Statutory Corporate tax rate tables. & Appendix \ref{taxr} \\

			\multicolumn{1}{c}{$t_{lh}$} & Tax havens' tax rate. OECD's Country-by-Country reporting. & Appendix \ref{taxr} \\

			\multicolumn{1}{c}{$\Pi_{l}$} & Profits recorded in $l$. National Accounts, methodology from \citet{torslov_missing_2022}. & Appendix \ref{app_data_profits} \\

			\multicolumn{1}{c}{$\iota_{l}$} & Profits-sales gap. Computed using: $\iota_{l}=\sigma \frac{\Pi_l}{\sum_i X_{ill}}$.  & Section \ref{sec:data} \\

			\multicolumn{1}{c}{$\sigma$} & Elasticity of substitution. Set to 6.88 following a 17\% markup in French firm-level data (\citealp{deloecker2012markups} methodology). & Section \ref{sec:data} \\

			\multicolumn{1}{c}{$\tilde{\upsilon}_{1}$} & Elasticity of the tax base. Estimated following equation (\ref{gravitynoPSmodel_ppml}) using Orbis data. Set to 2.05 & Section \ref{main_elasticities} \\		

			\multicolumn{1}{c}{$\tilde{\upsilon}_{2}$} & Elasticity of profit shifting. Estimated following equation (\ref{gravityPSmodel_ppml}) and Orbis data. Set to 6.83 & Section \ref{main_elasticities} \\	
			\multicolumn{1}{c}{$k$} & Non-linear elasticity of profit shifting shape parameter. Calibrated to match data moments. Set to 0.66 & Section \ref{app_nonlin} \\	 
			\noalign{\smallskip}\noalign{\smallskip}	
			\textit{Frictions} & & \\ \cline{1-1}
			\noalign{\smallskip}\noalign{\smallskip} 
			\multicolumn{1}{c}{$\gamma_{il}$} & Multinational production frictions. Backed out from $X_{ill}$ shares.  & Appendix \ref{appendix_ps_frictions} \\	

			\multicolumn{1}{c}{$\tau_{ln}$} & Trade frictions. Backed out from $X_{ln}$ shares.  & Appendix \ref{appendix_ps_frictions} \\		

			\multicolumn{1}{c}{$\alpha_{lh}$} & Profit shifting frictions. Backed-out from $X_{ilh}$.  & Section \ref{main_frictions}, Appendix \ref{appendix_ps_frictions}  \\

			 \bottomrule
	\end{tabular}}

\end{table}

\clearpage

\begin{figure}[htbp]
	\begin{subfigure}[t]{.495\textwidth}
		\centering
		\includegraphics[width=\linewidth]{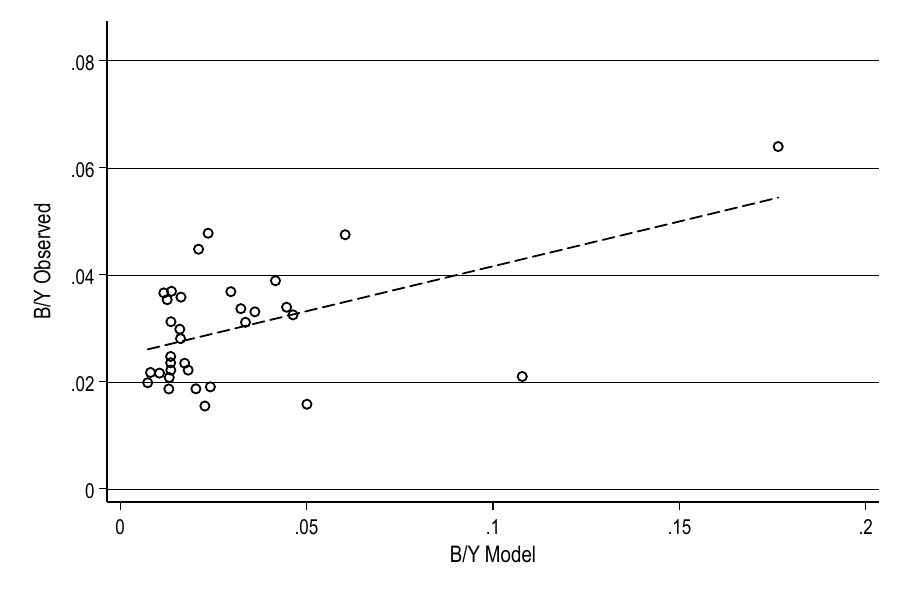} 
		\caption{In non-haven countries} 
	\end{subfigure}
\hfill
 \begin{subfigure}[t]{.495\textwidth}
		\centering
		\includegraphics[width=\linewidth]{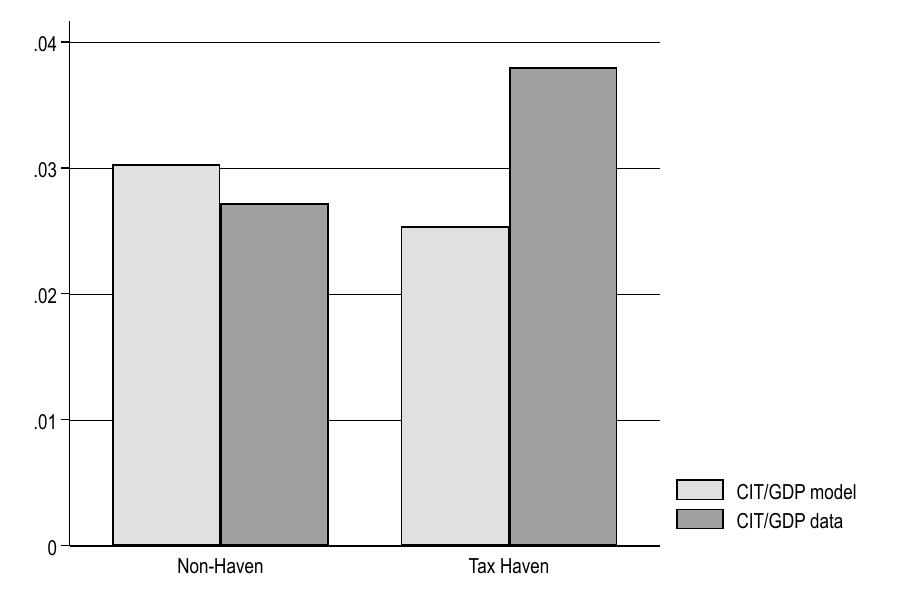} 
		\caption{Tax havens vs. non-havens} 
	\end{subfigure}
 \caption{Tax revenues over GDP ($B/Y$): data versus model}
 \label{validation_tax_gdp}
	\caption*{\footnotesize Note: Data on corporate tax revenues over GDP is obtained from UNU-WIDER's Government Revenue Dataset. We select the variable ``Taxes on income, profits and capital gains from corporation'' (corresponding to OECD item 1200). The figure in Panel (a) is drawn for the sample of non-haven countries. Panel (b) compares tax havens and non-havens}
\end{figure}

\begin{figure}[htbp]
		\centering
		\includegraphics[width=0.675\linewidth]{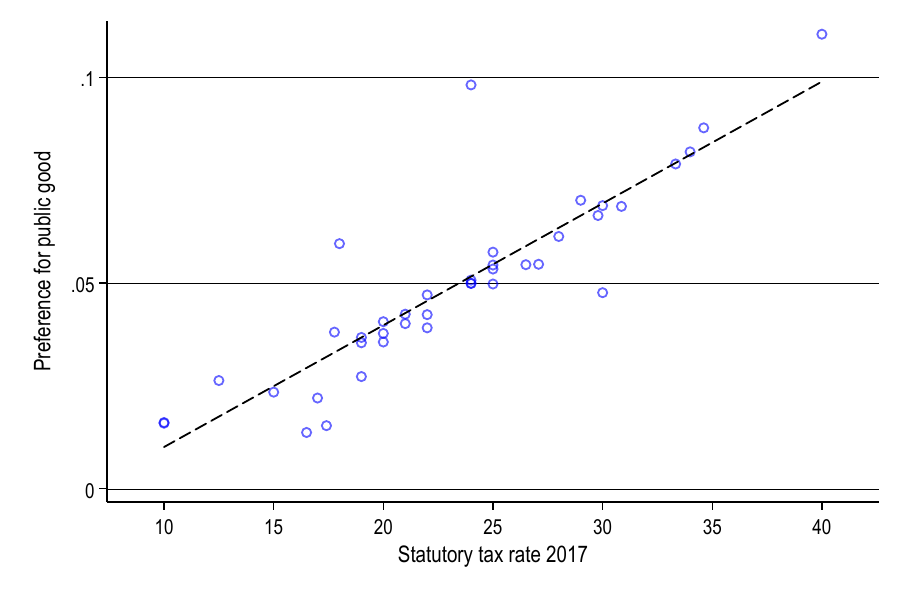} 
 \caption{Prefernce for public goods and the tax rate}
 \label{beta_and_taxrate}
	\caption*{\footnotesize Note: This plots shows the correlation between the estimated preference for public goods and the statutory tax rate in 2017. The correlation between both variables is 0.88.}
\end{figure}

\subsection{Illustrating model mechanisms} \label{app_mechanism}

Table \ref{scenario_appendix} illustrates the impact of different scenarios on tax revenues, profit shifting, real production, real income and welfare.

\begin{table}[htbp]
	\centering
	\caption{Impact of different counterfactual scenarios}\label{scenario_appendix}
\scalebox{.9}{	\begin{tabular}{lc ccccc} \toprule
		&&\multicolumn{5}{c}{\% change in ...} \\ 
   \noalign{\smallskip}  \cline{2-7}   \noalign{\smallskip}

	\textbf{Scenario}   &		&	 Tax         & Profit 	&	Real 	&	Consumer 	& Welfare\\
	                               	&		&	 revenues   &  Shifting	&		Production        &	 Real Income	&\\
 \noalign{\smallskip}  \cline{1-2}   \noalign{\smallskip}
5\% decrease statutory tax rate	&&	-4.21	&	-6.36	&	0.34	&	0.44	&	-0.04	\\
Closing Singapore	&&	0.29	&	-3.99	&	-0.02	&	-0.02	&	0.01	\\
Effective anti-abuse regulations	&&	7.24	&	-100	&	-0.41	&	-0.52	&	0.25 \\     
\bottomrule
	\end{tabular}}
\end{table}

\paragraph{Unilateral tax reform.}
We illustrate the percentage change of a unilateral reduction of 5\% in the U.S. corporate tax rate (from 40\% to 38\%) on five outcomes in Table \ref{scenario_appendix}. We show that it increases real income by 0.44\% while slightly reducing welfare by 0.04\%. The changes in tax revenues, profit shifting, and real production are presented in Table \ref{scenario_appendix}.

\begin{figure}[htbp]
	\begin{subfigure}[t]{.495\textwidth}
		\centering
		\includegraphics[width=\linewidth]{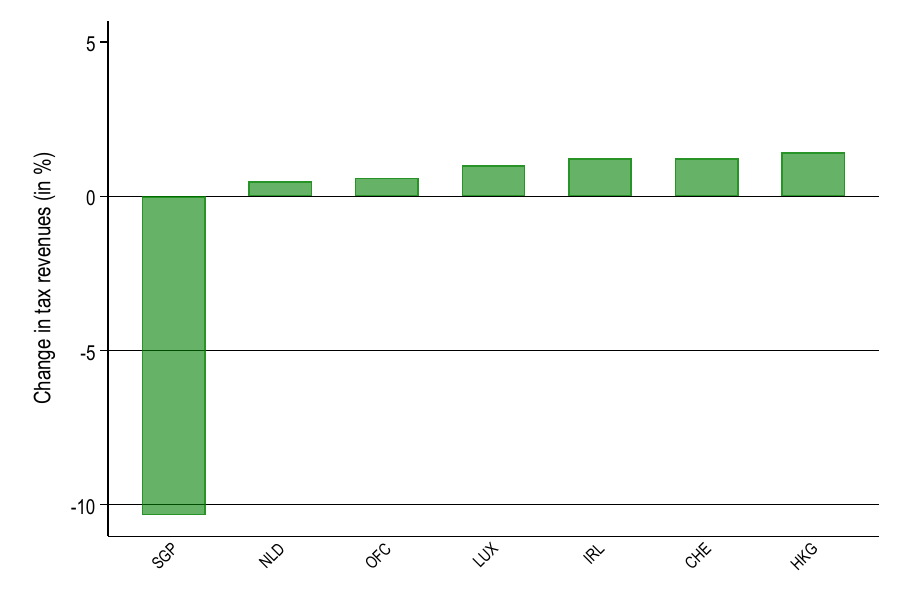} 
		\caption{Impact on tax revenues in tax havens} 
  	\label{fig:close_sgp_dtax_havens}
	\end{subfigure}
\hfill
 \begin{subfigure}[t]{.495\textwidth}
		\centering
		\includegraphics[width=\linewidth]{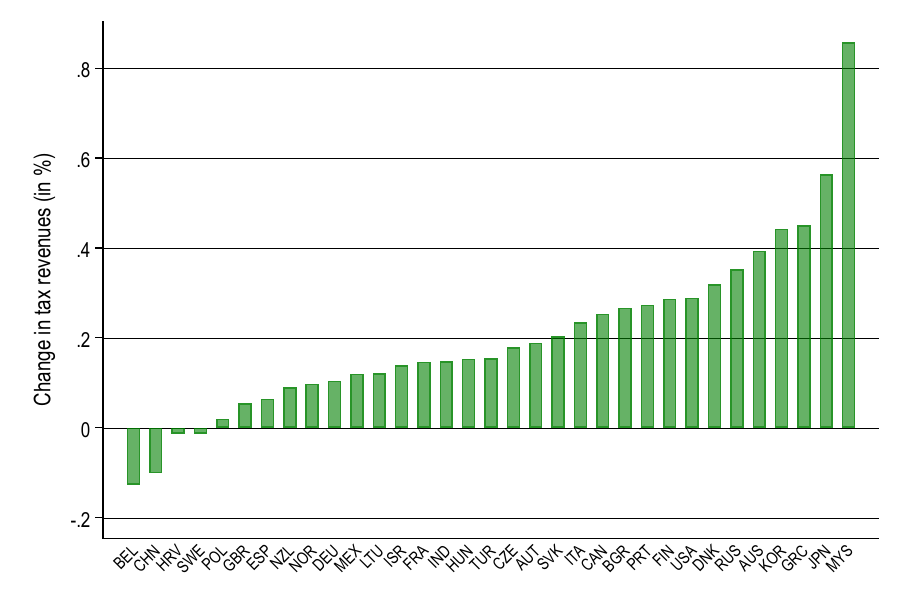} 
		\caption{Impact on tax revenues in non-havens} 
  	\label{fig:close_sgp_dtax_nonhavens}
	\end{subfigure}
 \caption{Effect of closing Singapore as a tax haven}
 \label{fig:close_sgp}
	\caption*{\footnotesize Note: These two histograms illustrate the impact of closing the access to Singapore as a tax haven. Panel (a) shows how this reform would impact tax revenues in tax havens. Panel (b) shows how this reform would impact tax revenues in non-havens}
\end{figure}

\paragraph{Closing a tax haven.}
In Table \ref{scenario_appendix}, we examine the impact of closing Singapore on U.S. tax revenues, GDP, profit shifting, consumers’ real income, and welfare. Figure \ref{fig:close_sgp} shows the impact of this reform on i) tax revenues across tax havens (Panel a) and on ii) tax revenues across non-tax havens (Panel b). 

\paragraph{Effective anti-abuse laws.}
\emph{What are the effects of implementing multilateral effective anti-abuse laws in non-haven countries}? 
 
Assuming costless implementation, Table \ref{scenario_appendix} shows the results for the U.S. policy. 
The policy raises the U.S. effective tax rate and generates an 7.24\% increase in tax revenues while reducing production by 0.41\%. 
Consumers' real income decreases by 0.52\%, but the welfare effect is positive (0.25\%) due to a large increase in corporate tax revenues. 

The increase in tax revenues is due to reduced profit shifting and reallocation of production. 
High-tax countries benefit from tax havens, and non-haven countries might use lax enforcement of anti-abuse laws to attract mobile firms (for instance \citealp{altshuler2005three}, \citealp{praise_hong_2010} or \citealp{multinationals_dharmapala_2020}).

\subsection{Quantification of equilibrium effects \label{app_equilibrium_effects}}

This subsection illustrates the quantification of equilibrium effects in long-run minimum taxation scenarios. 
We compute what would have been the effect of these reforms if we did not allow the tax base to adjust. 
This is tantamount to forcing production choices, including location and profit shifting to remain unchanged after the introduction of a minimum tax. The goal of this exercise is to quantify the mismeasurement of the reforms' impact if we were just considering the mechanical tax rate effects.

Post-reform tax revenues of country $k$ are given by
\begin{align*}
	B'_{k}=\sum_{i,l,h}t_{ilh}^{\prime g_k} \mathcal{N}_{i}' \mathbb{P}_{ilh}' \frac{w'_{i}f_E}{1-t'_{ilh}}.
\end{align*}
This includes a mechanical adjustment of the tax rate $t_{ilh}^{\prime g_k}$ and an equilibrium response of the tax base $\mathcal{N}_{i}' \mathbb{P}_{ilh}' \frac{w'_{i}f_E}{1-t'_{ilh}}$. 
We define a counterfactual tax revenue stream in which we force the tax base not to move. 
Formally
\begin{align*}
	B^{\prime TRE}_{k}=\sum_{i,l,h}t_{ilh}^{\prime g_k} \mathcal{N}_{i} \mathbb{P}_{ilh} \frac{w_{i}f_E}{1-t'_{ilh}},
\end{align*}
where we use the superscript $TRE$ to denote the tax rate effect. For clarity, we separate the change in tax revenues between those coming from the corporate income tax base and those coming from the minimum tax base. 

Along similar lines, we note that the real income of country $k$ post-reform is given by $\frac{Y'_{k}}{P'_k}	=	\frac{w'_{k}L_{k}+B'_{k} + \Delta'_k}{P'_k} $. 
This is clearly driven by a tax revenues effect, $B'_k$, as well as the rest of the equilibrium adjustment, for example, the changes in wages and prices. 
We can then define a mechanical real income response as $\frac{w_{k}L_{k}+B^{\prime TRE}_{k} + \Delta_k}{P_k}$, where only the tax revenues are allowed to move and only through mechanical tax rates effects.

\paragraph{Results}

Table \ref{ge_revenues_welfare} breaks down the change in tax revenues for different scenarios under a 15\% minimum tax rate. 
We present results for the implementation of a unilateral, residence-based minimum tax by the U.S. Our model predicts a 3.89\% increase in tax revenues. 
Since the corporate tax rate remains unchanged, the direct effect on the CIT base is zero. However, the reduction in profit shifting increases tax revenues by 2.72\%, all else being equal. This is partially offset by a real effect on the CIT base of -0.16\%.

New revenues are also collected on the minimum tax base, with the mechanical tax rate effect raising revenues by 2.2\%. This effect is reduced by -0.86\% due to decreased profit shifting after the reform. Additionally, changes in production choices slightly lower tax revenues on this base (-0.01\%). 

When source-based minimum taxation is applied, the additional tax revenues from reduced profit shifting are larger than in the residence-based scenario.

When tax havens adjust their tax rate in response to the implementation of the minimum tax, no MNE is taxed below the minimum rate and then no minimum tax is levied. The change in tax revenues only comes from changes in the profit shifting behavior and in the production choices of MNEs. This change is the same as in other multilateral scenarios (+2.84\%) as the world distribution of tax rates is the same.


\begin{table}[H]
	\centering
	\caption{Profit-shifting and GE effects of a 15\% minimum tax on tax revenues}
	\label{ge_revenues_welfare}
	\begin{tabular}{@{\extracolsep{.1em}}lp{2.8cm}p{1.7cm}p{1cm}p{1.3cm}p{1.7cm}p{1cm}p{1.3cm}} \toprule
		
		\multirow{3}{*}{Counterfactual} & Change in real tax revenues (in \%)& \multicolumn{3}{p{4.4cm}}{CIT Base - Contribution (in \%)} & \multicolumn{3}{p{4.4cm}}{Min. tax Base - Contribution (in \%)} \\ \noalign{\smallskip}   \cline{3-5}   \cline{6-8} \noalign{\smallskip}
		&& Tax Rate Effect &  PS effect & Real effect & Tax Rate Effect &  PS effect & Real effect \\
		& \multicolumn{1}{c}{(1)} & \multicolumn{1}{c}{(2)} & \multicolumn{1}{c}{(3)} & \multicolumn{1}{c}{(4)} & \multicolumn{1}{c}{(5)} & \multicolumn{1}{c}{(6)}& \multicolumn{1}{c}{(7)} \\ \midrule 

		\textit{15\% min. tax} &&&&&&& \\ \noalign{\smallskip} \cline{1-1} \noalign{\smallskip}
Unil. Residence	&	\multicolumn{1}{c}{3.89}	&	\multicolumn{1}{c}{0}	&	\multicolumn{1}{c}{2.72}	&	\multicolumn{1}{c}{-0.16}	&	\multicolumn{1}{c}{2.20}	&	\multicolumn{1}{c}{-0.86}	&	\multicolumn{1}{c}{-0.01}		\\
Unil. Source	&	\multicolumn{1}{c}{3.97}	&	\multicolumn{1}{c}{0	}&	\multicolumn{1}{c}{2.88} &	\multicolumn{1}{c}{-0.23}	&	\multicolumn{1}{c}{2.16}	&	\multicolumn{1}{c}{-0.84}	&	\multicolumn{1}{c}{0.00}		\\
Multi.  Residence	&	\multicolumn{1}{c}{4.05	}&	\multicolumn{1}{c}{0	}&	\multicolumn{1}{c}{2.84}	&	\multicolumn{1}{c}{-0.12}	&	\multicolumn{1}{c}{2.20}	&	\multicolumn{1}{c}{-0.86}	&	\multicolumn{1}{c}{0.00	}	\\
Multi. Source	&	\multicolumn{1}{c}{4.04	}&	\multicolumn{1}{c}{0} &	\multicolumn{1}{c}{2.84}	&	\multicolumn{1}{c}{-0.12}	&	\multicolumn{1}{c}{2.16}	&	\multicolumn{1}{c}{-0.83}	&	\multicolumn{1}{c}{-0.01}		\\
TH adjustment	&	\multicolumn{1}{c}{2.71}	&	\multicolumn{1}{c}{0}&	\multicolumn{1}{c}{2.84} &	\multicolumn{1}{c}{-0.12}	&	\multicolumn{1}{c}{0} &	\multicolumn{1}{c}{0}	&	\multicolumn{1}{c}{0}	\\
		\bottomrule
		
\end{tabular}
\caption*{\footnotesize Note: Results in this table are provided for the United States. Column (1) corresponds to the effect computed using our quantitative model.``Tax Rate Effect'' in columns (2), and (5) indicates the reform's effect as computed assuming no change in profit-shifting activity or production location. ``PS effect'' in columns (3), and (6) indicates the change in tax revenues due to the change in the profit-shifting strategy of MNEs all other things being equal. ``Real effect'' in column (4), and (7) indicates the change in tax revenues due to the change in the location strategy of MNEs all other things being equal.}
\end{table}		

The table highlights the importance of considering profit-shifting and real effects when predicting the impact of tax reforms on tax revenues and real income. 
It is also worth mentioning that unilateral and multilateral scenarios lead to identical results concerning the change in tax revenues from the minimum tax base in estimations that do not consider the reallocation of real and paper profits (those estimates would rely on column 5 only). 
These tables capture relevant channels that a pure "accounting" exercise would miss.

\newpage 
\subsection{DBCFT}\label{dbcft_app}

\paragraph{Implementation}
We first update the definition of market market access as follows:

\[
\Xi_{l}^{1-\sigma}=\sum_{n}\ensuremath{\Xi_{ln}^{1-\sigma}}=\sum_{n}\tau_{ln}^{1-\sigma}\left(1+s_{ln}\right)^{\sigma-1}\left(\frac{1}{1+tr_{ln}}\right)^{\sigma}Y_{n}P_{n}^{\sigma-1}.
\]

Simplifying with 
\[
\Xi_{l}^{1-\sigma}=\sum_{n}\ensuremath{\Xi_{ln}^{1-\sigma}}=\sum_{n}\tau_{ln}^{1-\sigma}\left(1+s_{l}\right)^{\sigma-1}\left(\frac{1}{1+tr_{n}}\right)^{\sigma}Y_{n}P_{n}^{\sigma-1},
\]
we obtain that post-tax profits are given by \[
\left(1-t_{ilh}\right)\frac{\iota_{l}}{\sigma}\left(\frac{\sigma}{\sigma-1}\frac{\gamma_{il}\alpha_{lh}}{\varphi}w_{l}\Xi_{l}\right)^{1-\sigma}\quad\mbox{ where } \Xi_{l}=\sum_{n}\frac{\tau_{ln}^{1-\sigma}\left(1+s_{l}\right)^{\sigma-1}}{\left(1+tr_{n}\right)^{\sigma}}\frac{Y_{n}}{P_{n}^{1-\sigma}},
\]
and where $n=l\Rightarrow tr_{n}=s_{n}$.

The change in trade costs here are given by $\hat{\tau}_{ln}=\frac{1}{1+s_{l}}\left(\frac{1}{1+tr_{n}}\right)^{\frac{\sigma}{1-\sigma}}$.
It determines the new $\mathbb{P}_{ilh}$ as a function of $Y_{n}$,$P_{n}$, $w_{l}$.

The share of production by firms from $i$ in $l$ shifting
in $h$ is undistorted with regards to $\mathbb{P}_{ilh}$:
\[
\beta_{ilh}=\frac{\mathbb{P}_{ilh}/(\iota_{l}(1-t_{ilh}))}{\mathbb{P}_{ilh}/(\iota_{l}(1-t_{ilh}))}.
\]

In the labor market, we have
\[
w_{i}L_{i}=\ensuremath{N_{i}w_{i}f_{E}}+\ensuremath{\frac{\sigma-1}{\sigma}}(1+\ensuremath{s_{i}})\ensuremath{Q_{i}}.
\]

Using 
\[
\frac{X_{ln}}{X_{l}}=\frac{(1+tr_{n})^{-\sigma}(1+s_{l})^{\sigma-1}\tau_{ln}^{1-\sigma}\left(Y_{n}/P_{n}^{1-\sigma}\right)}{\Xi_{l}^{1-\sigma}}
\],
the price index is (implicitly) given by:
\[
P_{n}^{1-\sigma}=(1+t\ensuremath{r_{n})^{1-\sigma}\sum_{l}\frac{X_{l}}{\Xi_{l}^{1-\sigma}}\tau_{ln}^{1-\sigma}\left(1+s_{l}\right)^{\sigma-1}},
\]
where the value of production in $k$ is
\[
Q_{l}=\ensuremath{\sigma}\ensuremath{\sum_{k,h}N_{k}\frac{\mathbb{P}_{klh}w_{k}f_{E}}{\left(1-t_{klh}\right)\iota_{l}}}.
\]

\paragraph{Additional results}
Table \ref{dbcft_breakdown} decomposes the change in real tax revenues (column 1) and the change in real GDP (column 4) when DBCFT is implemented. 

The change in real tax revenues is separated between the taxation of domestic sales (when $l=n$) and the border adjustment. 
The border adjustment corresponds to the difference between additional tax revenues from taxing imports and tax expenses from subsidizing exports. Columns (2) and (3) add up to the change in tax revenues in column (1).

The change in real GDP is decomposed between the contribution of domestic sales, and the contribution of foreign multinationals' sales to real GDP. Columns (5) and (6) add up to the change in GDP in column (4).

\begin{table}[!ht]
	\centering
	\caption{Breakdown of the increase in real tax revenues}
	\label{dbcft_breakdown}
\scalebox{0.9}{\begin{tabular}{l|ccc|ccc} \toprule
   DBCFT rate	&	\% Change &	Contrib.&	Contrib. Border	&	\% Change &	Contrib.&	Contrib. 	\\ 
    $(tr)$&real tax rev.&domestic tax& Adjustment&real GDP&domestic sales& Foreign MP \\
   & (1) & (2) & (3) & (4) & (5) & (6) \\ \midrule
\textbf{USA} &&&&&& \\
5\%	&	-78.4	&	-86.95	&	8.55	&	4.49	&	4.44	&	0.06	\\
10\%	&	-57.85	&	-74.04	&	16.19	&	4.06	&	3.62	&	0.43	\\
20\%	&	-19.62	&	-48.75	&	29.12	&	2.95	&	1.46	&	1.49	\\
30\%	&	14.82	&	-24.35	&	39.17	&	1.54	&	-1.71	&	3.24	\\
33.3\%$^\star$	&	25.36	&	-16.46	&	41.82	&	0.99	&	-3.1	&	4.09	\\
40\% (BAT)	&	44.61	&	-1.12	&	45.73	&	-0.21	&	-6.68	&	6.46	\\ \midrule
\textbf{Japan} &&&&&& \\
25\%$^\star$	&	-38.99	&	-25.96	&	-13.02	&	1.58	&	-2.66	&	4.25	\\
31\% (BAT)	&	-29.97	&	-9.27	&	-20.7	&	1.22	&	-5.16	&	6.38	\\
\bottomrule
 \end{tabular}}
 \caption*{\footnotesize Note: This table breaks down the change in real tax revenues and in real GDP. The change in tax revenues is broken down between the contribution of domestic revenues (as compared with B, the tax revenues collected at the initial equilibrium) and the contribution of the border adjustment. The later is presented as the net effect between the revenues coming from the taxation of imports and the revenues spent by subsidizing exports.  Columns (2) and (3) add up to the change in tax revenues in column (1). The change in GDP is broken down between the contribution of domestic sales and the contribution of foreign multinationals' sales. Columns (5) and (6) add up to the change in GDP in column (4). The $^\star$ represents the optimal rate.}
\end{table}

\printbibliography[heading=subbibintoc]
\end{refsection}

\end{document}